\documentclass[aps,prx,twocolumn,superscriptaddress,
groupedaddress,10pt]{revtex4}
\usepackage{amsmath}
\usepackage{amssymb}
\usepackage{graphicx}
\usepackage{epsfig}
\usepackage{dcolumn}
\usepackage{bm}
\usepackage{bm}
\usepackage{blindtext}
\usepackage{natbib}
\usepackage{siunitx} 

\usepackage{multirow}
\usepackage{bbm}

\usepackage[usenames,dvipsnames]{xcolor}
\usepackage{amsthm}
\usepackage{booktabs}
\usepackage{hyperref}
\usepackage{tikz}
\usetikzlibrary{calc}
\usepackage[printwatermark]{xwatermark}
\usepackage{mathtools}

\def\be{\begin{equation}} \def\ee{\end{equation}}
\def\bea{\begin{eqnarray}} \def\eea{\end{eqnarray}}

\def\nn{\nonumber}

\begin{document}
\title{
Counter-rotating spiral, zigzag,  and 120$^\circ$ orders from coupled-chain analysis of Kitaev-Gamma-Heisenberg model, and relations to honeycomb iridates
}

\author{Wang Yang}
\affiliation{Department of Physics and Astronomy and Stewart Blusson Quantum Matter Institute,
University of British Columbia, Vancouver, B.C., Canada, V6T 1Z1}

\author{Alberto Nocera}
\affiliation{Department of Physics and Astronomy and Stewart Blusson Quantum Matter Institute, 
University of British Columbia, Vancouver, B.C., Canada, V6T 1Z1}

\author{Chao Xu}
\affiliation{Kavli Institute for Theoretical Sciences, University of Chinese Academy of Sciences, Beijing 100190, China}

\author{Hae-Young Kee}
\affiliation{Department of Physics, University of Toronto, Toronto, Ontario M5S 1A7, Canada}
\affiliation{Canadian Institute for Advanced Research, CIFAR Program in Quantum Materials, Toronto, Ontario M5G 1M1, Canada}

\author{Ian Affleck}
\affiliation{Department of Physics and Astronomy and Stewart Blusson Quantum Matter Institute, 
University of British Columbia, Vancouver, B.C., Canada, V6T 1Z1}

\begin{abstract}

We study the nearest neighboring spin-1/2 Kitaev-Heisenberg-Gamma ($KJ\Gamma$) model on the honeycomb lattice
in the parameter region of ferromagnetic (FM) Kitaev and antiferromagnetic (AFM) Heisenberg couplings relevant for honeycomb iridates, using a coupled-chain analysis. 
Starting from the gapless Luttinger liquid phase of a decoupled $KJ\Gamma$ chain, 
the inter-chain interactions in the two-dimensional model is treated within a self-consistent mean field approach based on the Luttinger liquid theory. 
In the FM Gamma region, our analysis recovers the  reported 120$^\circ$ magnetic order,
previously obtained by classical analysis and exact diagonalization method. 
On the other hand, new physics is revealed in the AFM Gamma region, where three magnetic orders are found, 
including 120$^\circ$, commensurate counter-rotating spiral, and zigzag orders. 
Interestingly, the two first order phase transition lines separating these three magnetic orders merge at a single point at $K = -2 \Gamma$ and $J=0$, which is predicted to be a quantum critical point. 
The current theory captures the experimentally observed counter-rotating spiral order in $\alpha$-Li$_22$IrO$_3$ and the zigzag order in Na$_2$IrO$_3$, 
thereby indicating that the spin-1/2 $KJ\Gamma$  model may serve as a minimal model for honeycomb iridates. 
Limitations of the mean field theory presented in this work and the $J \rightarrow 0$ regime are also discussed.

\end{abstract}

\maketitle

\section{Introduction}
\label{sec:intro}

The Kitaev spin-1/2 model on the honeycomb lattice is an exactly solvable spin model with bond-dependent Ising interactions \cite{Kitaev2006}.
The braiding statistics of the fractionalized excitations emerging from this model can be used for realizing topological quantum computations \cite{Kitaev2006,Nayak2008}.
For this reason,  material realizations of the Kitaev model have attracted intense research attentions in the past decade \cite{Witczak-Krempa2014,Rau2016,Winter2017,Hermanns2018} 
on both theoretical and experimental sides 
\cite{Jackeli2009,Chaloupka2010,Singh2010,Liu2011,Kimchi2011,Price2012,Singh2012,Choi2012,Ye2012,Chaloupka2013,Gao2013,Foyevtsova2013,Gretarsson2013,Plumb2014,Rau2014,Biffin2014,Biffin2014_2,Manni2014,Kimchi2014,Sizyuk2014,Reuther2014,Rau2014b,Kim2015,Johnson2015,Chaloupka2015,Kimchi2015,Chun2015,Winter2016,Williams2016,Kimchi2016,Baek2017,Leahy2017,Sears2017,Wolter2017,Zheng2017,Rousochatzakis2017,Ran2017,Wang2017,Kasahara2018,Catuneanu2018,Gohlke2018,Motome2020,Chern2020,Gohlke2020,Liu2021,Rayyan2021,Liu2022}.
Honeycomb iridates are a class of Kitaev materials, including
Na$_2$IrO$_3$ \cite{Singh2010} and $\alpha$-Li$_2$IrO$_3$ \cite{Kobayashi2003} among others.
So far, the proposed candidate materials  are experimentally observed to be magnetically ordered at sufficiently low temperatures.
For example, zigzag magnetic order has been found in Na$_2$IrO$_3$ \cite{Choi2012,Liu2011,Ye2012},
whereas $\alpha$-Li$_2$IrO$_3$ has a counter-rotating spiral order \cite{Williams2016}. 
Hence, one of the central questions in the field of Kitaev materials is to understand why different magnetic orders appear in these materials \cite{Kimchi2011,Singh2012,Foyevtsova2013,Sizyuk2014,Reuther2014,Kimchi2015}. 

On the theory side, a variety of generalized Kitaev spin models have been used to model and analyze Kitaev materials, which contain interactions beyond the pure Kitaev coupling \cite{Jackeli2009,Chaloupka2010,Rau2014,Kimchi2014,Wang2017}.
These additional interactions arise from exchange processes among the spin-orbit coupled orbitals in the underlying lattices, 
and are natural from a symmetry point of view, since in principle, any interaction compatible with the lattice symmetries is inevitable in real materials. 
The simplest generalized Kitaev spin  models are those which only contain interactions up to the nearest neighboring level.
The minimal model compatible with the lattice symmetries and having only nearest neighboring  interactions
is the Kitaev-Heisenberg-Gamma ($KJ\Gamma$) model  \cite{Rau2014}, which in addition to the Kitaev and Heisenberg couplings, contains an off-diagonal symmetric Gamma term. 
In the Kitaev candidate materials, theories and experiments have established the facts that the Kitaev interaction is ferromagnetic (FM).
In a recent work of Ref. \cite{Liu2022}, 
it has been proposed that while Gamma  is antiferromagnetic (AFM), the Heisenberg coupling in the $\alpha$-Li$_2$IrO$_3$ material is  AFM in nature,
different from another candidate $\alpha$-RuCl$_3$ with FM Heisenberg interaction. 

\begin{figure*}[htbp]
\centering
\includegraphics[width=8.9cm]{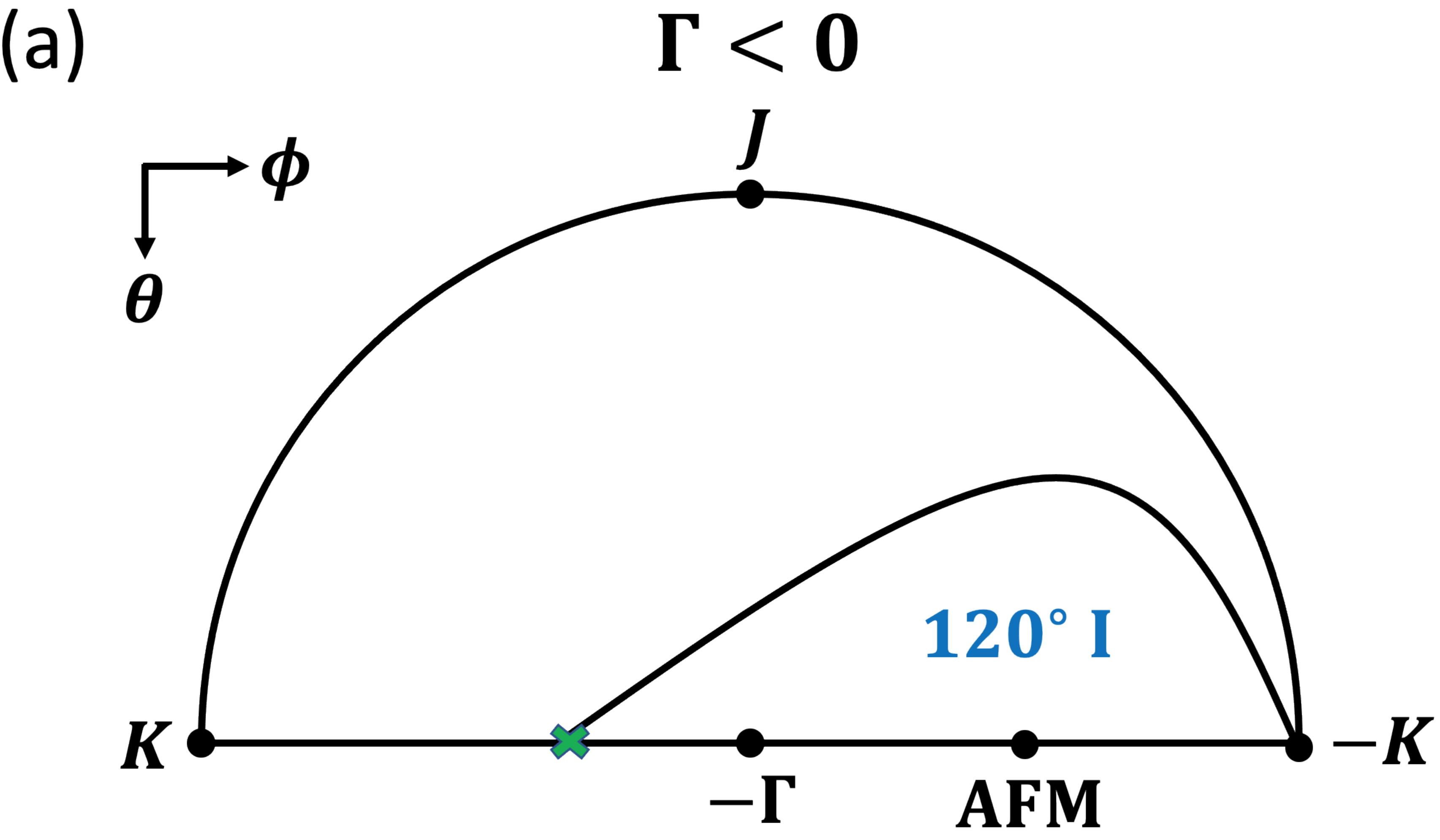}
\includegraphics[width=8.9cm]{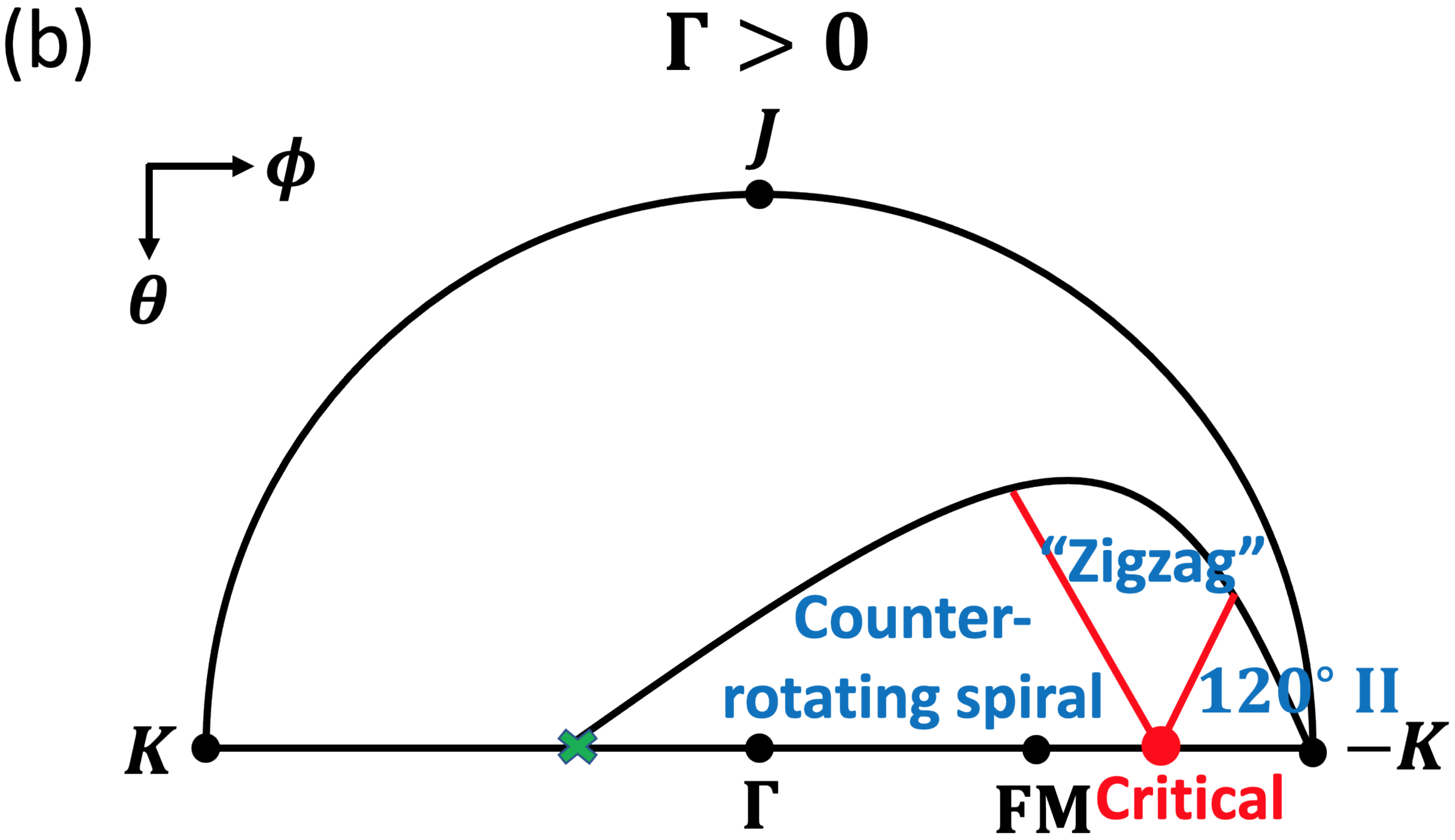}
\caption{Schematic plot of the  phase diagram of the anisotropic  spin-1/2  Kitaev-Heisenberg-Gamma model on the 2D honeycomb lattice in the region (a) $K<0$, $\Gamma<0$, $J>0$,
(b) $K<0$, $\Gamma>0$, $J>0$.
By assuming an absence of phase transition from the anisotropic to isotropic cases, the results can be applied to the isotropic spin-1/2  Kitaev-Heisenberg-Gamma model  as well. 
In (a,b), the parametrization is $J=\cos(\theta)$, $K=\sin(\theta)\cos(\phi)$, $\Gamma=\sin(\theta)\sin(\phi)$; 
the AFM and FM  points have hidden SU(2) symmetries  revealed by the six-sublattice rotation \cite{Chaloupka2015}. 
In (b), a quotation is put on ``zigzag" since this phase arises from a subdominant channel of instability (see Sec. \ref{subsec:uC_0} for detailed discussions); 
 the phase transition lines represented by the two red lines are first order phase transitions;
the two red lines merge to a single point $K=-2\Gamma$ on the equator (i.e., $J=0$),
which is predicted to be a quantum critical point.
} 
\label{fig:phase-Gamma}
\end{figure*}

The determination of models and parameters for real Kitaev materials  has been a challenge in the community, 
and one approach was taken from investigating the corresponding quasi-one-dimensional (1D) models \cite{Sela2014,Gruenewald2017,Agrapidis2018,Agrapidis2019,Catuneanu2019,You2020,Yang2019,Yang2020,Yang2020b,Yang2021b,Luo2021,Luo2021b,Sorensen2021,Yang2022a,Yang2022_2,Yang2022,Yang2022d}, which may give insights into the two-dimensional (2D) limit.
Unlike the typical theoretical difficulties in 2D, 
1D has the advantage that there are many powerful analytical and numerical methods \cite{Haldane1981,Haldane1981a,Belavin1984,Knizhnik1984,Affleck1985,Affleck1988,Affleck1995a,White1992,White1993,Schollwock2011}.
Besides providing hints for 2D,  1D generalized Kitaev models are interesting on their own, since they contain rich strongly correlated physics, 
including emergent conformal symmetry \cite{Yang2019,Yang2022d}, nonlocal string orders \cite{Catuneanu2019,Sorensen2021}, and magnetic orders which break exotic nonsymmorphic symmetries \cite{Yang2019,Yang2020,Yang2020b,Yang2021b}.

In this work, we focus on the experimentally relevant parameter region of
FM Kitaev, AFM Gamma and AFM Heisenberg couplings of the spin-1/2 $KJ\Gamma$ model on the honeycomb lattice, and study both signs of the Gamma interaction. 
While our interest is in the AFM Gamma region, we will also present FM Gamma to make proper comparisons.
The strategy is to take the Luttinger liquid phase in a decoupled spin-1/2 $KJ\Gamma$ chain \cite{Yang2020} as the starting point, 
and consider a system of weakly coupled chains on the honeycomb lattice,
which form an anisotropic  $KJ\Gamma$ model
where the inter-chain interactions can be treated in a self-consistent mean field approach. 
The obtained phase diagram may be applied to the isotropic  $KJ\Gamma$ model 
by assuming an absence of phase transition from weak to intermediate inter-chain interactions,
which is worth for further numerical tests. 

\begin{figure*}[htbp]
\centering
\includegraphics[width=5.8cm]{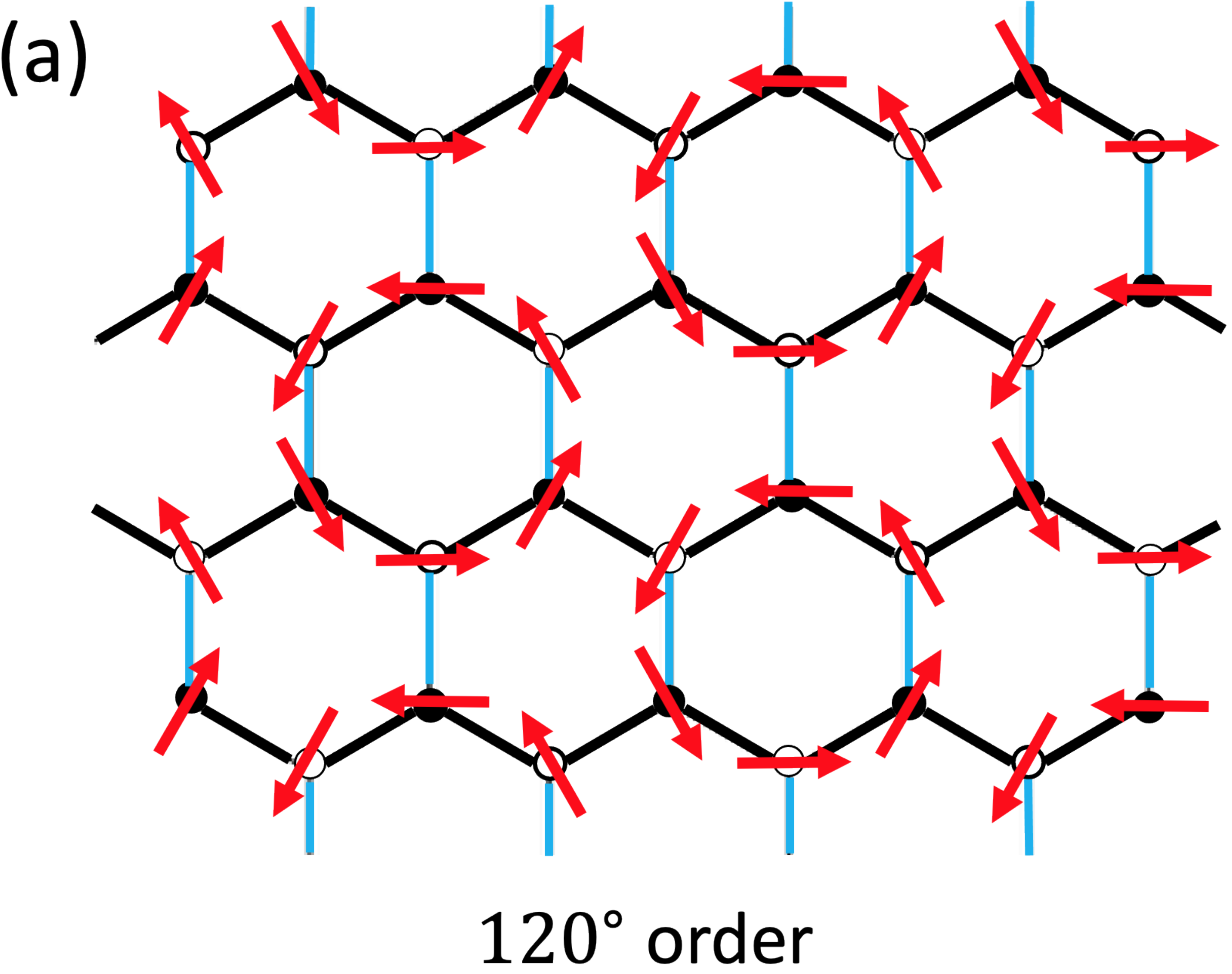}
\includegraphics[width=5.8cm]{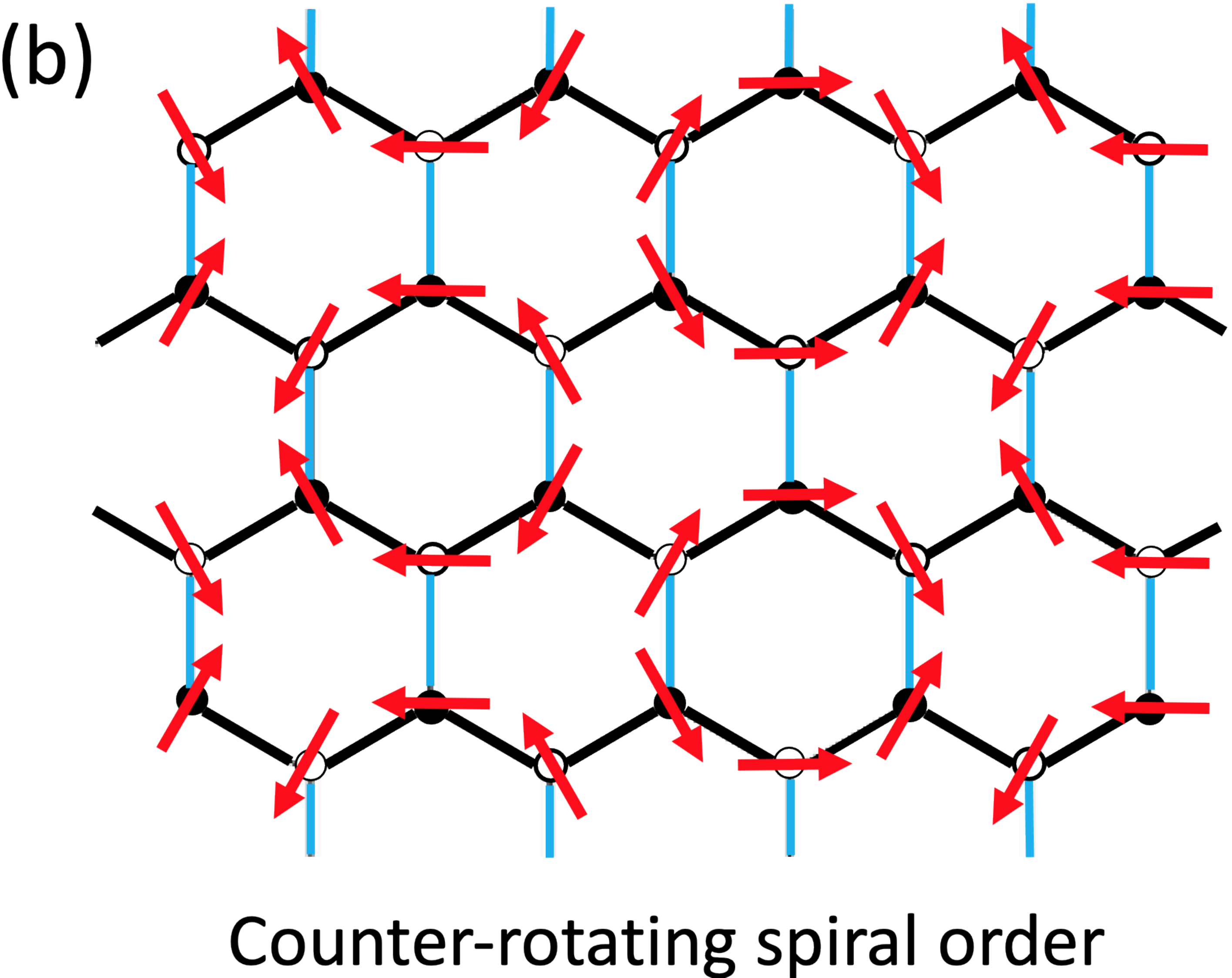}
\includegraphics[width=5.8cm]{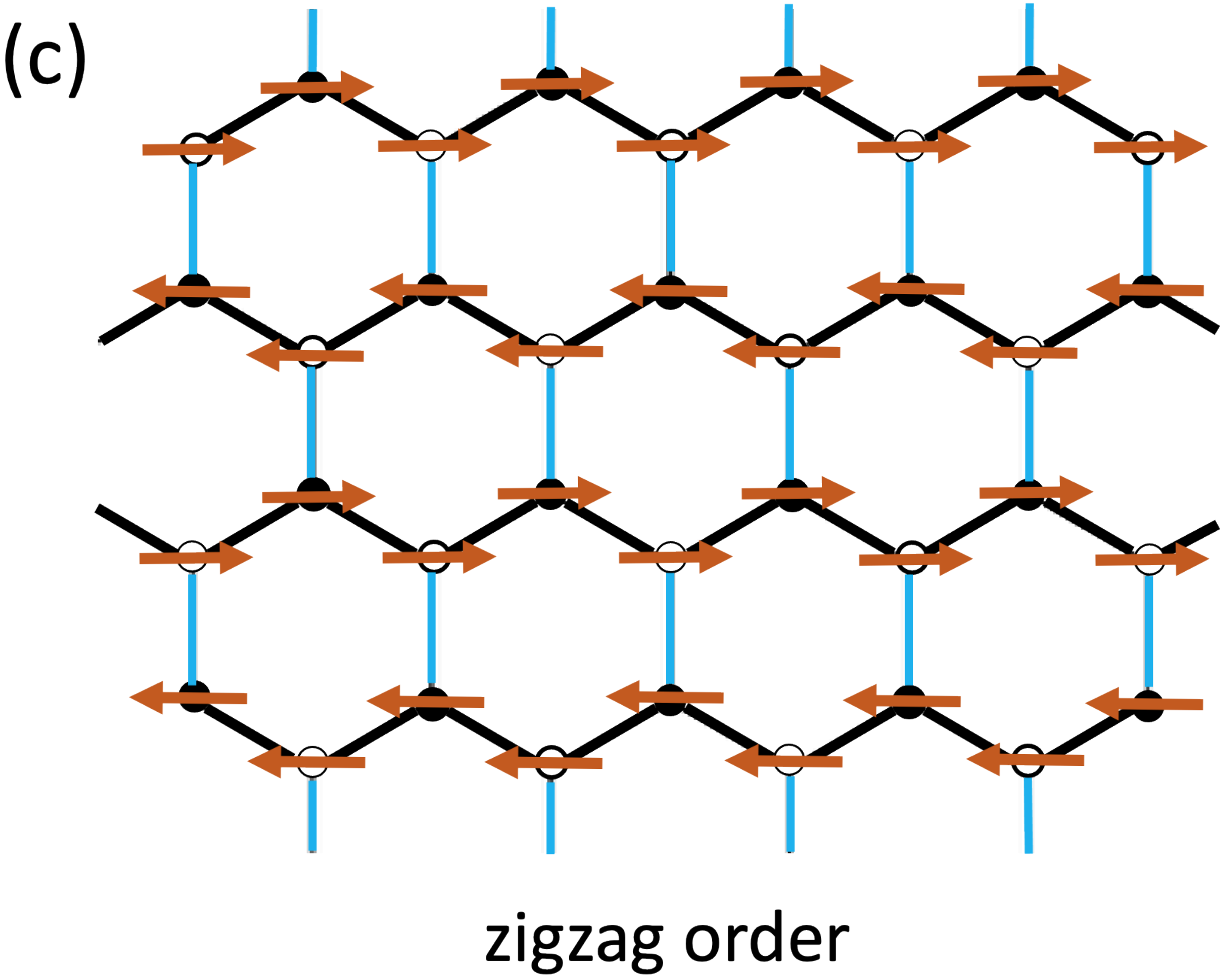}
\caption{Spin ordering patterns of (a) $120^\circ$ magnetic order, (b) counter-rotating spiral order, and (c) zigzag order.
For the 120$^\circ$ order in (a), the red arrows represent spin directions in the spin space, which are approximately coplanar, locating within the plane whose normal direction is along the $(1,1,1)$- and $(1,1,-1)$-directions in the small $J$ limit for the ``120$^\circ$ I" phase in Fig. \ref{fig:phase-Gamma} (a) and ``120$^\circ$ II" phase in Fig. \ref{fig:phase-Gamma} (b), respectively.
For the counter-rotating spiral order in (b), the red arrows represent spin directions in the spin space, which are approximately coplanar, locating within the plane whose normal direction is along the $(1,1,-1)$-direction in the small $J$ limit.
For the zigzag order in (c), the brown arrows represent spin directions in the spin space, which are approximately pointing along the $(1,1,-1)$-direction in the small $J$ limit.
In (a,b,c), 
the blue and black links represent the weak and strong bonds, respectively, where the bond pattern is shown in Fig. \ref{fig:honeycomb_original}; 
the black (white) circles represent the sites in the $A$ ($B$) sublattice of the honeycomb lattice; 
the spin orientations are only approximate since there are corrections due to the bosonization coefficients $\sigma_C$, $\delta_C$, $\rho_C$;
the $(1,1,1)$- or $(1,1,-1)$-directions can be changed when $J$ becomes large.
} 
\label{fig:orders}
\end{figure*}

Next we describe the main results in this work.
In the FM Gamma region, we find a 120$^\circ$ magnetic order named as ``120$^\circ$ I" in Fig. \ref{fig:phase-Gamma} (a) with spin textures plotted in Fig. \ref{fig:orders} (a),  which are consistent with the findings in Ref. \cite{Rau2014}.
More interestingly, in the AFM Gamma region as shown in Fig. \ref{fig:phase-Gamma} (b), 
our coupled-chain analysis  reveals three distinct types of magnetic orders,
including a 120$^\circ$ order named as ``120$^\circ$ II" in Fig. \ref{fig:phase-Gamma} (b) (see Fig. \ref{fig:orders} (a) for spin textures), 
a commensurate counter-rotating spiral order (see Fig. \ref{fig:orders} (b)),
and a zigzag order (see Fig. \ref{fig:orders} (c)).

In Fig. \ref{fig:phase-Gamma} (b), a quotation is put on the zigzag order since it arises from a subdominant channel of instability in the 1D Luttinger liquid theory,
in contrast to the other two orders which originate from dominant ones.
As a result of sub-dominance, the zigzag order only arises when the instability in the dominant channel becomes very weak,
or some other small interactions such as FM $\Gamma^\prime$ interaction \cite{Rau2014b} may support the zigzag order.
We note that our analysis may not be applicable  to  the 2D Kitaev-Gamma model (i.e., $J=0$) and the small $J$ limit, since the dominant and sub-dominant channels become degenerate or nearly degenerate in those cases. 
The $J\rightarrow 0$ regime may require an independent study which is worth for future considerations. 
Despite this, we will frequently take the small $J$ limit in analyzing the magnetic orders to gain better understanding of the structure of the patterns.

The counter-rotating spiral order appearing in the AFM Gamma region has the same magnetic ordering as the one observed in experiments in the $\alpha$-Li$_2$IrO$_3$ material, except that the wavevector in experiments is found to be slightly incommensurate (about $3\%$ away from commensuration). 
Based on detailed analysis of the exchange processes, Ref. \cite{Liu2022} has proposed that the nearest neighboring interactions $K$, $J$, $\Gamma$ in $\alpha$-Li$_2$IrO$_3$ satisfy $K<0$, $J>0$, $\Gamma>0$,
which is the parameter region where the counter-rotating spiral order in Fig. \ref{fig:phase-Gamma} (b) appears.
In addition, the zigzag order in the AFM Gamma region is also consistent with the experimentally observed pattern in Na$_2$IrO$_3$. 
By assuming that the relations $K<0$, $J>0$, $\Gamma>0$ also hold for Na$_2$IrO$_3$, 
the zigzag order in Fig. \ref{fig:phase-Gamma} (b) shares the same parameter region as the Na$_2$IrO$_3$ material.
Furthermore, the use of an anisotropic version of the generalized Kitaev models to model these two materials can be justified, 
since  the monoclinic lattice structures in Na$_2$IrO$_3$ and $\alpha$-Li$_2$IrO$_3$ lead to anisotropies in the bond strengths \cite{Singh2010,Gretarsson2013,Chun2015,Kimchi2015}. 
The above reasonings imply that  our results are potentially able to capture both the counter-rotating spiral order in $\alpha$-Li$_2$IrO$_3$ and the zigzag order in Na$_2$IrO$_3$,
thereby indicating that  the nearest neighboring spin-1/2 $KJ\Gamma$ model may be taken as a unified minimal model describing both $\alpha$-Li$_2$IrO$_3$ and Na$_2$IrO$_3$ materials.

It is worth to mention that while the coupled-chain analysis 
predicts two first order phase transition lines separating the zigzag order from the 120$^\circ$ and counter-rotating spiral orders in the AFM Gamma region (shown by the two red lines in Fig. \ref{fig:phase-Gamma} (b)),
these two lines are predicted to merge at a single quantum critical point at $K=-2\Gamma$, $J=0$ as shown by the red solid circle in  Fig. \ref{fig:phase-Gamma} (b). 
Therefore, the $K=-2\Gamma<0$ point in the Kitaev-Gamma model is a continuous phase transition point where several distinct magnetically ordered phases meet,
though the possibility of  an extended disordered phase in the 2D Kitaev-Gamma model
cannot be excluded according to previous works \cite{Rousochatzakis2017,Catuneanu2018,Gohlke2018,Liu2021}. 


Finally, we note that besides the 2D analysis in terms of coupled chains,
our work also contains a detailed study on the nonsymmorphic Luttinger liquid behaviors of a single chain,
which lays the foundation for the coupled-chain analysis. 
A single $KJ\Gamma$ chain has an intricate nonsymmorphic symmetry group structure,
 most easily formulated after a six-sublattice rotation \cite{Yang2020}.
In this work, the abelian bosonization formulas for the lattice spin operators are proposed, which break the emergent U(1) symmetry at low energies and only respects the exact nonsymmorphic symmetries of the model. 
These nonsymmorphic abelian bosonization formulas contain ten parameters, 
which are useful to determine the spin textures in  2D magnetically ordered phases 
and turn out to be crucial to explain  experiments. 
The analytical predictions for the chain are supported by our large-scale density matrix renormalization group (DMRG) numerical simulations.

The rest of the paper is organized as follows. 
In Sec. \ref{sec:Ham}, the model Hamiltonian is introduced.
Sec. \ref{sec:nonsym_bosonize} includes a detailed analysis of the Luttinger liquid theory in a decoupled single chain.
In Sec. \ref{sec:FM_Gamma}, the coupled-chain analysis is applied to the FM Gamma region,
which reveals a 120$^\circ$ order.
Sec. \ref{sec:AFM_Gamma} is devoted to analyzing the AFM Gamma region,
where the 120$^\circ$, counter-rotating spiral, and zigzag orders are found,
and the quantum critical point is discussed. 
In Sec. \ref{sec:material}, possible relations to the counter-rotating spiral order observed in $\alpha$-Li$_2$IrO$_3$ and the zigzag order in Na$_2$IrO$_3$ are  discussed. 
In Sec. \ref{sec:summary}, we briefly summarize the main results of the paper.

\section{Model Hamiltonian}
\label{sec:Ham}

In this section, we give the Hamiltonian studied in this work, and briefly describe the obtained phase diagram. 

\subsection{Hamiltonian of anisotropic  spin-1/2 $KJ\Gamma$ model}

\begin{figure}[h]
\begin{center}
\includegraphics[width=8cm]{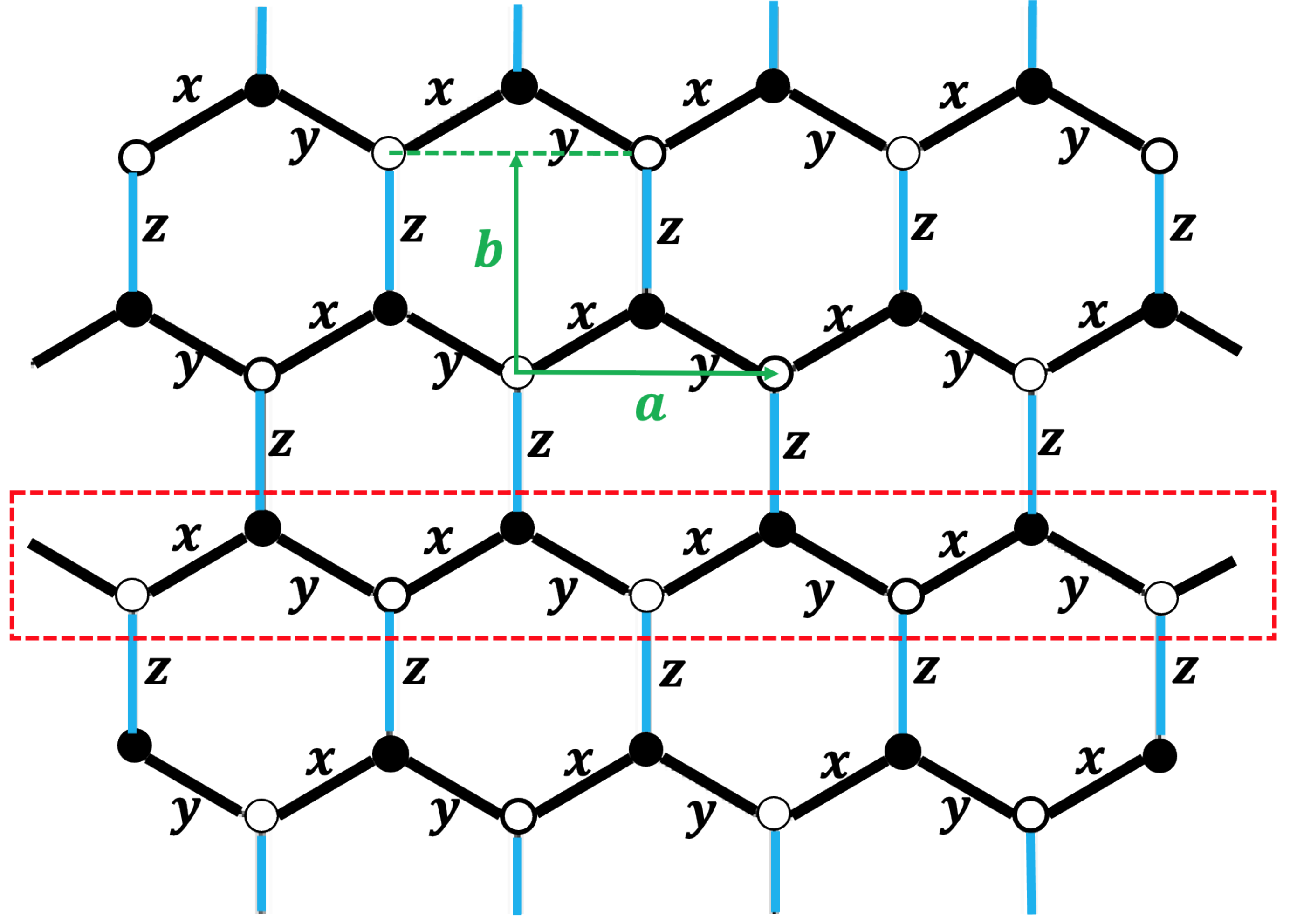}
\caption{Anisotropic  $KJ\Gamma$ model on the honeycomb lattice.
The black and blue links represent the strong and weak bonds, respectively. 
The green arrows represent a set of mutually perpendicular vectors spanning a unit cell of the honeycomb lattice.
The black (white) circles represent the sites in the $A$ ($B$) sublattice of the honeycomb lattice
} \label{fig:honeycomb_original}
\end{center}
\end{figure}

We consider an anisotropic  spin-1/2 Kitaev-Heisenberg-Gamma ($KJ\Gamma$) model on the honeycomb lattice  shown in Fig. \ref{fig:honeycomb_original}, in which each link is associated with a spin direction denoted by $\gamma$ where $\gamma=x,y,z$.
The interaction between two nearest neighboring sites $i,j$ connected by a link of bond type $\gamma$ is 
\begin{flalign}
 &K_\gamma S_i^\gamma S_j^\gamma+ J_\gamma \vec{S}_i\cdot \vec{S}_j+\Gamma_\gamma (S_i^\alpha S_j^\beta+S_i^\beta S_j^\alpha),
\label{eq:Ham_2D_orig}
\end{flalign}
in which  $\gamma$ is the spin direction associated with the $\gamma$ bond  in Fig. \ref{fig:honeycomb_original};
$\alpha\neq\beta$ are the two remaining spin directions other than $\gamma$;
$K_\gamma$, $J_\gamma$, and $\Gamma_\gamma$
are the Kitaev, Heisenberg, and Gamma couplings on the bond $\gamma$, respectively, given by 
\begin{flalign}
&K_x=K_y=K,~J_x=J_y=J,~\Gamma_x=\Gamma_y=\Gamma,
\end{flalign}
and
\begin{flalign}
&K_z=\alpha_z K,~ J_z=\alpha_z J,~ \Gamma_z=\alpha_z \Gamma,
\end{flalign}
where $\alpha_z$ is the anisotropy  parameter.

In this work, we consider the parameter region satisfying $K<0$, $J>0$, with both positive and negative signs of $\Gamma$.
Since it is known that there are real Kitaev materials in which $K$ is FM and $J$ is AFM,
the chosen parameter region is relevant to real situations.
For the anisotropy  parameter $\alpha_z$, we assume $0\leq \alpha_z\leq 1$ throughout this work,
which corresponds to an anisotropic system where the $z$-bonds are the weak bonds.
The Hamiltonian in Eq. (\ref{eq:Ham_2D_orig}) describes the isotropic   $KJ\Gamma$ model when $\alpha_z=1$,
whereas it represents  a system of decoupled 1D $KJ\Gamma$ chains when $\alpha_z=0$.

We will study the small $\alpha_z$ limit,
corresponding to a system of weakly coupled chains on the honeycomb lattice.
The strategy is to take the decoupled chains as the unperturbed system and treat the inter-chain interaction as a perturbation. 
By assuming an absence of phase transition from small $\alpha_z$ to $\alpha_z=1$,
the results can be used to understand the phase diagram of the isotropic  2D spin-1/2 $KJ\Gamma$ model as well.  

\subsection{Brief description of the phase diagram}

Here we give a brief description of the phase diagram in the $K<0$, $J>0$ region obtained in this work.
As shown in Fig. \ref{fig:phase-Gamma} (a,b), the 120$^\circ$ magnetic orders appear in both the FM and AFM Gamma regions,
which has a vortex-like magnetic structure shown in Fig. \ref{fig:orders} (a).
However, the 120$^\circ$ orders in the two regions are distinct in nature:
the spins in the ``120$^\circ$ I" phase in Fig. \ref{fig:phase-Gamma} (a) lie in a plane perpendicular to the $(1,1,1)$-direction in the small $J$ limit,
whereas the normal direction of the common plane shared by the spins in the ``120$^\circ$ II" phase in Fig. \ref{fig:phase-Gamma} (b) is along the $(1,1,-1)$-direction in the small $J$ limit. 
The characteristic feature of the 120$^\circ$ order is that
in both $A$ and $B$ sublattices of the honeycomb lattice,
the nearest neighboring spins (which are next-nearest neighbors in the honeycomb lattice) are at relative 120$^\circ$ angles, which is the origin of the name ``120$^\circ$ order" \cite{Rau2014}. 

On the other hand, the counter-rotating spiral order only appears in the AFM Gamma region as shown in Fig. \ref{fig:phase-Gamma} (b).
The spin orientations are plotted in Fig. \ref{fig:orders} (b), which does not exhibit a vortex structure. 
The characteristic feature of the counter-rotating spiral order is that within each zigzag chain formed by the black bonds in Fig. \ref{fig:orders} (b),
the spins rotate in a counter-clockwise manner for the upper sites in the zigzag chain,
whereas they rotate in a clockwise way for the lower sites,
which is the origin of the name ``counter-rotating spiral" for this magnetic order \cite{Kimchi2014}, since the spins in the upper and lower sites rotate in opposite directions. 

The zigzag order appears in the AFM Gamma region as shown in Fig. \ref{fig:phase-Gamma} (b),
which is separated from the counter-rotating spiral and 120$^\circ$ II phases by two first order transition lines,
represented by the two red lines in Fig. \ref{fig:phase-Gamma} (b).
The characteristic feature of the zigzag order is that the spins align ferromagnetically within the zigzag chains, but antiferromagnetically  among different chains. 
The spin texture in the zigzag order is shown in Fig. \ref{fig:orders} (c). 

Finally, the two first order phase transition lines in Fig. \ref{fig:phase-Gamma} (b) terminate at a common end point, which is a quantum critical point  represented by the solid red circle in Fig. \ref{fig:phase-Gamma} (b). 
This is critical point where several distinct ordered phases meet. 

\section{Luttinger liquid in a decoupled chain}
\label{sec:nonsym_bosonize}

We start by analyzing a decoupled single chain on the honeycomb lattice shown by the zigzag chain enclosed by the red dashed line in Fig. \ref{fig:honeycomb_original}, which will serve as the starting point for  our perturbative treatment in later sections.

\subsection{1D Hamiltonian and six-sublattice rotation}

\begin{figure}[h]
\begin{center}
\includegraphics[width=8cm]{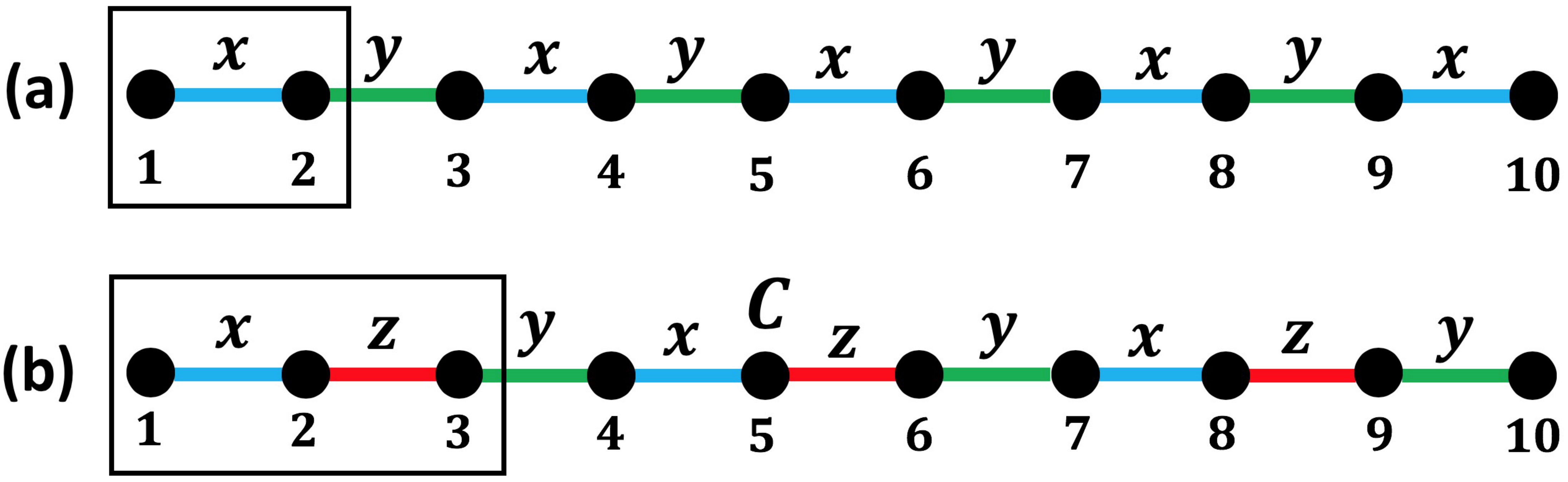}
\caption{
Bond patterns of the $KJ\Gamma$ chain 
(a) before and (b) after the six-sublattice rotation.
} \label{fig:bonds}
\end{center}
\end{figure}

\begin{figure*}[htbp]
\includegraphics[width=8.9cm]{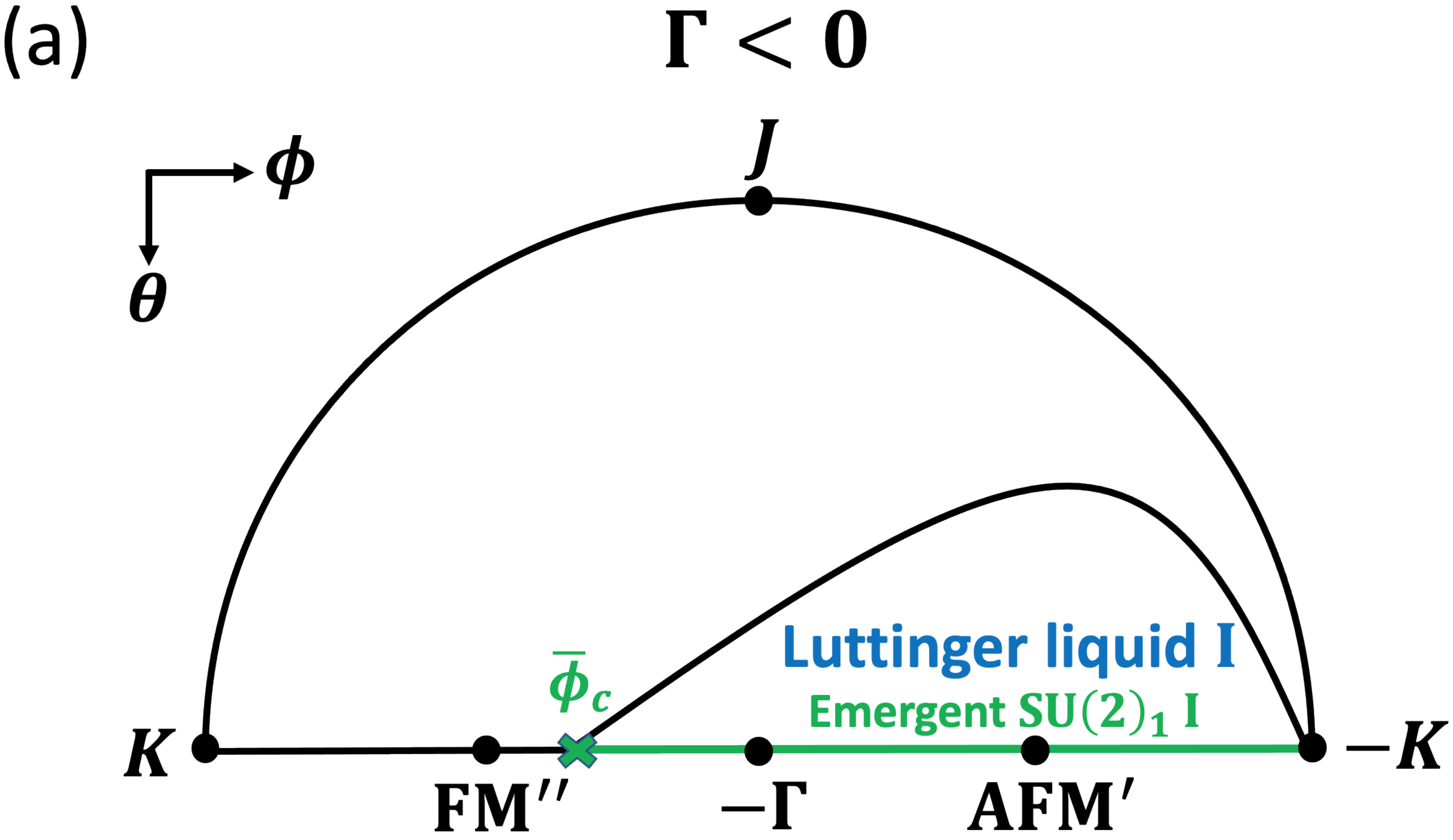}
\includegraphics[width=8.9cm]{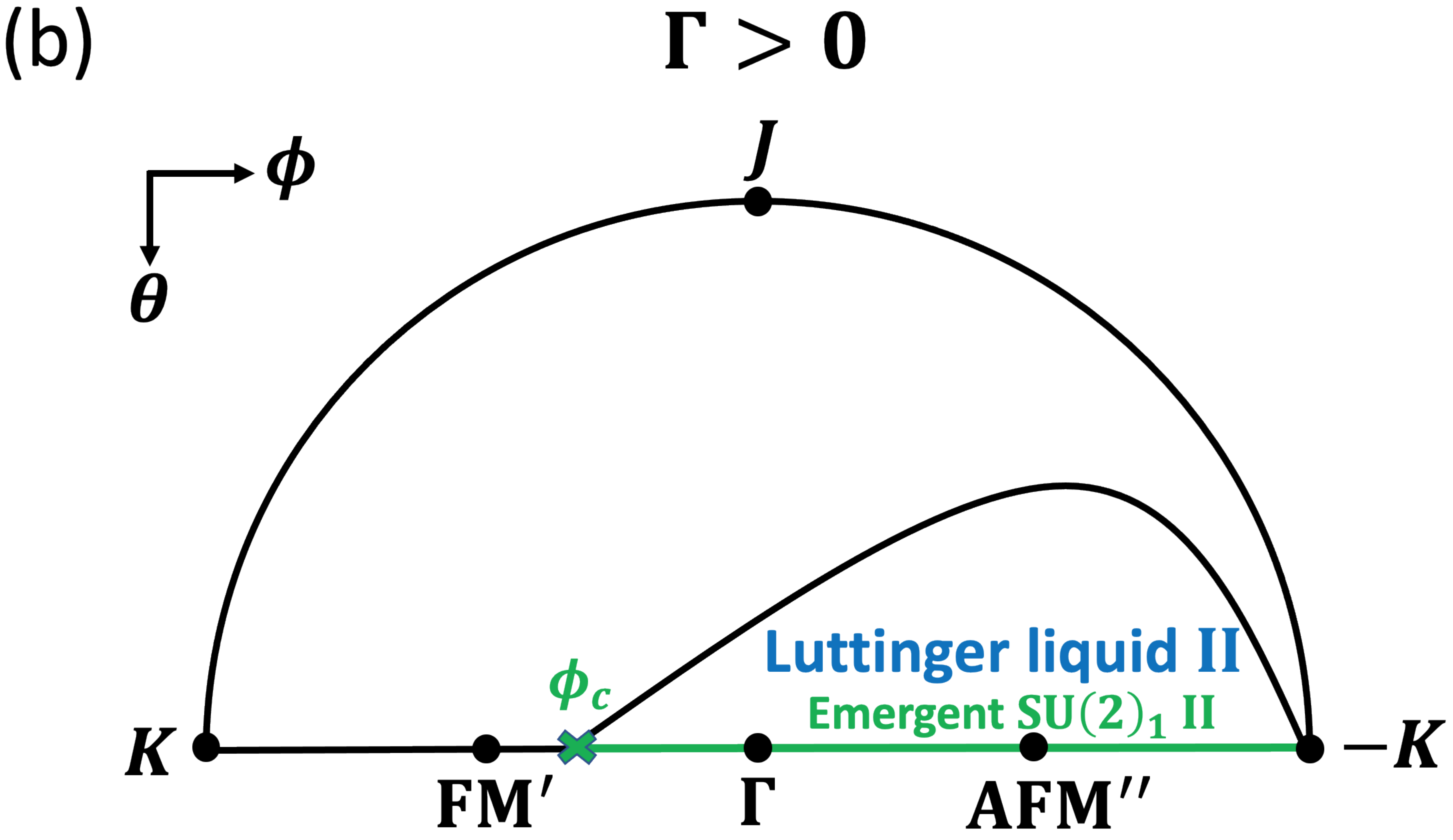}
\caption{
The Luttinger liquid phases (a) ``Luttinger liquid I" in the region $K<0$, $\Gamma<0$, $J>0$, and (b) ``Luttinger liquid II" in the region $K<0$, $\Gamma>0$, $J>0$, in the phase diagram of the 1D spin-1/2 Kitaev-Heisenberg-Gamma model.
The two regions are unitarily equivalent because of the equivalence in Eq. (\ref{eq:equiv_1D}).
The lines marked with green color for $J=0$ has an emergent SU(2)$_1$ conformal symmetry at low energies,  where $\bar{\phi}_c=1.67\pi$ in (a)
and $\phi_c=0.33\pi$ in (b),
 as discussed in Ref. \cite{Yang2019}.
The FM$^{\prime\prime}$ and AFM$^\prime$ points in (a), as well as the FM$^\prime$ and AFM$^{\prime\prime}$ points in (b)
 have hidden SU(2) symmetries as revealed by the six-sublattice rotation $U_6$ and the global spin rotation around the $z$-axis by an angle $\pi$.
} \label{fig:phase_KHG_1D}
\end{figure*}

In this subsection, we give the Hamiltonian for the spin-1/2 $KJ\Gamma$ chain,
and discuss the six-sublattice rotation $U_6$ \cite{Chaloupka2015,Yang2019,Yang2020} which is useful in studying the 1D generalized Kitaev spin models.

The Hamiltonian of the spin-1/2 $KJ\Gamma$ chain is
\bea
H_{1D}&=&\sum_{<ij>\in\gamma\,\text{bond}}\big[ KS_i^\gamma S_j^\gamma+ J\vec{S}_i\cdot \vec{S}_j\nn\\
&&+\Gamma (S_i^\alpha S_j^\beta+S_i^\beta S_j^\alpha)\big],
\label{eq:Ham_1D}
\eea
in which the bond pattern for $\gamma\in\{x,y\}$ is shown in Fig. \ref{fig:bonds} (a).
The explicit form of $H_{1D}$ is included in Appendix \ref{sec:ham_explicit}.

Since a global spin rotation around $z$-axis by $\pi$ (denoted as $R(\hat{z},\pi)$) changes the sign of $\Gamma$  in Eq. (\ref{eq:Ham_1D}) while keeping $K$ and $J$ unchanged,
 there is the  equivalent relation
\bea
(K,J,\Gamma)\simeq (K,J,-\Gamma).
\label{eq:equiv_1D}
\eea
As a result of Eq. (\ref{eq:equiv_1D}), it is enough to consider the parameter region $\Gamma>0$ for a single chain (note: this does not hold for 2D).
For later convenience,  we introduce the following parametrization
\bea
J&=&\cos(\theta),\nn\\
K&=&\sin(\theta)\cos(\phi),\nn\\
\Gamma&=&\sin(\theta)\sin(\phi).
\label{eq:parametrize_KJG}
\eea

There is a useful unitary transformation $U_6$ called six-sublattice rotation \cite{Chaloupka2015,Yang2019,Yang2020}, 
defined as
\begin{eqnarray}
\text{Sublattice $1$}: & (x,y,z) & \rightarrow (x^\prime,y^\prime,z^\prime),\nn\\ 
\text{Sublattice $2$}: & (x,y,z) & \rightarrow (-x^\prime,-z^\prime,-y^\prime),\nn\\
\text{Sublattice $3$}: & (x,y,z) & \rightarrow (y^\prime,z^\prime,x^\prime),\nn\\
\text{Sublattice $4$}: & (x,y,z) & \rightarrow (-y^\prime,-x^\prime,-z^\prime),\nn\\
\text{Sublattice $5$}: & (x,y,z) & \rightarrow (z^\prime,x^\prime,y^\prime),\nn\\
\text{Sublattice $6$}: & (x,y,z) & \rightarrow (-z^\prime,-y^\prime,-x^\prime),
\label{eq:6rotation}
\end{eqnarray}
in which "Sublattice $i$" ($1\leq i \leq 6$) represents all the sites $i+6n$ ($n\in \mathbb{Z}$) in the chain;
$S^\alpha_j$ and $S^{\prime\alpha}_j$ are used  to denote the spin operators within the original frame and $U_6$ frame, respectively;
and we have abbreviated $S^\alpha$ ($S^{\prime \alpha}$) as $\alpha$ ($\alpha^\prime$) in Eq. (\ref{eq:6rotation}) for short ($\alpha=x,y,z$).
The Hamiltonian $H^\prime_{1D}=U_6 H U_6^{-1}$ in the six-sublattice rotated frame can be derived as
\begin{flalign}
&H^\prime_{1D}=\sum_{<ij>\in \gamma\,\text{bond}}\big[ -KS_i^{\prime\gamma} S_j^{\prime\gamma}-\Gamma (S_i^{\prime\alpha} S_j^{\prime\alpha}+S_i^{\prime\beta} S_j^{\prime\beta})  \nn\\
&-J(S_i^{\prime\gamma} S_j^{\prime\gamma}+S_i^{\prime\alpha} S_j^{\prime\beta}+S_i^{\prime\beta} S_j^{\prime\alpha}),
\label{eq:6rotated}
\end{flalign}
in which $\gamma=x,z,y$ has a three-site periodicity as shown in Fig. \ref{fig:bonds} (b).
The explicit form of $H^\prime_{1D}$ is included in Appendix \ref{sec:ham_explicit}.

As can be seen from Eq. (\ref{eq:6rotated}), 
$H^\prime_{1D}$ is just the SU(2) symmetric Heisenberg model when $K=\Gamma$, $J=0$.
In addition, using Eq. (\ref{eq:equiv_1D}), it can be seen that the system is also SU(2) invariant when $K=-\Gamma$, $J=0$.
For vanishing $J$, $H^\prime_{1D}$ represents the AFM (FM) Heisenberg model when $K=-|\Gamma|<0$   ($K=|\Gamma|>0$).
These four hidden SU(2) symmetric points are denoted as FM$^{\prime\prime}$, AFM$^{\prime}$, FM$^{\prime}$, and AFM$^{\prime\prime}$ in Fig. \ref{fig:phase_KHG_1D} (a,b).  

\subsection{The Luttinger liquid phase in a single chain}

It has been established in Ref. \cite{Yang2020} that there is an extended gapless phase for $K<0$ in the phase diagram of the 1D spin-1/2 $KJ\Gamma$ model,
where the low energy physics can be described by the following Luttinger liquid Hamiltonian,
\bea
H_{LL}=\frac{v}{2} \int dx [\frac{1}{\kappa} (\nabla \varphi)^2 +\kappa (\nabla \theta)^2],
\label{eq:LL_liquid}
\eea
in which the $\theta,\varphi$ fields satisfy the commutation relation $[\varphi(x),\theta(x^\prime)]=\frac{i}{2}\text{sgn}(x^\prime-x)$,
 $v$ is the velocity, and $\kappa$ is the Luttinger liquid parameter.
 Fig. \ref{fig:phase_KHG_1D} (a) and  Fig. \ref{fig:phase_KHG_1D} (b) show the schematic plots of the phase diagram of the spin-1/2 $KJ\Gamma$ chain \cite{Yang2020} in the FM and AFM Gamma regions, respectively, 
in which the parameters $\theta$, $\phi$ are defined in Eq. (\ref{eq:parametrize_KJG}) (note: the distinction between the parameter $\theta$ and the $\theta$-field in Eq. (\ref{eq:LL_liquid}) should be clear from the context).
The phase diagram in Fig. \ref{fig:phase_KHG_1D} (b) is unitarily equivalent with the one  in Fig. \ref{fig:phase_KHG_1D} (a)
via the global spin rotation $R(\hat{z}^\prime,\pi)$ as discussed in  Eq. (\ref{eq:equiv_1D}).
 
It is helpful to study the structure of the symmetry group  of the model to better understand the Luttinger liquid phase.
In particular, the symmetry analysis can be used to determine the symmetry axis of the emergent U(1) symmetry in the low energy Luttinger liquid theory. 
Here we briefly review the symmetries of $H^\prime_{1D}$ in the $U_6$ frame \cite{Yang2020}. 

It can be checked that $H^\prime_{1D}$ in Eq. (\ref{eq:6rotated}) is invariant under the following symmetry transformations
\begin{eqnarray}
1.&T &:  (S_i^{\prime x},S_i^{\prime y},S_i^{\prime z})\rightarrow (-S_{i}^{\prime x},-S_{i}^{\prime y},-S_{i}^{\prime z})\nn\\
2.&R_I I&: (S_i^{\prime x},S_i^{\prime y},S_i^{\prime z})\rightarrow (-S_{10-i}^{\prime z},-S_{10-i}^{\prime y},-S_{10-i}^{\prime x})\nn\\
3.& R_aT_a&:  (S_i^{\prime x},S_i^{\prime y},S_i^{\prime z})\rightarrow (S_{i+1}^{\prime z},S_{i+1}^{\prime x},S_{i+1}^{\prime y}),
\label{eq:sym_Jneq0}
\end{eqnarray}
in which $T$ is time reversal; 
$T_a$ is the spatial translation by one lattice site;
$I$ is the spatial inversion around the point $C$ in Fig. \ref{fig:bonds} (b);
and $R_a=R(\hat{z}^{\prime\prime},-2\pi/3)$, $R_I=R(\hat{y}^{\prime\prime},\pi)$, 
where $R(\hat{n},\beta)$ represents a global spin rotation by angle $\beta$ and
\begin{eqnarray}
\hat{z}^{\prime\prime}&=&\frac{1}{\sqrt{3}}(1,1,1)^T, \nn\\
\hat{y}^{\prime\prime}&=&\frac{1}{\sqrt{2}}(-1,0,1)^T.
\label{eq:yz_prime}
\end{eqnarray}
The symmetry group $G$ is generated by the symmetry operations in Eq. (\ref{eq:sym_Jneq0}) as
\begin{eqnarray}
G=\mathopen{<}T, R_aT_a, R_I I\mathclose{>},
\label{eq:group_G}
\end{eqnarray} 
where $\mathopen{<}...\mathclose{>}$ represents the group generated by the elements in the brackets. 

\begin{figure}[h]
\includegraphics[width=7.0cm]{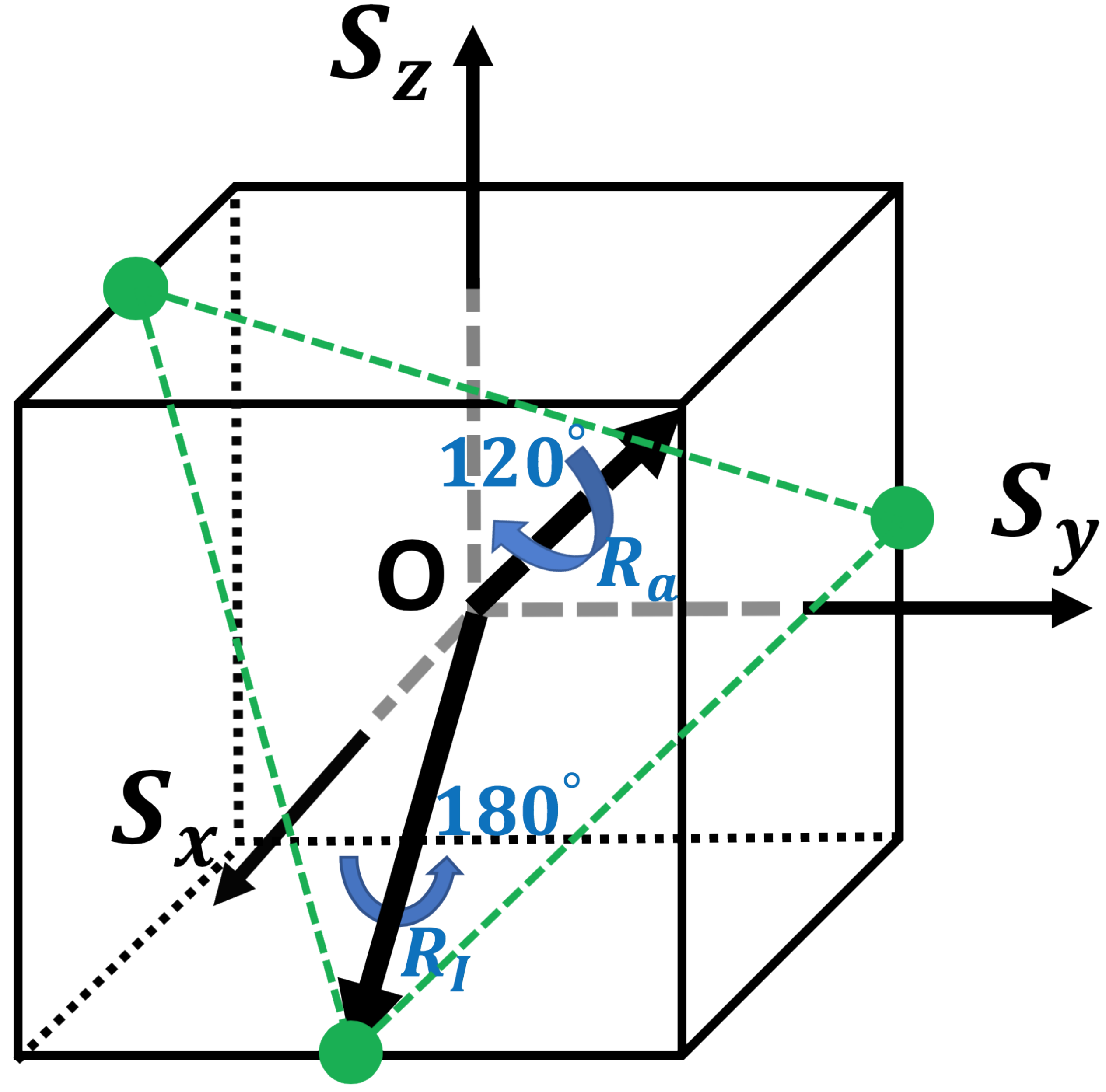}
\caption{$\mathopen{<}R_a,R_I\mathclose{>}$ as the symmetry group of the regular triangle formed by the green dashed lines in spin space.
The three solid green circles are located at the middle points of the corresponding edge of the spin cube. 
This figure is taken from Ref. \cite{Yang2020}.
} \label{fig:D3d_geometry}
\end{figure}

We note that $G$ is nonsymmorphic because of the group element $R_aT_a$.
It has been proved in Ref. \cite{Yang2020} that the group structure satisfies
\bea
G/\mathopen{<}T_{3a}\mathclose{>}  \cong D_{3d},
\label{eq:group_structure}
\eea
in which $D_{3d}\cong D_3\times \mathbb{Z}_2^T$ where $D_n$ is the dihedral group of order $2n$,
and $\mathbb{Z}_2^T$ is the $\mathbb{Z}_2$ group generated by the time reversal operation. 
Here we will not go into details of the group structure, but only give an intuitive understanding about  Eq. (\ref{eq:group_structure}).
If  the spatial components $T_a$ and $I$ are neglected in $R_aT_a$  and $R_II$,  then the actions of $R_a$ and $R_I$ in the spin space constitute a set of  generators for the symmetry group of the regular triangle formed by the green dashed lines in Fig. \ref{fig:D3d_geometry}.
On the other hand, time reversal operation commutes with spin rotations, hence $\mathopen{<}T,R_a,R_I\mathclose{>}$ is isomorphic to $D_{3d}=D_3\times\mathbb{Z}_2^T$ where $\mathbb{Z}_2^T=\mathopen{<}T\mathclose{>}$.
In fact, it can be proved that Eq. (\ref{eq:group_structure}) is satisfied when  $T_a$ and $I$ are taken into account \cite{Yang2020}.

There is a quick and intuitive way to understand why the low energy physics of the 1D spin-1/2 $KJ\Gamma$ model is the same as that of a spin-1/2 XXZ chain with an easy-plane anisotropy. 
In the long wavelength limit, the three spins within a unit cell in the Hamiltonian $H^\prime_{1D}$ in Eq. (\ref{eq:6rotated}) get smeared and can no longer be clearly distinguished. 
Denoting $\vec{\mathcal{S}}^\prime(x)$ to be the smeared spin operator for the three spins in the unit cell at position $x$
and summing over all the terms in the Hamiltonian within the unit cell, we obtain the smeared local term $\mathcal{H}^\prime_{1D}(x)$ in the Hamiltonian at unit cell $x$ as
\begin{flalign}
\mathcal{H}_{1D}^\prime(x)=-(K+2\Gamma) \vec{\mathcal{S}}^\prime(x)\cdot \vec{\mathcal{S}}^\prime(x^\prime) - 3J \mathcal{S}^{\prime\prime z}(x)\cdot \mathcal{S}^{\prime\prime z}(x^\prime),
\label{eq:smeared_H}
\end{flalign}
in which $x^\prime=x+\Delta x$ is a point adjacent to $x$,
and $\mathcal{S}^{\prime\prime z}=\hat{z}^{\prime\prime}\cdot \vec{\mathcal{S}}$, where $\hat{z}^{\prime\prime}$ is defined in Eq. (\ref{eq:yz_prime}). 
As can be seen from Eq. (\ref{eq:smeared_H}), the smeared Hamiltonian  $\sum_x\mathcal{H}^\prime_{1D}(x)$ has a U(1) rotational  symmetry around $z^{\prime\prime}$-axis.
In particular, $\sum_x\mathcal{H}^\prime_{1D}(x)$ has an easy-plane anisotropy when $K<0$, $\Gamma<0$ and $J>0$,
which explains why the system can be described by a Luttinger liquid theory at low energies.
The analysis is equally applicable for $K<0$, $\Gamma>0$, $J>0$,
because of the equivalent relation  in Eq. (\ref{eq:equiv_1D}).

However, we emphasize that Eq. (\ref{eq:smeared_H}) is only a hand-waving argument which is not rigorous, 
since in addition to the uniform component, the smeared spin operator also contains a staggered component with a sign alternation between adjacent sites,
which is not taken into account in the derivation of Eq. (\ref{eq:smeared_H}).
A rigorous proof of the existence of an extended Luttinger liquid phase has been performed in Ref. \cite{Yang2020} 
based on a symmetry analysis of the low energy field theory (see Appendix \ref{app:sym_analysis_low} for a brief review). 
On the other hand, it is rather clear from Fig. \ref{fig:D3d_geometry} that the symmetry axis for the emergent U(1) symmetry is along the normal direction of the green regular triangle, i.e., the $(1,1,1)$-direction,
which originates from the symmetry structure, unrelated to the argument given in Eq. (\ref{eq:smeared_H}).
Our large-scale DMRG simulations are consistent with a symmetry axis being along the $(1,1,1)$-direction to a high degree of accuracy, as discussed in details in Appendix \ref{app:DMRG_sym_axis}.

\subsection{The nonsymmorphic abelian bosonization formulas}
\label{subsec:nonsym_bosonize_formula}

In this subsection, we derive the nonsymmorphic abelian bosonization formulas for the lattice spin operators, which are compatible with the nonsymmorphic symmetry group of the model but break the emergent U(1) symmetry. 
Since the rotation axis for the emergent U(1) symmetry is along the $(1,1,1)$-direction, we define a new spin coordinate system $\{x^{\prime\prime},y^{\prime\prime},z^{\prime\prime}\}$ via the following orthogonal transformation 
\bea
O=\left(\begin{array}{ccc}
-\frac{1}{\sqrt{6}} & -\frac{1}{\sqrt{2}} & \frac{1}{\sqrt{3}}\\
\sqrt{\frac{2}{3}} & 0 & \frac{1}{\sqrt{3}}\\
-\frac{1}{\sqrt{6}} & \frac{1}{\sqrt{2}} & \frac{1}{\sqrt{3}}
\end{array}
\right).
\label{eq:O_rotate}
\eea
such that 
\bea
(x^{\prime\prime}~y^{\prime\prime}~z^{\prime\prime})=(x^\prime~y^\prime~z^\prime) O.
\label{eq:transform_O}
\eea
Notice that the unit vectors $\hat{y}^{\prime\prime}$, $\hat{z}^{\prime\prime}$ are defined in Eq. (\ref{eq:yz_prime}), and $\hat{x}^{\prime\prime} $ is given by 
\bea
\hat{x}^{\prime\prime}=\frac{1}{\sqrt{6}}(-1,2,-1).
\eea
We will call the spin coordinate system after a further $O$ transformation superimposed on $U_6$ as the $OU_6$ frame.

If the system has an exact U(1) symmetry around the $\hat{z}^{\prime\prime}$ axis, then the abelian  bosonization formulas for the spin operators are given by 
\begin{flalign}
S^{\prime\prime z}(x)&= -\frac{1}{\sqrt{\pi}} \nabla \varphi(x) + \text{const.} \frac{1}{a}(-)^n \cos(2\sqrt{\pi} \varphi(x)),\nn\\
S^{\prime\prime +}(x)&=\text{const.} \frac{1}{\sqrt{a}}e^{-i\sqrt{\pi}\theta(x)} \big[(-)^n+\cos(2\sqrt{\pi}\varphi(x))\big],
\label{eq:abelian_bosonize}
\end{flalign}
in which $S^{\prime\prime +}=S^{\prime\prime x}+iS^{\prime\prime y}$,
and $x=na$ ($n\in \mathbb{Z}$) is the spatial  coordinate in the continuum limit.
However, Eq. (\ref{eq:abelian_bosonize}) ceases to apply in the Luttinger liquid phase shown in Fig. (\ref{fig:phase_KHG_1D}), 
since the U(1) symmetry is only emergent at low energies and the symmetry group of the microscopic Hamiltonian is discrete and nonsymmorphic.

We propose the following abelian bosonization formulas in the $OU_6$ frame
\begin{flalign}
&(S^{\prime\prime x}_{i+3n}~S^{\prime\prime y}_{i+3n}~S^{\prime\prime z}_{i+3n})=\nn\\
&(\mathcal{J}^{ x}~\mathcal{J}^{ y}~\mathcal{J}^{ z})\mathcal{D}_i
+(-)^{i+n}(\mathcal{N}^{ x}~\mathcal{N}^{ y}~\mathcal{N}^{ z})\mathcal{C}_i,
\label{eq:abelian_LL1_matrix}
\end{flalign}
in which $\mathcal{D}_i,\mathcal{C}_i$ ($i=1,2,3$) are $3\times 3$ matrices,
and $\mathcal{J}^{\alpha},\mathcal{N}^{\alpha}$ are defined as
\bea
\mathcal{J}^{ x}&=&\frac{1}{a}\cos(\sqrt{4\pi} \varphi) \cos(\sqrt{\pi} \theta),\nn\\
\mathcal{J}^{ y}&=&\frac{1}{a}\cos(\sqrt{4\pi} \varphi) \sin(\sqrt{\pi} \theta),\nn\\
\mathcal{J}^{ z}&=&-\frac{1}{\sqrt{\pi}}\nabla \varphi,
\label{eq:bosonize_J}
\eea
and 
\bea
\mathcal{N}^{ x}&=&\frac{1}{a}\cos(\sqrt{\pi}\theta),\nn\\
\mathcal{N}^{ y}&=&\frac{1}{a}\sin(\sqrt{\pi}\theta),\nn\\
\mathcal{N}^{ z}&=&\frac{1}{a}\sin(\sqrt{4\pi}\varphi).
\label{eq:bosonize_N}
\eea

Notice that  the low energy fields $\mathcal{J}^{ \alpha}$, $\mathcal{N}^{ \alpha }$ ($\alpha=x,y,z$) remain in the low energy sector 
when a symmetry operation is performed.
As a result, the bosonization coefficients  are not all independent, 
since the left and right hand sides of  Eq. (\ref{eq:abelian_LL1_matrix}) have to be covariant under symmetry transformations of the system. 
The symmetry constraints lead to (for details, see Appendix \ref{app:bosonization_formulas})
\begin{flalign}
&\mathcal{C}_1=M_z^{-1}\mathcal{C}_2M_z,~\mathcal{C}_3=M_z\mathcal{C}_2M_z^{-1},\nn\\
&\mathcal{D}_1=M_z^{-1}\mathcal{D}_2M_z,~\mathcal{D}_3=M_z\mathcal{D}_2M_z^{-1},
\label{eq:Cs_z}
\end{flalign}
in which 
\bea
\mathcal{C}_2&=&\left(\begin{array}{ccc}
\lambda_C & 0 & \sigma_C\\
0 & \lambda_C+\delta_C & 0\\
\rho_C&0&\nu_C
\end{array}
\right),\nn\\
\mathcal{D}_2&=&\left(\begin{array}{ccc}
\lambda_D & 0 & \sigma_D\\
0 & \lambda_D+\delta_D & 0\\
\rho_D&0&\nu_D
\end{array}
\right),
\label{eq:CD_mat_mathcal}
\eea
and
\bea
M_z=\left(\begin{array}{ccc}
-\frac{1}{2} & \frac{\sqrt{3}}{2} & 0\\
-\frac{\sqrt{3}}{2} & -\frac{1}{2} & 0\\
0&0&1
\end{array}
\right).
\eea

Performing the inverse of the transformation in Eq. (\ref{eq:transform_O}),
 the nonsymmorphic bosonization formulas in the $U_6$ frame for the spin operators  $S^{\prime \alpha}_j$ ($\alpha=x,y,z$) can be obtained as 
\begin{flalign}
&(S^{\prime x}_{i+3n}~S^{\prime y}_{i+3n}~S^{\prime z}_{i+3n})=\nn\\
&(\mathcal{J}^{ x}~\mathcal{J}^{ y}~\mathcal{J}^{ z}) D_i
+(-)^{i+n}(\mathcal{N}^{ x}~\mathcal{N}^{ y}~\mathcal{N}^{ z})C_i,
\label{eq:abelian_LL1_matrix_B}
\end{flalign}
in which 
\bea
D_i=\mathcal{D}_iO^T,~C_i=\mathcal{C}_iO^T.
\label{eq:new_C_D}
\eea
Detailed derivations and explicit forms of the nonsymmorphic bosonization formulas for $S^{\prime\prime\alpha}_j$ in Eq. (\ref{eq:abelian_LL1_matrix}) in the $OU_6$ frame 
and $S^{\prime\alpha}_j$ in Eq. (\ref{eq:abelian_LL1_matrix_B}) in the $U_6$ frame
are included in Appendix \ref{app:bosonization_formulas}.
There are in total ten free parameters $\lambda_\Lambda$, $\nu_\Lambda$, $\sigma_\Lambda$, $\rho_\Lambda$, $\delta_\Lambda$ ($\Lambda=C,D$) in the nonsymmorphic bosonization formulas,
which turn out to play crucial roles to compare with experiments as discussed in Sec. \ref{sec:material}.
These ten free parameters can in principle be determined by comparing numerical results on correlation functions with analytical predictions, though not easy in practice. 

We note that in the $J\rightarrow 0$ limit, the Hamiltonian in Eq. (\ref{eq:6rotated}) has an additional $\mathbb{Z}_2\times \mathbb{Z}_2$ symmetry, corresponding to the global spin rotations $R(\hat{\alpha}^\prime,\pi)$ ($\alpha=x,y,z$). 
In this case, the off-diagonal matrix elements in $\mathcal{C}_i$, $\mathcal{D}_i$  ($i=1,2,3$) vanish, 
i.e., six out of ten parameters $\sigma_C$, $\rho_C$, $\delta_C$, $\sigma_D$,  $\rho_D$ and $\delta_D$ approach zero due to the additional $\mathbb{Z}_2\times \mathbb{Z}_2$ symmetry. 
When $J\ll 1$, these six parameters are very small though not vanishing. 

\section{$\Gamma<0$ region: 120$^\circ$ order from weakly coupled chains}
\label{sec:FM_Gamma}

In this section, we consider the spin-1/2 $KJ\Gamma$ model, and show that the $120^{\circ}$ order on the 2D honeycomb lattice can be obtained from an infinite number of weakly coupled  spin-1/2 chains. 
We study the case of an FM Gamma interaction in this section, while the AFM Gamma interaction will be discussed in Sec. \ref{sec:AFM_Gamma}.

\begin{figure*}[htbp]
\begin{center}
\includegraphics[width=7.5cm]{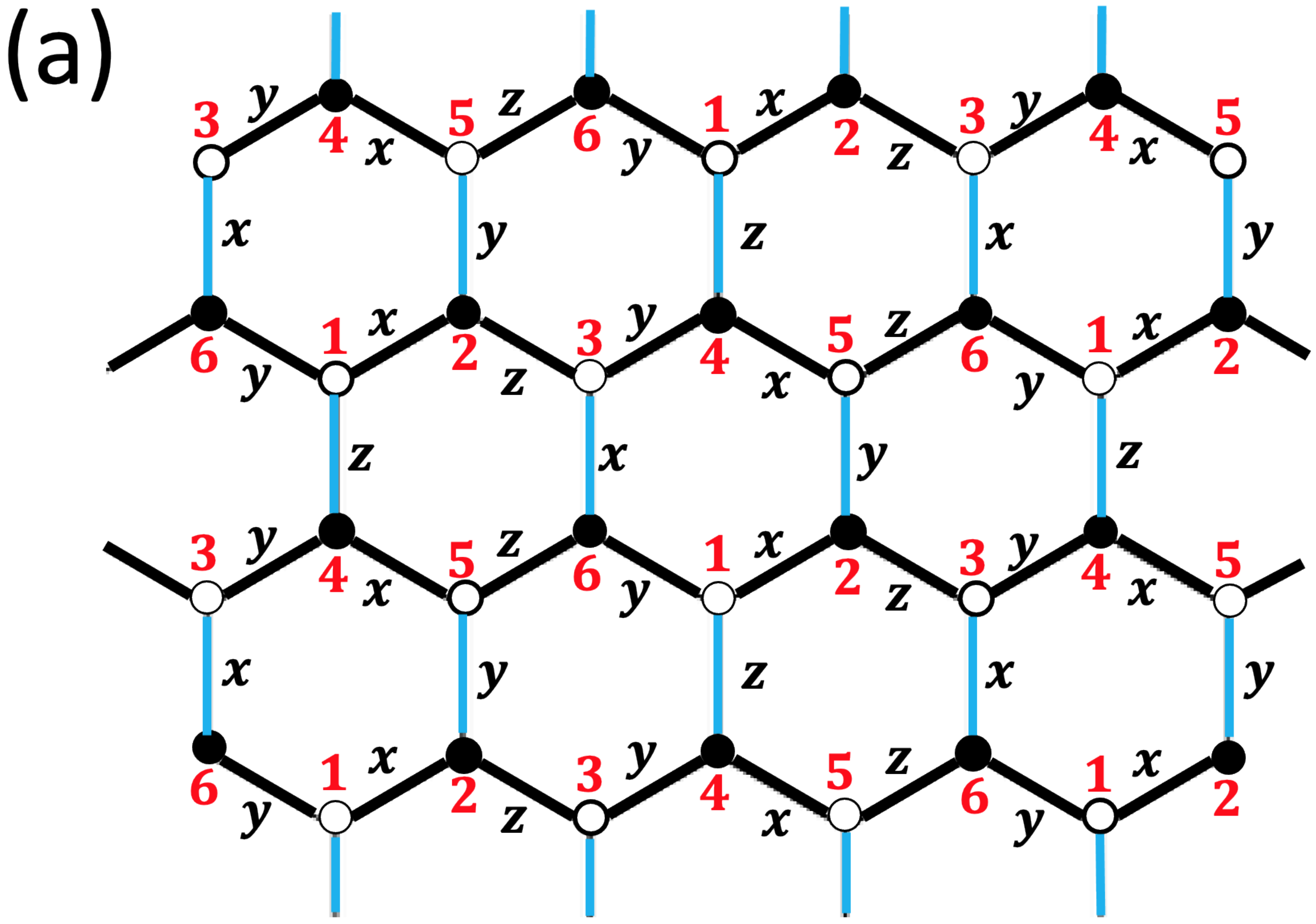}
\includegraphics[width=9.5cm]{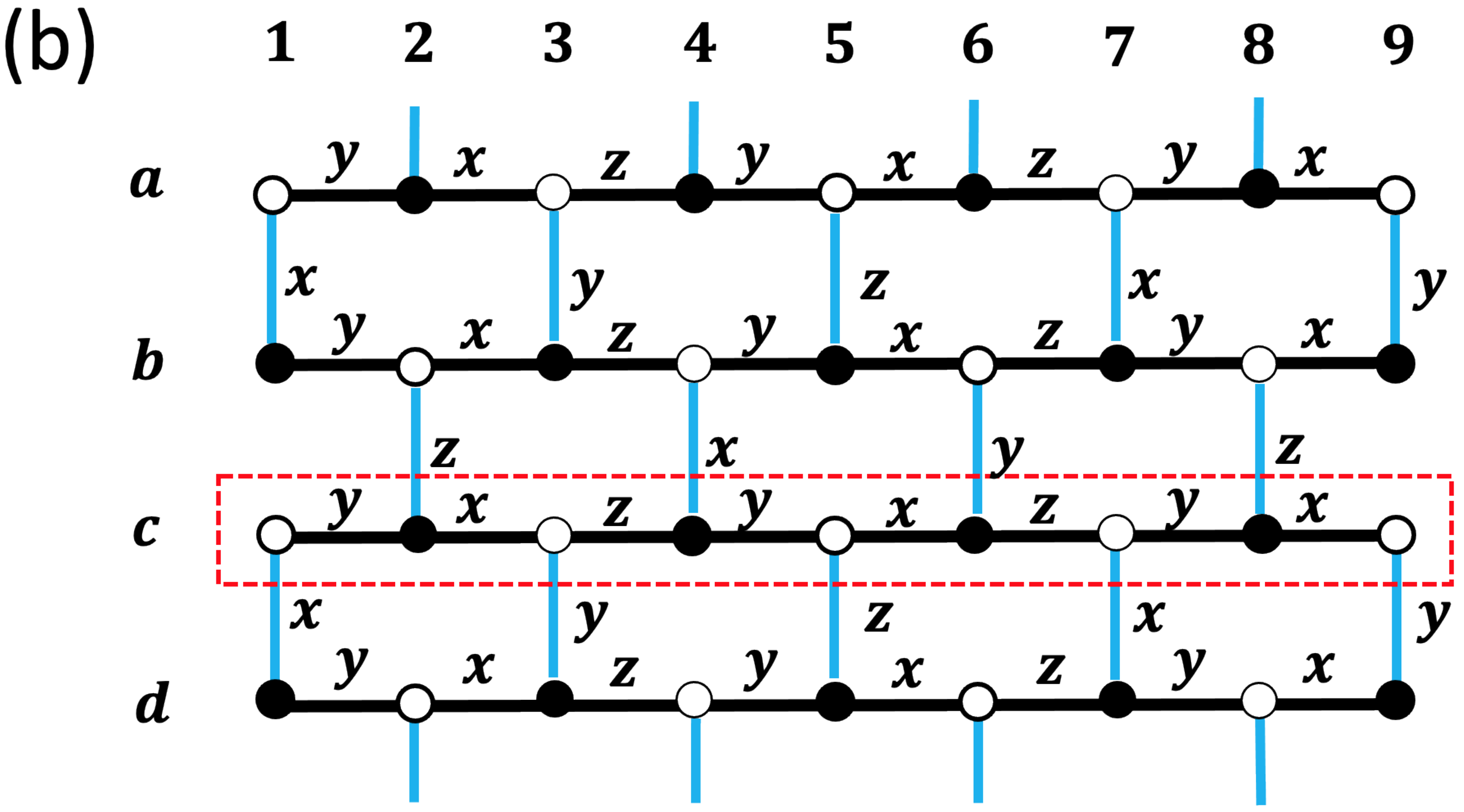}
\caption{(a) Six-sublattice division and bond pattern for the anisotropic  Kitaev-Heisenberg-Gamma model on the honeycomb lattice in the $U_6$ frame,
(b) the two-dimensional brick wall lattice which is equivalent with the honeycomb lattice. 
In (a), the red numbers denote the sublattices in the definition of the $U_6$ transformation in the 2D case.
In (b), the letters and numbers represent the labels for the rows and columns of the brick wall lattice, respectively. 
In both (a,b), $x,y,z$ denote the bond directions; 
the black and blue links denote the strong and weak bonds, respectively; 
the black (white) circles represent the sites in the $A$ ($B$) sublattice of the honeycomb lattice.
} \label{fig:honeycomb}
\end{center}
\end{figure*}

\subsection{Spin ordering in the chain due to inter-chain coupling}
\label{sec:spin_ordering_chain}

We consider an anisotropic  spin-1/2 $KJ\Gamma$ model on the honeycomb lattice defined in Eq. (\ref{eq:Ham_2D_orig}). 
We will study the limit $0<\alpha_z\ll 1$, so that the chains are weakly coupled. 

The six-sublattice rotation $U_6$ can be equally defined on the 2D honeycomb lattice \cite{Chaloupka2015}, and in fact, it was first proposed in 2D and later verified that the transformation is also generalizable  to 1D \cite{Yang2019}. 
The pattern for the local spin rotations in the $U_6$ transformation on the honeycomb lattice is shown in Fig. \ref{fig:honeycomb} (a),
in which the red numbers represent the indices of the sublattices,
and the transformations of the spin operators belonging to the six different types of sublattices are the same as the 1D case defined in Eq. (\ref{eq:6rotation}). 
Throughout this paper, we refer to the division of the honeycomb lattice into six sublattices shown in Fig. \ref{fig:honeycomb} (a) as ``six-sublattice division"  in 2D.

After the six-sublattice rotation, the interaction on bond $\gamma$ becomes
\bea
H_{ij}^\prime&=&-K_\gamma S_i^{\prime\gamma} S_j^{\prime\gamma}-\Gamma_\gamma (S_i^{\prime\alpha} S_j^{\prime\alpha}+S_i^{\prime\beta} S_j^{\prime\beta})\nn\\
&&  -J_\gamma (S_i^{\prime\gamma} S_j^{\prime\gamma}+S_i^{\prime\alpha} S_j^{\prime\beta}+S_i^{\prime\beta} S_j^{\prime\alpha}),
\label{eq:2D_Hamiltonian}
\eea
where the pattern for the bond $\gamma$  is shown in Fig. \ref{fig:honeycomb} (a).
It can be seen that the six-sublattice rotation in 2D reveals two hidden SU(2) symmetric points $(K<0,\Gamma<0,J=0)$ and $(K>0,\Gamma>0,J=0)$ 
corresponding to AFM and FM Heisenberg models, respectively, shown by the ``AFM" and ``FM" points in Fig. \ref{fig:phase-Gamma} (a,b).

The honeycomb lattice is equivalent with the brick wall lattice which is more convenient for our purposes. 
The brick wall lattice is plotted in Fig. \ref{fig:honeycomb} (b) where the zigzag chains formed by strong bonds are stretched into straight horizontal lines. 
The numberings for the rows and columns in the brick wall lattice are represented as letters and numbers, respectively, in Fig. \ref{fig:honeycomb} (b). 
It is clear from Fig. \ref{fig:honeycomb} (b) that the anisotropic  $K J\Gamma$ model can be viewed as an infinite number of weakly coupled chains stacked in  the vertical direction in the figure,
where the interactions along horizontal and vertical links give rise to the intra- and inter-chain couplings, respectively.  

We consider row ``$c$" as a representative row in the brick wall lattice in Fig. \ref{fig:honeycomb} (b).
When $\alpha_z=0$, 
row $c$ is completely decoupled from the other rows and the low energy theory is described by the Luttinger liquid Hamiltonian 
\bea
H_{LL,c}=\frac{v}{2} \int dx [\kappa^{-1} (\nabla \varphi_c)^2 +\kappa (\nabla \theta_c)^2],
\eea
where the subscript ``$c$" in $\varphi_c$ and $\theta_c$ is used to emphasize the fact that these fields only act on the degrees of freedom in row $c$.
In what follows, the subscript $c$ will be neglected  for simplicity. 
We will assume that the Luttinger parameter $\kappa$ satisfies $0.5<\kappa< 1$. 
It is known that $\kappa=0.5$ (up to a logarithmic correction)
in  the emergent SU(2)$_1$ phase in the spin-1/2 Kitaev-Gamma model \cite{Yang2019},
corresponding to the green lines with $J=0$ in Fig. \ref{fig:phase_KHG_1D} (a,b).
Hence $0.5<\kappa< 1$ is satisfied at least for small enough $J$.

When $\alpha_z\neq 0$, magnetic orders may be induced in row $c$ due to the coupling to the other rows. 
In 1+1 dimension, the operators which have scaling dimensions smaller (larger) than two are relevant (irrelevant) in the sense of renormalization group. 
Notice that in the $OU_6$ frame,  $\mathcal{N}^{ x}$ and $\mathcal{N}^{ y}$ have the smallest scaling dimension, both equal to $1/(4\kappa)$.
Therefore, the leading instability in $H_{LL,c}$ is the Neel ordering in the $S^{\prime\prime x}S^{\prime\prime y}$-plane  in the $OU_6$ frame,
and we expect that a long range Ne\'el order in row $c$ is developed within  the $S^{\prime\prime x}S^{\prime\prime y}$-plane when a nonzero $\alpha_z$ is introduced into the system.
Here we emphasize that the restriction of the N\'eel order in the $S^{\prime\prime x}S^{\prime\prime y}$-plane only strictly holds in the limit $J\ll 1$.
For general values of $J$, there are out-of-plane corrections originating from the off-diagonal bosonization coefficients as will be clear from later discussions. 

Since the Luttinger liquid theory has an emergent U(1) symmetry, different directions in the $S^{\prime\prime x}S^{\prime\prime y}$-plane are degenerate, 
and naively it seems that the direction of the Ne\'el order can be along any direction in the $S^{\prime\prime x}S^{\prime\prime y}$-plane.
However, when the system develops a magnetic order, a spin gap opens which provides an infrared cutoff for the RG flow.
In this case, the irrelevant couplings can no longer be completely neglected since they do not flow to zero due to the infrared cutoff set by the spin gap. 
These irrelevant couplings break the emergent U(1) symmetry in the low energy Hamiltonian, and as a result, some special directions are preferred for the Ne\'el order.

Next, we figure out the leading irrelevant coupling and determine the direction of the Ne\'el order in the $OU_6$ frame based on a symmetry analysis.
Before that, it is useful to write down the symmetry transformation properties of the fields $\theta,\varphi$ in the Luttinger liquid theory.
The actions of the operations $T$, $T_a$, $I$, $R(\hat{y}^{\prime\prime},\pi)$ and $R(\hat{z}^{\prime\prime},\beta)$ on the fields $\varphi$ and $\theta$ are given by
\bea
T &:& \theta(t,x)\rightarrow \theta(-t,x)+\sqrt{\pi},~ \varphi(t,x)\rightarrow -\varphi(-t,x),\nn\\
T_a&:& \theta\rightarrow \theta +\sqrt{\pi},~\varphi\rightarrow \varphi+\frac{\sqrt{\pi}}{2},\nn\\
I &:& \theta(t,x)\rightarrow \theta(t,-x),~\varphi(t,x)\rightarrow -\varphi(t,-x),\nn\\
R(\hat{y}^{\prime\prime},\pi)&:&\theta\rightarrow -\theta +\sqrt{\pi}, ~\varphi\rightarrow -\varphi,\nn\\
R(\hat{z}^{\prime\prime},\beta)&:& \theta\rightarrow \theta+\frac{\beta}{\sqrt{\pi}},~\varphi\rightarrow \varphi,
\label{eq:transformation_theta_phi}
\eea
in which $\beta\in\mathbb{R}$ is an angle.


The scaling fields in the Luttinger liquid theory are the vertex operators $e^{i\lambda \varphi}$, $e^{i\lambda \theta}$, and their combinations. 
In the bosonization formulas in Eq. (\ref{eq:abelian_LL1_matrix}), the spin operators are invariant under the transformation $\varphi\rightarrow \varphi+\sqrt{\pi}$, $\theta\rightarrow \theta+2\sqrt{\pi}$. 
Hence, the Hamiltonian is also invariant under these transformations.
This means that only vertex operators of the form $e^{i\sqrt{\pi}(2n \varphi+m\theta)}$ can appear in the low energy Hamiltonian  where $n,m\in \mathbb{Z}$.
Since a N\'eel order within $S^{\prime\prime x}S^{\prime\prime y}$-plane is most easily induced by inter-chain couplings, 
we focus on the vertex operators $\cos(m\sqrt{\pi}\theta)$ and  $\sin(m\sqrt{\pi}\theta)$. 


The transformation properties in Eq. (\ref{eq:transformation_theta_phi}) lead to the following transformations of the $\cos(m\sqrt{\pi}\theta)$ and  $\sin(m\sqrt{\pi}\theta)$ fields under the generators of the symmetry group $G$ in Eq. (\ref{eq:group_G}), 
\bea
T&:& \cos(m\sqrt{\pi}\theta)\rightarrow (-)^m \cos(m\sqrt{\pi}\theta),\nn\\
&&\sin(m\sqrt{\pi}\theta)\rightarrow (-)^m \sin(m\sqrt{\pi}\theta),
\eea
\begin{flalign}
&R(\hat{z}^{\prime\prime},-\frac{2\pi}{3})T_a:\nn\\
& \cos(m\sqrt{\pi}\theta)\rightarrow \nn\\
&(-)^m [\cos(\frac{2\pi m}{3}) \cos(m\sqrt{\pi}\theta)+\sin(\frac{2\pi m}{3}) \sin(m\sqrt{\pi}\theta)],\nn\\
&\sin(m\sqrt{\pi}\theta)\rightarrow \nn\\
&(-)^m [\cos(\frac{2\pi m}{3}) \sin(m\sqrt{\pi}\theta)-\sin(\frac{2\pi m}{3}) \cos(m\sqrt{\pi}\theta)],
\end{flalign}
\bea
R(\hat{y}^{\prime\prime},\pi)I&:&\cos(m\sqrt{\pi}\theta)\rightarrow(-)^m \cos(m\sqrt{\pi}\theta) \nn\\
&&\sin(m\sqrt{\pi}\theta)\rightarrow(-)^{m+1} \sin(m\sqrt{\pi}\theta).
\label{eq:tranformation_cos_sin_theta}
\eea
Combining the transformation properties under $T$ and $R(\hat{y}^{\prime\prime},\pi)I$ in Eq. (\ref{eq:tranformation_cos_sin_theta}), it can be seen that  $\sin(m\sqrt{\pi}\theta)$ is not allowed,
and $m$ has to be an even integer for $\cos(m\sqrt{\pi}\theta)$. 
Then using the transformation under $R(\hat{z}^{\prime\prime},-\frac{2\pi}{3})T_a$, we see that the smallest $m$ which renders $\cos(m\sqrt{\pi}\theta)$ invariant is $m=6$.
Therefore, the following term is allowed in the low energy Hamiltonian
\bea
g\cos(6\sqrt{\pi}\theta),
\label{eq:u}
\eea
in which $g$ is the coupling constant.
Clearly, the scaling dimension of $\cos(6\sqrt{\pi}\theta)$ is $9/\kappa$, which is irrelevant when $0.5<\kappa<1$.

Now we are prepared to figure out what directions can be selected by the irrelevant coupling $\cos(6\sqrt{\pi}\theta)$. 
We need to distinguish between two scenarios, namely $g>0$ and $g<0$.
In principle, the sign of $g$ can be determined from the microscopic $KJ\Gamma$ Hamiltonian.
However, the determination of the sign of $g$ requires a third order perturbation, which is a difficult calculation.
Because of the difficulty, we will  not perform such calculation, and instead discuss both two possibilities. 

If $g$ is positive, then $g\cos(6\sqrt{\pi}\theta)$ is minimized for 
\bea
\theta_n=\frac{2n+1}{6}\sqrt{\pi},~n\in\mathbb{Z}.
\label{eq:solution_n_u_positive}
\eea
Since $\mathcal{N}^{ \pm}\sim e^{\pm i\sqrt{\pi}\theta}$ (where $\mathcal{N}^{ \pm}$ is defined as $\mathcal{N}^x\pm i\mathcal{N}^y$), 
there are six independent values of $n$ in Eq. (\ref{eq:solution_n_u_positive}) given by $1\leq n\leq 6$.
Let's take $n=1$ as an example. 
Plugging $\theta_1=\sqrt{\pi}/2$ into $\mathcal{N}^{ x}$ and $\mathcal{N}^{ y}$, we obtain
\bea
\langle \mathcal{N}^{ x}\rangle =0,~\langle \mathcal{N}^{ y}\rangle \neq 0. 
\label{eq:Neel_y}
\eea
Hence, this is a Ne\'el order along $\hat{y}^{\prime\prime}$-direction in the $OU_6$ frame, dubbed ``the Ne\'el-$y^{\prime\prime}$ order".
In the $U_6$ frame, the spin ordering is along the $(1,0,-1)$-direction.
For the purpose of finding the other five degenerate solutions, it is useful to figure out the broken and unbroken symmetries of the Ne\'el-$y^{\prime\prime}$ order. 
Once this is done, other degenerate Ne\'el orders can be obtained by performing the broken symmetry operations on Eq. (\ref{eq:Neel_y}).

It can be clearly seen that the Ne\'el-$y^{\prime\prime}$ order is invariant under $R(\hat{y}^{\prime\prime},\pi)I$, and in fact, this is the only unbroken symmetry of the Ne\'el-$y^{\prime\prime}$ order. 
Hence, the symmetry breaking pattern is
\bea
D_{3d}\rightarrow \mathbb{Z}_2^{(y)},
\label{eq:sym_break_y}
\eea
where $\mathbb{Z}_2^{(y)}=\mathopen{<} R(\hat{y}^{\prime\prime},\pi)I \mathclose{>}$ is a $\mathbb{Z}_2$ group. 
The other five degenerate ground states can be obtained by applying representative group elements in the cosets of the quotient $D_{3d}/\mathbb{Z}_2^{(y)}$ on the Ne\'el-$y^{\prime\prime} $ order. 
More explicitly, $R(\hat{z}^{\prime\prime},-2\pi/3)T_a$ and $T$ can be chosen as representative  broken symmetries.
Since they generate a group containing six elements, the six degenerate ground states can be obtained by applying the six symmetry operations in the group  $\mathopen{<}T,R(\hat{z}^{\prime\prime},-2\pi/3)T_a\mathclose{>}$ on the Ne\'el-$y^{\prime\prime} $ order.

If $g$ is negative, then $g\cos(6\sqrt{\pi}\theta)$ is minimized for 
\bea
\theta_n=\frac{n}{3}\sqrt{\pi},~n\in\mathbb{Z},
\label{eq:solution_n_u_negative}
\eea
where again $1\leq n\leq 6$.
This time we take $n=6$ as an example, which gives
\bea
\langle \mathcal{N}^{ x}\rangle \neq 0,~\langle \mathcal{N}^{ y}\rangle  = 0. 
\label{eq:Neel_x},
\eea
corresponding to a Ne\'el-$x^{\prime\prime} $ order.
Since the Ne\'el-$x^{\prime\prime} $ order is invariant under $TR(\hat{y}^{\prime\prime},\pi)I$, the symmetry breaking pattern is
\bea
D_{3d}\rightarrow \mathbb{Z}_2^{(x)},
\label{eq:sym_break_x}
\eea
where $\mathbb{Z}_2^{(x)}=\mathopen{<} TR(\hat{y}^\prime,\pi)I \mathclose{>}$ is a $\mathbb{Z}_2$ group. 
The other five degenerate ground states can be obtained by applying representative group elements in the cosets of the quotient $D_{3d}/\mathbb{Z}_2^{(x)}$ on the Ne\'el-$x^{\prime\prime} $ order. 
This time, $R(\hat{z}^{\prime\prime},-2\pi/3)T_a$ and $T$ are still broken symmetries,
hence the six degenerate ground states can be obtained by applying the  six symmetry operations in the group $\mathopen{<}T,R(\hat{z}^{\prime\prime},-2\pi/3)T_a\mathclose{>}$ on the Ne\'el-$x^{\prime\prime} $ order.

\subsection{Self-consistent mean field analysis}
\label{eq:mean_field}

We still need to check whether it is consistent to
assume  a Ne\'el order in the $OU_6$ frame.
In what follows, we take the Ne\'el-$y^{\prime\prime}$ order as an example to perform a self-consistent mean field analysis, which is controllable in the limit $\alpha_z\ll 1$.
The discussion for the Ne\'el-$x^{\prime\prime}$ order is similar. 

For convenience, we denote
\bea
a&=&\frac{1}{\sqrt{2}}\lambda_C+\frac{1}{2\sqrt{2}}\delta_C-\frac{1}{2}\sigma_C\nn\\
b&=&\frac{1}{2\sqrt{2}}\delta_C+\frac{1}{2}\sigma_C\nn\\
c&=&\frac{1}{\sqrt{2}}\lambda_C+\frac{1}{2}\sigma_C\nn\\
d&=&\frac{1}{\sqrt{2}}(\lambda_C+\delta_C).
\label{eq:abcd}
\eea
Setting $\langle \mathcal{N}^{ y}\rangle \neq 0$ in Eq. (\ref{eq:abelian_LL1_matrix_B})  (for explicit expressions, see Eqs. \ref{eq:bosonize_LL1_S1_b},\ref{eq:bosonize_LL1_S2_b},\ref{eq:bosonize_LL1_S3_b}), we obtain the following replacement rule
\bea
\langle\vec{S}^\prime_1 \rangle&=&\langle \mathcal{N}^{ y}\rangle (a,-b,-c)^T\nn\\
\langle\vec{S}^\prime_2 \rangle&=&\langle \mathcal{N}^{ y}\rangle (-d,0,d)^T\nn\\
\langle\vec{S}^\prime_3 \rangle&=&\langle \mathcal{N}^{ y}\rangle (c,b,-a)^T\nn\\
\langle\vec{S}^\prime_4 \rangle&=&\langle \mathcal{N}^{ y}\rangle (-a,b,c)^T\nn\\
\langle\vec{S}^\prime_5 \rangle&=&\langle \mathcal{N}^{ y}\rangle (d,0,d)^T\nn\\
\langle\vec{S}^\prime_6 \rangle&=&\langle \mathcal{N}^{ y}\rangle (-c,-b,a)^T,
\label{eq:spin_pattern_y}
\eea
in which $i$ ($1\leq i \leq 6$) is the sublattice index in the six-sublattice division,
or to say, the spin orientation in Fig. \ref{fig:honeycomb}  (a) on a site with red number $i$ is given by $\langle \vec{S}^\prime_i\rangle$ in Eq. (\ref{eq:spin_pattern_y}).

For row $c$ in Fig. \ref{fig:honeycomb} (b),  the mean field Hamiltonian is 
\bea
H_c=H_{cc}+H_{cb}+H_{cd},
\label{eq:H_c}
\eea
in which $H_{cc}$ is the intra-chain Luttinger liquid Hamiltonian, 
and the interchain interactions between row $c$ and rows $b$, $d$ are
\begin{flalign}
&H_{cb}=\alpha_z\sum_{n}\sum_{j=1,2,3}\big[\nn\\
&-(K+J)S^{\prime \gamma_{2j}}_{c,2j+6n}\langle S^{\prime \gamma_{2j}}_{b,2j+6n}\rangle \nn\\
&-\Gamma (S^{\prime \alpha_{2j}}_{c,2j+6n}\langle S^{\prime \alpha_{2j}}_{b,2j+6n}\rangle +S^{\prime \beta_{2j}}_{c,2j+6n} \langle S^{\prime \beta_{2j}}_{b,2j+6n} \rangle)\nn\\
&-J (S^{\prime \alpha_{2j}}_{c,2j+6n}\langle S^{\prime \beta_{2j}}_{b,2j+6n}\rangle+S^{\prime \beta_{2j}}_{c,2j+6n}\langle S^{\prime \alpha_{2j}}_{b,2j+6n})\rangle\big],
\label{eq:Hcb}
\end{flalign}
and
\begin{flalign}
&H_{cd}=\alpha_z\sum_{n}\sum_{j=1,2,3}\big[\nn\\
&-(K+J)S^{\prime \gamma_{2j-1}}_{c,2j-1+6n}\langle S^{\prime \gamma_{2j-1}}_{b,2j-1+6n}\rangle \nn\\
&-\Gamma (S^{\prime \alpha_{2j-1}}_{c,2j-1+6n}\langle S^{\prime \alpha_{2j-1}}_{b,2j-1+6n}\rangle+S^{\prime \beta_{2j-1}}_{c,2j-1+6n} \langle S^{\prime \beta_{2j-1}}_{b,2j-1+6n} \rangle)\nn\\
&-J (S^{\prime \alpha_{2j-1}}_{c,2j-1+6n}\langle S^{\prime \beta_{2j-1}}_{b,2j-1+6n}\rangle+S^{\prime \beta_{2j-1}}_{c,2j-1+6n}\langle S^{\prime \alpha_{2j-1}}_{b,2j-1+6n})\rangle\big],
\label{eq:Hcd}
\end{flalign}
in which $\lambda$ ($=b,c,d$) and $m$ ($\in\mathbb{Z}$) in  $\vec{S}_{\lambda,m}$
are the row and column indices of the spin operator $\vec{S}^\prime_{\lambda,m}$ in the brick wall lattice shown in Fig. \ref{fig:honeycomb};
 $\alpha_i\neq \beta_i$ ($1\leq i \leq 6$) are the two spin directions other than $\gamma_i$ among $\{x,y,z\}$;
and  $\gamma_1=\gamma_4=x$, $\gamma_2=\gamma_5=z$, $\gamma_3=\gamma_6=y$.
Notice that in Eqs. (\ref{eq:Hcb},\ref{eq:Hcd}), we have replaced the spin operators on rows $b$ and $d$ by their expectation values, in accordance with a mean field treatment.

Next we simplify Eq. (\ref{eq:Hcb}) and Eq. (\ref{eq:Hcd}),
which can be achieved by replacing the spin operators $\vec{S}^\prime_{\lambda,i+6n}$ with the low energy degrees of freedom $\vec{\mathcal{J}}$ and $\vec{\mathcal{N}}$ using the bosonization formulas in Eq. (\ref{eq:abelian_LL1_matrix_B}), where $\lambda=b,c,d$ and $1\leq i\leq 6$.
Recall that there are six different sets of bosonization formulas which apply  to the six sublattices (note: although Eqs. (\ref{eq:bosonize_LL1_S1_b},\ref{eq:bosonize_LL1_S2_b},\ref{eq:bosonize_LL1_S3_b}) seem to give three different sets of formulas,  one needs to further distinguish between odd and even $n$, leading to six sets of formulas).
The sublattice index for the sites in the $b,c,d$ rows should be read from Fig. \ref{fig:honeycomb} (a).
For example, we need to use 
$\vec{S}_3^\prime$  for the spin operator $\vec{S}^\prime_{c,1}$ in Fig. \ref{fig:honeycomb} (b).

Making the above mentioned replacements, we obtain,
\bea
H_{cb}+H_{cd}=\alpha_z \frac{1}{6a}\int dx \mathcal{H},
\label{eq:interchain_MF}
\eea
in which 
\bea
\mathcal{H}=\langle \mathcal{N}^{ y}\rangle[ (\mathcal{J}^{ x},\mathcal{J}^{ y},\mathcal{J}^{ z})A+(\mathcal{N}^{ x},\mathcal{N}^{ y},\mathcal{N}^{ z})B],
\label{eq:H_AB}
\eea
and the column vectors $A$  and $B$ are given by
\bea
A&=&(0,u_D,0)^T,\nn\\
B&=&(0,u_C,0)^T,
\label{eq:AB_column}
\eea
where
\bea
u_D&=&0,\nn\\
u_C&=&(2K+4\Gamma)(\lambda_C)^2\nn\\
&&+[6(\Gamma-J)\delta_C+2\sqrt{2}(K-\Gamma)\sigma_C]\lambda_C\nn\\
&&+3(\Gamma-J)(\delta_C)^2+(3J+K+2\Gamma)(\sigma_C)^2.
\label{eq:uC_uD_Neely}
\eea
Detailed derivations of $A$ and $B$ are included in Appendix \ref{app:derivation_low_MF}.

Therefore, the low energy mean field Hamiltonian for row $c$ in the $U_6$ frame is given by
\bea
H_{MF}&=&\frac{v}{2} \int dx [\kappa^{-1} (\nabla \varphi)^2 +\kappa (\nabla \theta)^2]\nn\\
&&+\frac{1}{6a}\alpha_zu_C\langle \mathcal{N}^{ y}\rangle\int dx \mathcal{N}^{ y}.
\label{eq:H_low_2}
\eea
Notice that $\lambda_C$ is positive, 
and dominates over $\delta_c$ and $\sigma_C$ when $J\ll 1$ according to the discussion by the end of Sec. \ref{subsec:nonsym_bosonize_formula}.
Since $K$ and $\Gamma$ are both FM (i.e., negative), we see that $u_C$ is negative when $J$ is small, and as result, the energy is lowered if $\langle \mathcal{N}^{ y}\rangle \neq 0$. 
Furthermore, it can be seen from the structure of the six-sublattice division in Fig. \ref{fig:honeycomb} (a)
that the mean field Hamiltonians for different rows in Fig. \ref{fig:honeycomb} (b) have the same form. 
Hence, it is fully consistent to assume a nonzero expectation value of $\mathcal{N}^{ y}$ from the start,
and in fact, $\langle \mathcal{N}^{ y}\rangle$ can be determined in a self-consistent way.

The self-consistent equation of the type in Eq. (\ref{eq:H_low_2}) has been solved in Ref. \cite{Yang2022_2},
which is briefly reviewed in Appendix \ref{app:self_consistent}.
Here we only quote the result for the self-consistent solution, which gives
\bea
\langle \mathcal{N}^y \rangle = \frac{1}{a}\big[\frac{\pi \alpha_z |u_C|}{6v\kappa \Lambda^2a^3}\big]^{\frac{1}{8\kappa-2}}.
\label{eq:value_Ny}
\eea
where $\Lambda$ is the momentum cutoff for the low energy Luttinger liquid theory.

\subsection{120$^\circ$ order on the honeycomb lattice}
\label{subsec:120_order_FM_Gamma}

\subsubsection{Approximate 120$^\circ$ order for $\langle \mathcal{N}^{ y}\rangle\neq 0$}

\begin{figure}[h]
\begin{center}
\includegraphics[width=7cm]{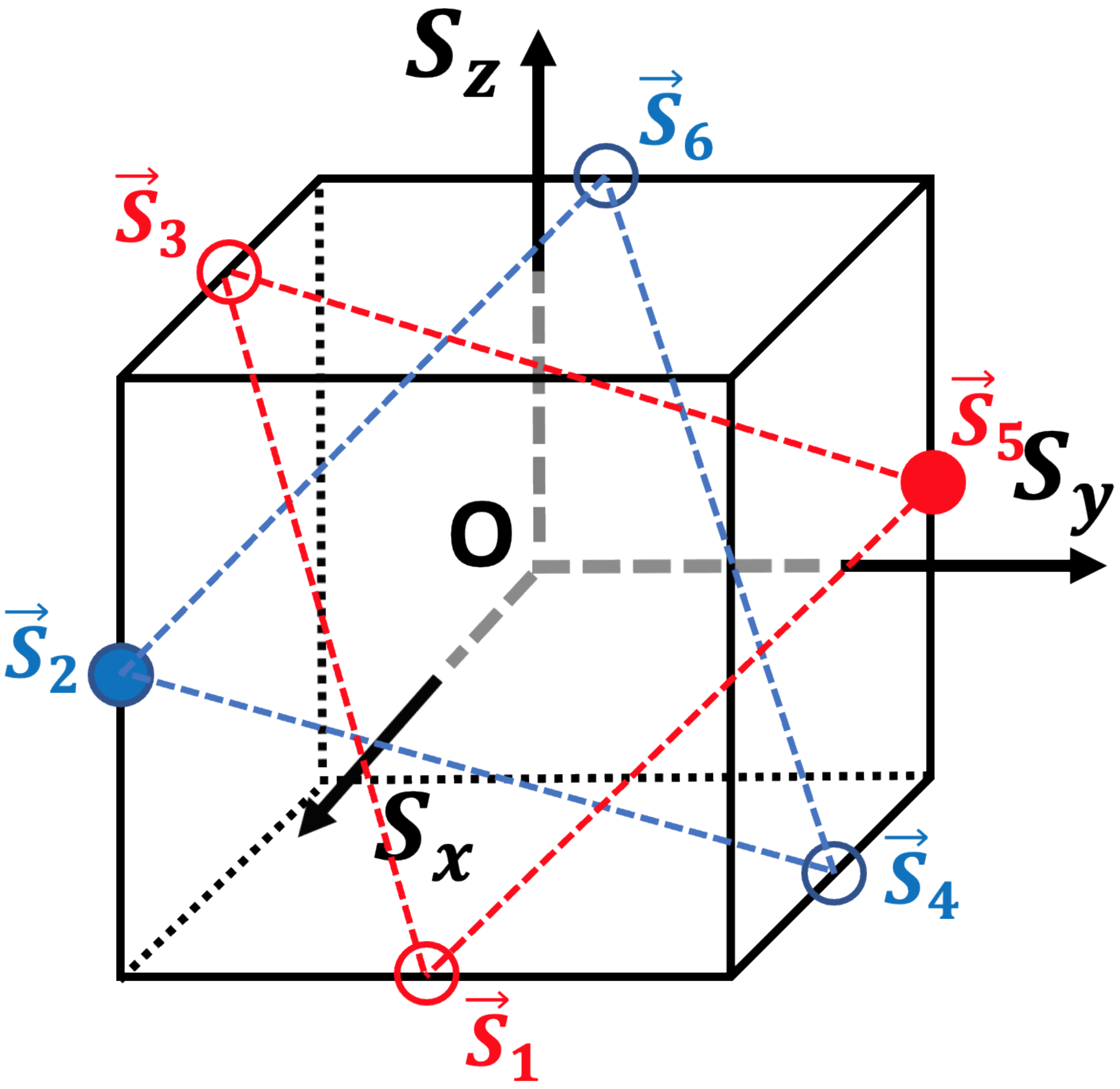}
\caption{Directions of the spin orientations $\vec{S}_i$ within the original frame in sublattice $i$ ($1\leq i \leq 6$) of the six-sublattice division (see Fig. \ref{fig:honeycomb} (a))   for $g>0$ (where $g$ is defined in Eq. (\ref{eq:u})).
The directions for the two solid circles are exact, whereas the remaining directions for the hollow circles are approximate.  
The three spins connected by either red or blue dashed lines are approximately at relative $120^\circ$  angles. 
} \label{fig:Ny_order_original}
\end{center}
\end{figure}

When $g>0$,  $\langle \mathcal{N}^{ y}\rangle$ is non-vanishing, 
and  the spin expectation value in sublattice $i$ ($1\leq i \leq 6$) in the $U_6$ frame is given by Eq. (\ref{eq:spin_pattern_y}).
Performing $(U_6)^{-1}$, we obtain the spin orientations in the original frame, as
\bea
\vec{S}_1&=&\langle \mathcal{N}^{ y}\rangle(a,-b,-c)^T\nn\\
\vec{S}_2&=&\langle \mathcal{N}^{ y}\rangle(d,-d,0)^T\nn\\
\vec{S}_3&=&\langle \mathcal{N}^{ y}\rangle(b,-a,c)^T\nn\\
\vec{S}_4&=&\langle \mathcal{N}^{ y}\rangle(-b,a,-c)^T\nn\\
\vec{S}_5&=&\langle \mathcal{N}^{ y}\rangle(-d,d,0)^T\nn\\
\vec{S}_6&=&\langle \mathcal{N}^{ y}\rangle(-a,b,c)^T,
\label{eq:original_Ny_order_1}
\eea
in which $i$ ($1\leq i\leq 6$) is the sublattice index shown in Fig. \ref{fig:honeycomb} (a),
and $a,b,c,d$ are given by Eq. (\ref{eq:abcd}).

Notice that when $J$ is small, both $\delta_C$ and $\sigma_C$ are small, hence
\bea
a \approx c\approx d,~ b\approx 0,~\text{when $J\ll 1$}.
\eea
In the limit $|\delta_C|,|\sigma_C|\ll1$, Eq. (\ref{eq:original_Ny_order_1}) becomes
\bea
\vec{S}_1&\approx&\langle \mathcal{N}^{y}\rangle(d,0,-d)^T\nn\\
\vec{S}_2&\approx&\langle \mathcal{N}^{y}\rangle(d,-d,0)^T\nn\\
\vec{S}_3&\approx&\langle \mathcal{N}^{y}\rangle(0,-d,d)^T\nn\\
\vec{S}_4&\approx&\langle \mathcal{N}^{y}\rangle(0,d,-d)^T\nn\\
\vec{S}_5&\approx&\langle \mathcal{N}^{y}\rangle(-d,d,0)^T\nn\\
\vec{S}_6&\approx&\langle \mathcal{N}^{y}\rangle(-d,0,d)^T.
\label{eq:original_Ny_order_2}
\eea
Clearly, all the vectors in Eq. (\ref{eq:original_Ny_order_2}) are coplanar and perpendicular to the $(1,1,1)$-direction.
In addition, the approximate expressions of $\vec{S}_1$, $\vec{S}_3$, $\vec{S}_5$  in Eq. (\ref{eq:original_Ny_order_2}) are at relative angles 
$120^\circ$, so do $\vec{S}_2$, $\vec{S}_4$, $\vec{S}_6$.
On the other hand, the angle between $\vec{S}_1$ and $\vec{S}_2$ is approximately $60^\circ$.
As can be checked from the pattern of the six-sublattice division in Fig. \ref{fig:honeycomb} (a), 
the next-nearest neighboring spins are at relative $120^\circ$ approximate angles  in the $J\ll 1$ limit, 
whereas the angles between nearest neighbor spins are  approximately $60^\circ$.

Fig. \ref{fig:Ny_order_original} shows the directions of the spin orientations in the original frame for the Ne\'el-$y^{\prime\prime}$ order, where $\vec{S}_i$ represents the spin direction in the $i$'th sublattice in the six-sublattice division of the honeycomb lattice defined in Fig. \ref{fig:honeycomb} (a).
According to Eq. (\ref{eq:original_Ny_order_1}), the directions for $\vec{S}_2$ and $\vec{S}_5$ are exact as represented by the solid circles in Fig. \ref{fig:Ny_order_original}, whereas the directions for the remaining spins represented by hollow circles are approximate, which only hold in the small $J$ limit. 
In Fig. \ref{fig:Ny_order_original}, each $\vec{S}_i$ represents a sublattice.
The larger sublattice formed by the three sublattices corresponding to the three spins connected by either the red or the blue dashed lines in Fig. \ref{fig:Ny_order_original}
is the $A$ or $B$ sublattice of the honeycomb lattice.
Since the angle between any two of the three spins connected by either the red or blue dashed lines in Fig. \ref{fig:Ny_order_original} is approximately $120^\circ$,
we see that the nearest neighboring spins in the $A$ or $B$ sublattice are approximately at a relative 120$^\circ$ angle,
which is the origin of the name of ``$120^\circ$ order" proposed in Ref. \cite{Rau2014}.

\begin{figure}[h]
\begin{center}
\includegraphics[width=8cm]{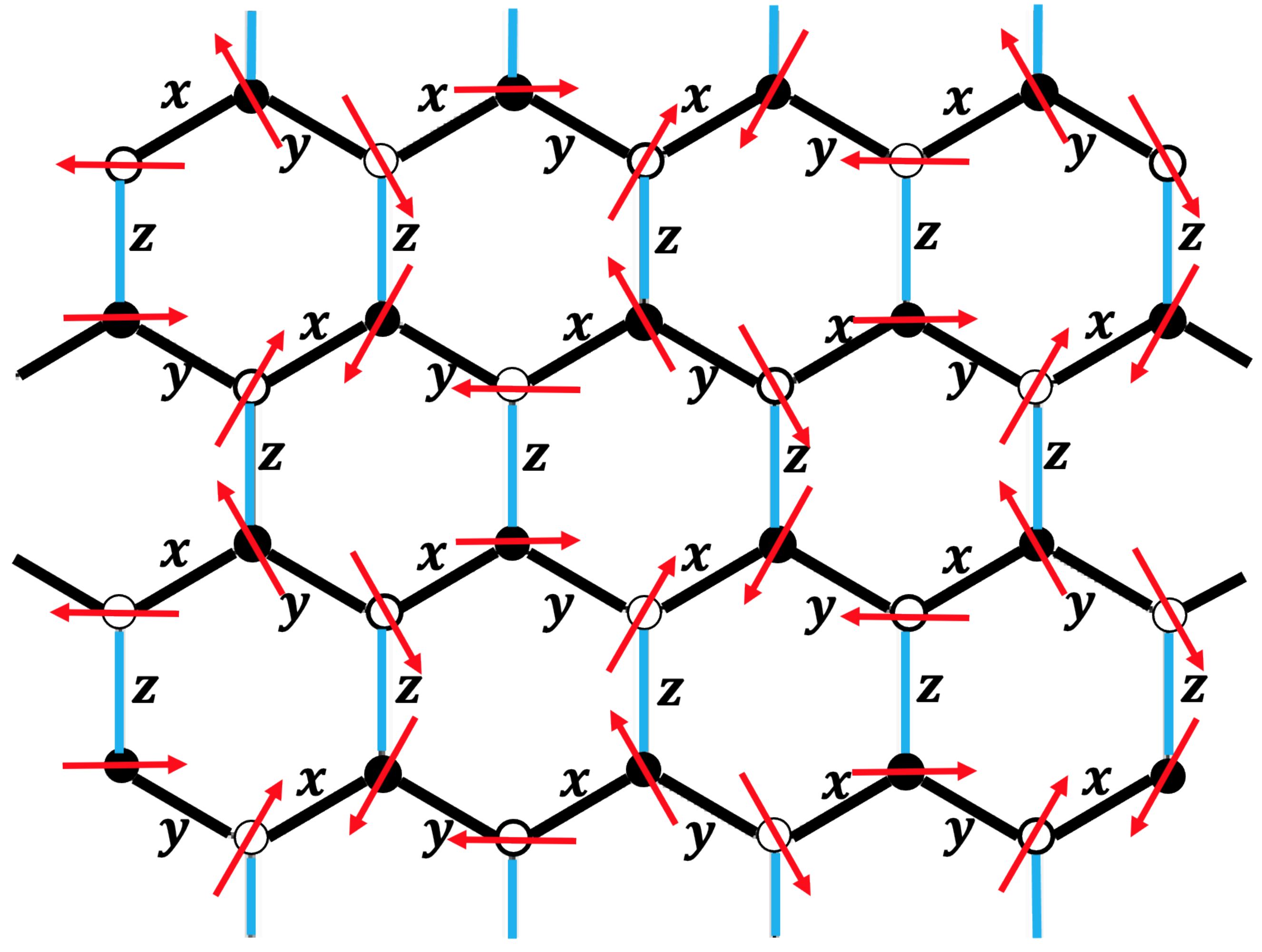}
\caption{
Spin ordering for the $120^\circ$ order in the original frame when $x$- and $y$-bonds are exchanged. 
The red arrows are approximately coplanar in spin space and  the shared plane is perpendicular to the $(1,1,1)$-direction.
The black (white) circles represent the sites in the $A$ ($B$) sublattice of the honeycomb lattice.
The spin orientations in the figure are only approximate since there are corrections due to the bosonization coefficients $\sigma_C$, $\delta_C$.
} \label{fig:reflect_xy}
\end{center}
\end{figure}

The spin pattern on the honeycomb lattice corresponding to Fig. \ref{fig:Ny_order_original} is shown in Fig. \ref{fig:orders} (a),
which exhibits a vortex-like structure and has a right-handed chirality.
The left-handed vortex can be obtained by performing the time reversal operation on the spin pattern in Fig. \ref{fig:orders} (a),
which is an energetically degenerate configuration.
It seems at first sight that the spin pattern in Fig. \ref{fig:orders} (a) is different from the one in Ref.  \cite{Rau2014}.
However, we note that they are essentially the same.
Fig. \ref{fig:reflect_xy} shows the spin ordering pattern when the $x$- and $y$-bonds in the original frame are switched relative to Fig. \ref{fig:honeycomb_original}.
It is clear that the pattern in Fig. \ref{fig:reflect_xy} is the same as that given in Ref. \cite{Rau2014}. 

On the other hand, there are in total six degenerate ground states.
The spin configurations in the other five ground states can be obtained by performing the broken symmetry transformations on the Ne\'el-$y^{\prime\prime}$ configuration in the $U_6$ frame, and then transforming to the original frame.
Recall that in the $U_6$ frame, the representative broken symmetries can be chosen as $T$ and $R(\hat{z}^{\prime\prime},-\frac{2\pi}{3})T_a$.
The corresponding operations in the original frame can be obtained from $(U_6)^{-1}TU_6$ and $(U_6)^{-1}R(\hat{z}^{\prime\prime},-\frac{2\pi}{3})T_aU_6$.
Straightforward calculations give
\bea
(U_6)^{-1}TU_6&=&T,\nn\\
(U_6)^{-1}R(\hat{z}^{\prime\prime},-\frac{2\pi}{3})T_aU_6&=&R(\hat{n}_N,\pi)T_a,
\label{eq:U6_T_RaTa}
\eea
in which $\hat{n}_N=\frac{1}{\sqrt{2}}(1,-1,0)$
and the action of $R(\hat{n}_N,\pi)$ is given by 
\bea
R(\hat{n}_N,\pi):(x,y,z)\rightarrow (-y,-x,-z).
\eea
By acting $(U_6)^{-1}TU_6$ and $(U_6)R(\hat{z}^{\prime\prime},-\frac{2\pi}{3})T_aU_6$ on the spin configuration in Fig. \ref{fig:Ny_order_original}, we obtain the spin orientations in the other five degenerate symmetry breaking ground states which are shown in 
Fig. \ref{fig:Ny_order_original_app} (b-f) in Appendix \ref{app:figures}.

In this way, we recover the $120^\circ$ order on the honeycomb lattice for the anisotropic  spin-1/2 $KJ\Gamma$ model in limit of the weak inter-chain couplings.
Assuming an absence of phase transition from $\alpha_z\ll1$ to $\alpha_z=1$,
the $120^\circ$ order  applies  to the isotropic  2D spin-1/2 $KJ\Gamma$ model as well.  
Hence, we see that the $120^\circ$ order has an essentially 1D nature. 

\subsubsection{The case for $\langle \mathcal{N}^{ x}\rangle\neq 0$}

Next we study the $g<0$ case, which leads to a nonzero $\langle \mathcal{N}^{ x}\rangle$.
For later convenience, we denote 
\bea
e&=&\frac{1}{\sqrt{6}}\lambda_C+\frac{\sqrt{3}}{2\sqrt{2}}\delta_C+\frac{1}{2\sqrt{3}}\sigma_C\nn\\
f&=&\sqrt{\frac{2}{3}} \lambda_C+\frac{\sqrt{3}}{2\sqrt{2}}\delta_C-\frac{1}{2\sqrt{3}}\sigma_C\nn\\
g&=&\frac{1}{\sqrt{6}}\lambda_C+\frac{1}{2\sqrt{3}} \sigma_C \nn\\
p&=&\frac{1}{\sqrt{6}} \lambda_C-\frac{1}{\sqrt{3}}\sigma_C\nn\\
q&=&\sqrt{\frac{2}{3}}\lambda_C+\frac{1}{\sqrt{3}}\sigma_C.
\label{eq:efgpq}
\eea
Notice that when $J$ is small, $\delta_C$ and $\sigma_C$ are small, and hence
\bea
f\approx q\approx 2e\approx 2g\approx 2p.
\eea

Assuming  $\langle \mathcal{N}^{ x}\rangle \neq 0$
and performing $(U_6)^{-1}$, we obtain the spin orientations in the original frame, as
\bea
\vec{S}_1&=&\langle \mathcal{N}^{ x}\rangle(-e,f,-g)^T\nn\\
\vec{S}_2&=&\langle \mathcal{N}^{ x}\rangle(-p,-p,q)^T\nn\\
\vec{S}_3&=&\langle \mathcal{N}^{ x}\rangle(f,-g,-e)^T\nn\\
\vec{S}_4&=&\langle \mathcal{N}^{ x}\rangle(f,-e,-g)^T\nn\\
\vec{S}_5&=&\langle \mathcal{N}^{ x}\rangle(-p,-p,q)^T\nn\\
\vec{S}_6&=&\langle \mathcal{N}^{ x}\rangle(-g,f,-e)^T.
\label{eq:original_Nx_order_1}
\eea
In the limit $J\ll 1$, Eq. (\ref{eq:original_Nx_order_1}) reduces to
\begin{flalign}
&\vec{S}_1\approx \vec{S}_6\approx\lambda_C\langle \mathcal{N}^{ x}\rangle\frac{1}{\sqrt{6}}(1,-2,1)^T\nn\\
&\vec{S}_2\approx\vec{S}_5\approx\lambda_C\langle \mathcal{N}^{ x}\rangle\frac{1}{\sqrt{6}}(1,1,-2)^T\nn\\
&\vec{S}_3\approx\vec{S}_4\approx\lambda_C\langle \mathcal{N}^{ x}\rangle\frac{1}{\sqrt{6}}(-2,1,1)^T.
\label{eq:original_Nx_order_approx}
\end{flalign}

\begin{figure}[h]
\begin{center}
\includegraphics[width=7cm]{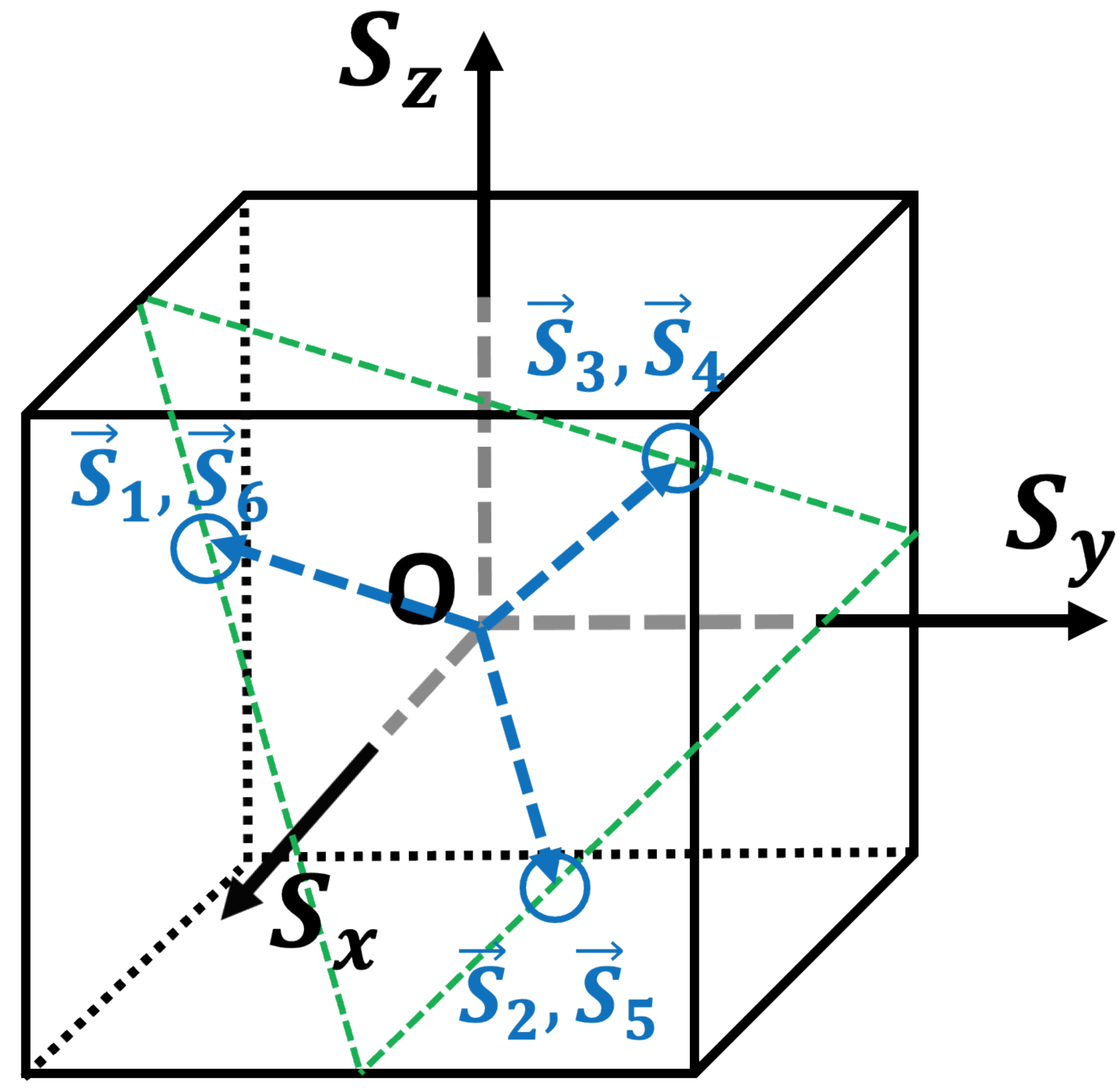}
\caption{
Directions of the spin orientations $\vec{S}_i$  within the original frame for the N\'eel-$x^{\prime\prime}$ order in sublattice $i$ ($1\leq i \leq 6$) of the six-sublattice division (see Fig. \ref{fig:honeycomb} (b))  for $g<0$ (where $g$ is defined in Eq. (\ref{eq:u})).
All the directions of the spins  are approximate due to the bosonization coefficients $\delta_C$, $\sigma_C$.
} \label{fig:Nx_order_original}
\end{center}
\end{figure}

\begin{figure}[h]
\begin{center}
\includegraphics[width=8cm]{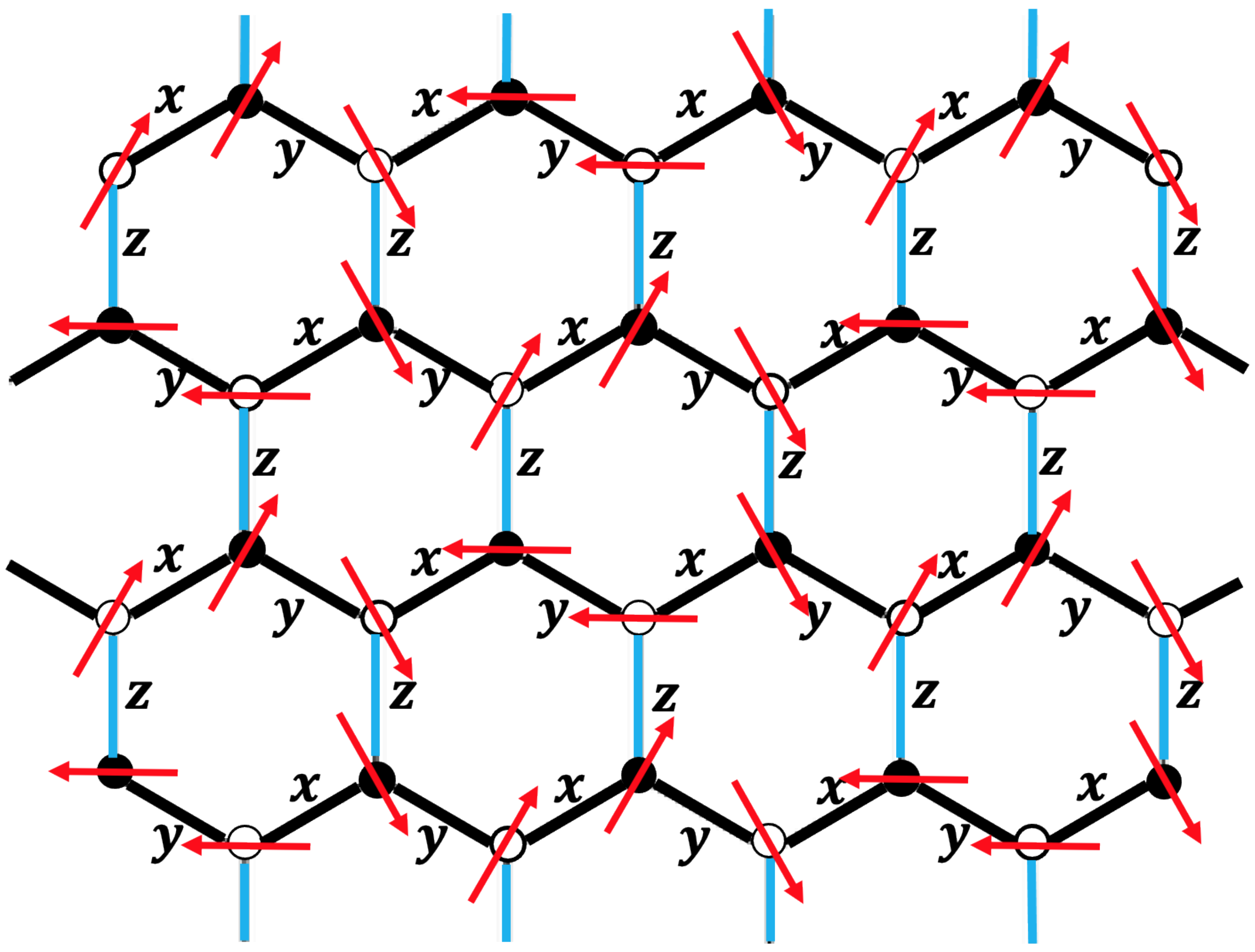}
\caption{Spin ordering pattern of the $120^\circ$ order in the original frame  for the Ne\'el-$x^{\prime\prime}$ order,
corresponding to Fig. \ref{fig:Nx_order_original}. 
The red arrows are approximately within the plane perpendicular to the $(1,1,1)$-direction in the spin space. 
The black (white) circles represent the sites in the $A$ ($B$) sublattice of the honeycomb lattice.
The spin orientations in the figure are only approximate since there are corrections due to the bosonization coefficients $\sigma_C$, $\delta_C$.
} \label{fig:vortex_Neel_x_1}
\end{center}
\end{figure}

Fig. \ref{fig:Nx_order_original}  shows the directions of the spin orientations in the original frame for the Ne\'el-$x^{\prime\prime}$ order, where $\vec{S}_i$ represents the $i$'th sublattice in the six-sublattice division of the honeycomb lattice defined in Fig. \ref{fig:honeycomb} (a).
All the spin directions are approximate which only hold in the small $J$ limit. 
The angle between any two of the  hollow circles in Fig. \ref{fig:Nx_order_original} is approximately $120^\circ$,
hence it is again a type of $120^\circ$ order, though different from the one for the Ne\'el-$y^{\prime\prime}$ case. 
The spin ordering pattern on the honeycomb lattice corresponding to Fig. \ref{fig:Nx_order_original}
again exhibits a vortex-like structure in the original frame,
which is shown in Fig. \ref{fig:vortex_Neel_x_1}.

We note that the other five degenerate ground states can be obtained by performing the broken symmetry transformations on the Ne\'el-$x^{\prime\prime}$ configuration in the $U_6$ frame, and then transforming back to the original frame.
The representative broken symmetry operations in the original frame can again be chosen as $(U_6)^{-1}TU_6$ and $(U_6)^{-1}R(\hat{z}^{\prime\prime},-\frac{2\pi}{3})T_aU_6$, whose expressions
 are given in Eq. (\ref{eq:U6_T_RaTa}).
By acting $(U_6)^{-1}TU_6$ and $(U_6)R(\hat{z}^{\prime\prime},-\frac{2\pi}{3})T_aU_6$ on the spin configuration in Fig. \ref{fig:Nx_order_original}, we obtain the spin configurations in the  other five degenerate symmetry breaking ground states which are shown in 
Fig. \ref{fig:Nx_order_original_app} (b-f) in Appendix \ref{app:figures}.
 
\section{$\Gamma>0$ region: 120$^\circ$, counter-rotating spiral, and zigzag orders}
\label{sec:AFM_Gamma}

Our previous discussion applies to an FM Gamma interaction.
In this section, we analyze the $(K<0, \Gamma>0,J>0)$ region in the phase diagram of the spin-$1/2$ $KJ\Gamma$ model by weakly coupling an infinite number of 1D chains.
The analysis reveals three distinct magnetic orders in the AFM Gamma region, including another type of 120$^\circ$ order, a commensurate counter-rotating spiral order, and a zigzag order. 
The two phase transition lines between the three orders are predicted to be first order phase transitions.
Interestingly, these two first order phase transition lines terminate at a common end point at $J=0$, $K=-2\Gamma$, which is predicted to be a quantum critical point, being a point of continuous  phase transition  in nature. 

We note that the planes in the spin space for the coplanar spin orientations in the 120$^\circ$ orders
are different between the $\Gamma>0$ and $\Gamma<0$ cases.
The normal direction of the plane is along the $(1,1,-1)$-direction for $\Gamma>0$, whereas it is $(1,1,1)$ for $\Gamma<0$,
both in the small $J$ limit. 
Hence the 120$^\circ$ orders in the $\Gamma<0$ and $\Gamma>0$ regions are essentially distinct, denoted as ``120$^\circ$ I" and ``120$^\circ$ II"
in Fig. \ref{fig:phase-Gamma} (a) and (b), respectively.

In this section, we again  assume that  the $x$- and $y$-bonds are the strong bonds and the $z$-bonds are the weak bonds.
The results for the cases when either $x$- or $y$-bonds are considered as the weak bonds
can be obtained by performing a spin-orbit coupled $C_3$ rotation.


\subsection{Single chain analysis}

According to  Eq. (\ref{eq:equiv_1D}), 
for the $\Gamma>0$ case, 
the 1D analysis for a single chain requires an additional global $\pi$-rotation around $z$-axis in the original frame, before applying the $U_6$ transformation.
Throughout this section, $\vec{S}_j^\prime$ will be used to denote the spin operators in the $U_6R(\hat{z},\pi)$ frame. 

Performing $U_6R(\hat{z},\pi)$, the intra-chain Hamiltonian becomes
\bea
&-KS_i^{\prime\gamma} S_j^{\prime\gamma}+\Gamma (S_i^{\prime\alpha} S_j^{\prime\alpha}+S_i^{\prime\beta} S_j^{\prime\beta})  \nn\\
&-J(S_i^{\prime\gamma} S_j^{\prime\gamma}+S_i^{\prime\alpha} S_j^{\prime\beta}+S_i^{\prime\beta} S_j^{\prime\alpha})
\label{eq:dimer_ham_U6Z_1}
\eea
where $\gamma=x,y$,
whereas the inter-chain Hamiltonian is
\bea
&\alpha_z[-KS_i^{\prime z} S_j^{\prime z}-\Gamma (S_i^{\prime x} S_j^{\prime x}+S_i^{\prime y} S_j^{\prime y}) \nn\\
& -J(S_i^{\prime z} S_j^{\prime z}+S_i^{\prime x} S_j^{\prime y}+S_i^{\prime y} S_j^{\prime x})]
\label{eq:dimer_ham_U6Z_2}
\eea
which differs from the intra-chain Hamiltonian by a sign in $\Gamma$. 
Notice that the intra-chain Hamiltonian reduces to the SU(2) symmetric AFM Heisenberg model when $K=-\Gamma<0$, $J=0$,
which is the advantage for applying a global $R(\hat{z},\pi)$ rotation in addition to $U_6$.

The nonsymmorphic bosonization formulas in Eq. (\ref{eq:abelian_LL1_matrix_B}) (for explicit forms, see Eqs. (\ref{eq:bosonize_LL1_S1_b},\ref{eq:bosonize_LL1_S2_b},\ref{eq:bosonize_LL1_S3_b})) equally applies to the  $\Gamma>0$ case, except that the spin operators $\vec{S}_j^\prime$ are defined in the $U_6R(\hat{z},\pi)$ frame. 
The analysis in Sec. \ref{sec:spin_ordering_chain}  for a decoupled chain also  applies to the current situation  in the $U_6R(\hat{z},\pi)$ frame,
namely: the system is in a Luttinger liquid phase for small enough $J$;  the symmetry axis of the emergent U(1) symmetry is along  the $(1,1,1)$-direction;  
and the system is most sensitive to a Ne\'el order in the $S^{\prime\prime x}S^{\prime\prime y}$-plane, where 
$\vec{S}^{\prime\prime}$ denote the spin operators in the $OU_6R(\hat{z},\pi)$ frame.

There are again two types of six-fold degenerate Ne\'el orders in the $U_6R(\hat{z},\pi)$ frame, represented by Ne\'el-$x^{\prime\prime}$ and  Ne\'el-$y^{\prime\prime}$ orders,
with symmetry breaking patterns given by Eq. (\ref{eq:sym_break_y}) and Eq. (\ref{eq:sym_break_x}), respectively.
Whether the Ne\'el-$x^{\prime\prime}$ or Ne\'el-$y^{\prime\prime}$ order is favored depends on the sign of the coupling $g$ (see Eq. (\ref{eq:u})).

Next, we will consider a system of weakly coupled chains on the honeycomb lattice. 
We take the  $\langle \mathcal{N}^y\rangle\neq 0$ case as an example,
and the analysis for the $\langle \mathcal{N}^x\rangle\neq 0$ case is similar. 

Assuming the same spin order as Eq. (\ref{eq:spin_pattern_y}) in the $U_6R(\hat{z},\pi)$ frame,
we still arrive at Eq. (\ref{eq:H_low_2}),
where the coupling $u_C$ is exactly given by Eq. (\ref{eq:uC_uD_Neely}).
We note that the expression of $u_C$ remains unchanged, since the interchain couplings are defined on $z$-bonds,
which remains the same form as the $\Gamma<0$ case according to Eq. (\ref{eq:2D_Hamiltonian}) and Eq. (\ref{eq:dimer_ham_U6Z_2}).
However, we now need to distinguish between two scenarios: $u_C<0$ and $u_C>0$.


\subsection{$u_C<0$ and the $120^\circ$ order}
\label{subsec:uC_negative}



If $u_C<0$, our previous analysis in Sec. \ref{subsec:120_order_FM_Gamma}  can be directly borrowed to the current situation.
The spin expectation values in the original frame can be obtained by performing $R(\hat{z},\pi)$ to Eq. (\ref{eq:original_Ny_order_1}),
as
\bea
\vec{S}_1&=&\langle \mathcal{N}^{ y}\rangle(-a,b,-c)^T\approx\langle \mathcal{N}^{ y}\rangle(-d,0,-d)^T\nn\\
\vec{S}_2&=&\langle \mathcal{N}^{ y}\rangle(-d,d,0)^T\approx\langle \mathcal{N}^{ y}\rangle(-d,d,0)^T\nn\\
\vec{S}_3&=&\langle \mathcal{N}^{ y}\rangle(-b,a,c)^T\approx\langle \mathcal{N}^{ y}\rangle(0,d,d)^T\nn\\
\vec{S}_4&=&\langle \mathcal{N}^{ y}\rangle(b,-a,-c)^T\approx\langle \mathcal{N}^{ y}\rangle(0,-d,-d)^T\nn\\
\vec{S}_5&=&\langle \mathcal{N}^{ y}\rangle(d,-d,0)^T\approx\langle \mathcal{N}^{ y}\rangle(d,-d,0)^T\nn\\
\vec{S}_6&=&\langle \mathcal{N}^{ y}\rangle(a,-b,c)^T\approx\langle \mathcal{N}^{ y}\rangle(d,0,d)^T,
\label{eq:original_Ny_order_2_AFM_Gamma}
\eea
in which the approximate  expressions in the small $J$ limit are also presented. 
As can be seen from the expression of $u_C$ in Eq. (\ref{eq:uC_uD_Neely}), the condition $u_C<0$  reduces to
$K<-2\Gamma$ in the small $J$ limit (i.e., $|\lambda_C|\gg |\delta_c|, |\sigma_C|$)

Clearly, the spins $\vec{S}_1,\vec{S}_3,\vec{S}_5$ are approximately  at mutual  relative  angle 120$^\circ$, and so do the spins $\vec{S}_2,\vec{S}_4,\vec{S}_6$.
All the spins are coplanar, but the normal direction of the plane is along the $(1,1,-1)$-direction this time. 
Hence, we still obtain a type of 120$^\circ$ order, but the spins lie in a plane different from the $\Gamma<0$ case. 
There are in total six degenerate spin configurations,
in which the spin orientations in the six sublattices of the six-sublattice division (defined in Fig. \ref{fig:honeycomb} (a)) in the original frame
can be obtained by applying $R(\hat{z},\pi)$ to the spin configurations in Fig. \ref{fig:Ny_order_original_app} (a-f).
The spin pattern on the honeycomb lattice is the same as Fig. \ref{fig:phase-Gamma} (c) except that the red arrows approximately lie in the plane perpendicular to the $(1,1,-1)$-direction in the spin space. 


As for the $g<0$ case (where $g$ is defined in Eq. (\ref{eq:u})), 
we have a N\'eel-$x^{\prime\prime}$ order in the $OU_6R(\hat{z},\pi)$ frame, as well as five other degenerate solutions.
The six degenerate spin configurations in the original frame can be obtained by performing the $R(\hat{z},\pi)$ rotation to the configurations in Fig. \ref{fig:Nx_order_original_app} (a-f).

The above analysis holds for the situation where the $x$- and $y$-bonds are the strong bonds, whereas the $z$-bonds are the weak bonds.
On the other hand, if the $x$- or $y$-bonds are the weak bonds, the $(1,1,-1)$-direction should be replaced with the $(-1,1,1)$- or $(1,-1,1)$-direction, which is the normal direction of the plane of the (approximately) coplanar 120$^\circ$ order in the small $J$ limit.
Assuming an absence of phase transition from anisotropic to isotropic cases, 
the three types of 120$^\circ$ orders (i.e., with normal directions of the plane shared by the spin orientations being along $(1,1,-1)$-, $(-1,1,1)$- and $(1,-1,1)$-directions) are degenerate in energies for the isotropic  spin-1/2 $KJ\Gamma$ model on the honeycomb lattice.   


\subsection{$u_C>0$ and the counter-rotating spiral order}
\label{subsec:uC_positive}

\begin{figure*}[htbp]
\begin{center}
\includegraphics[width=7cm]{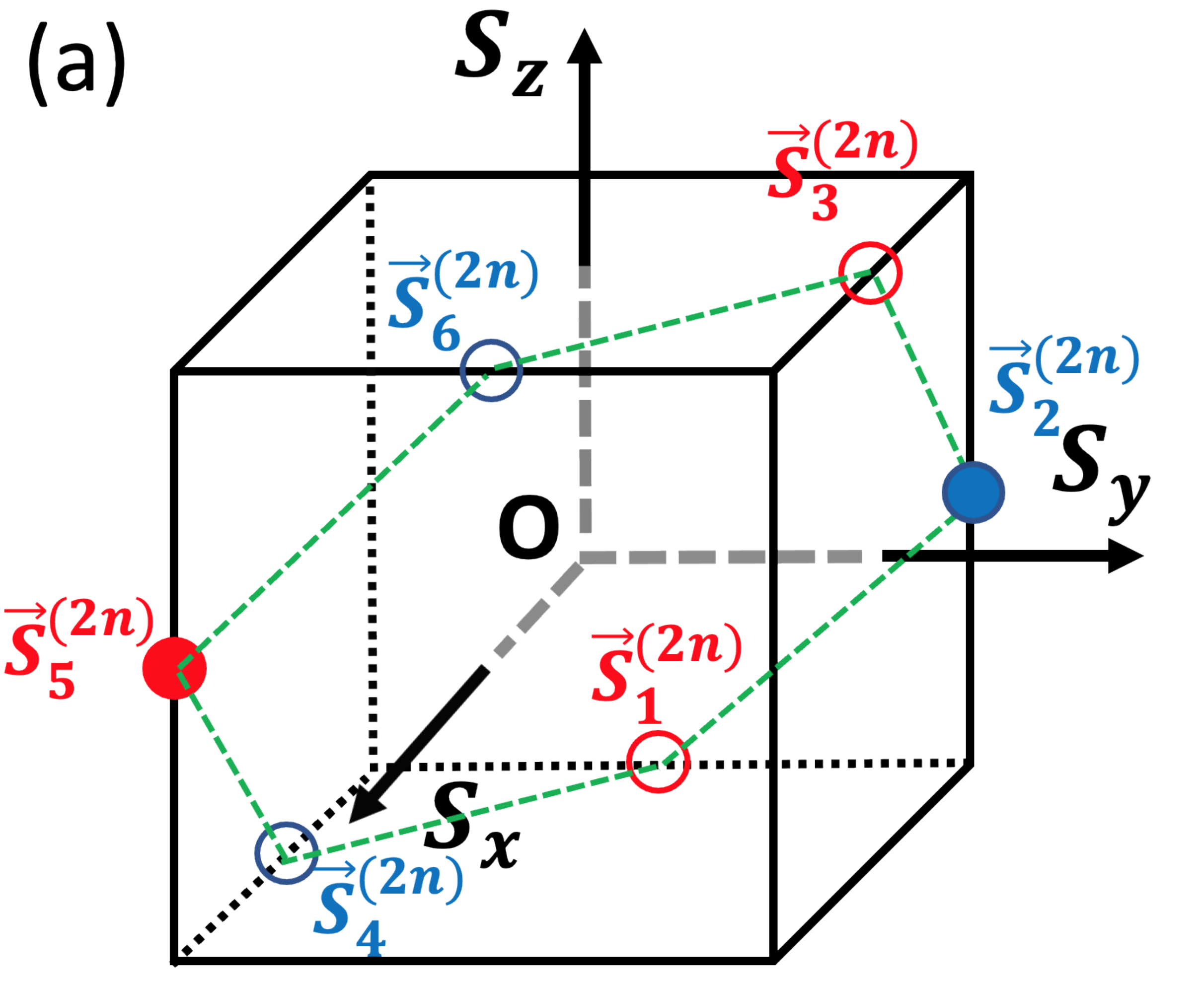}
\includegraphics[width=7cm]{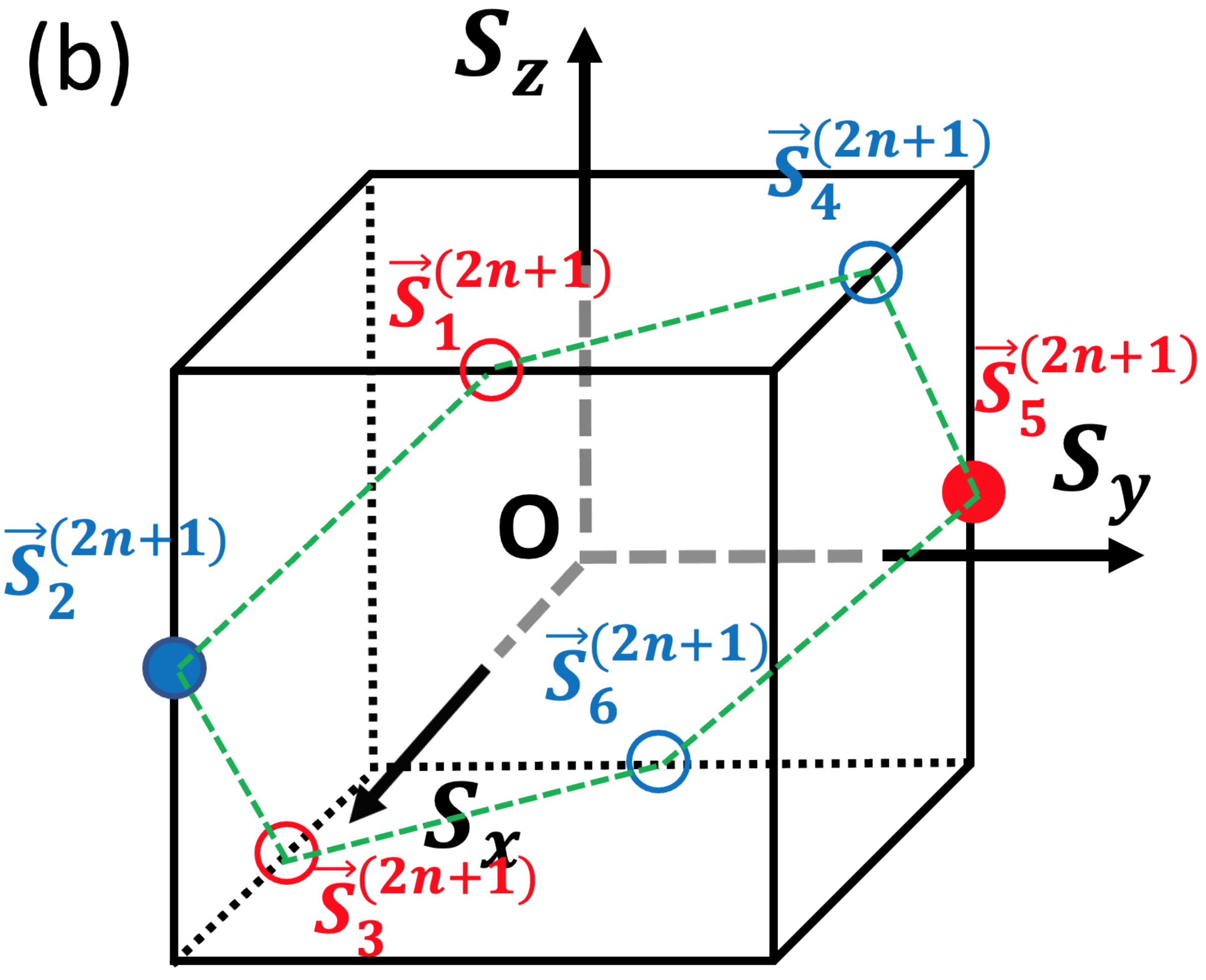}
\caption{
Directions of the spin orientations $\vec{S}^{(m)}_i$ within the original frame in row $m$, sublattice $i$ ($1\leq i \leq 6$) of the six-sublattice division (see Fig. \ref{fig:honeycomb} (a))  for the counter-rotating spiral order in the $\Gamma>0$ case for (a) even $m$ ($=2n$), and (b) odd $m$ ($=2n+1$).
In both (a,b), the hexagon formed by the green dashed line lies in the plane perpendicular to the $(1,1,-1)$-direction. 
The spin orientations in the figure are only approximate since there are corrections due to the bosonization coefficients $\sigma_C$, $\delta_C$.
} \label{fig:spiral}
\end{center}
\end{figure*}

If $u_C>0$, the analysis of the  $120^\circ$ order in Sec. \ref{subsec:120_order_FM_Gamma} does not apply, since a nonzero expectation value of $\langle  \mathcal{N}^{ y}\rangle$ increases rather than decreases the energy. 
The condition $u_C>0$  reduces to
$K>-2\Gamma$
in the small $J$ limit.

Now instead of Eq. (\ref{eq:spin_pattern_y}), we consider the following spin pattern in the $U_6R(\hat{z},\pi)$ frame
\bea
\vec{S}^{\prime(m)}_1&=&(-)^m\langle \mathcal{N}^{ y}\rangle(a,-b,-c)^T\nn\\
\vec{S}^{\prime(m)}_2&=&(-)^m\langle \mathcal{N}^{ y}\rangle(d,-d,0)^T\nn\\
\vec{S}^{\prime(m)}_3&=&(-)^m\langle \mathcal{N}^{ y}\rangle(b,-a,c)^T\nn\\
\vec{S}^{\prime(m)}_4&=&(-)^m\langle \mathcal{N}^{ y}\rangle(-b,a,-c)^T\nn\\
\vec{S}^{\prime(m)}_5&=&(-)^m\langle \mathcal{N}^{ y}\rangle(-d,d,0)^T\nn\\
\vec{S}^{\prime(m)}_6&=&(-)^m\langle \mathcal{N}^{ y}\rangle(-a,b,c)^T,
\label{eq:spin_pattern_y_AFM_Gamma}
\eea
in which 
$a,b,c,d$ are defined in Eq. (\ref{eq:abcd});
$\vec{S}^{\prime(m)}_i$ ($m\in \mathbb{Z}$, $1\leq i \leq 6$) is the spin operator in row $m$, sublattice $i$ in the brick wall lattice in Fig. \ref{fig:honeycomb} (b),
where $i$ is the sublattice index in the six-sublattice division defined in Fig. \ref{fig:honeycomb} (a).
Clearly, because of the staggered sign between adjacent rows in Eq. (\ref{eq:spin_pattern_y_AFM_Gamma}), $u_C$ should be replaced by $-u_C$ in the self-consistent mean field Hamiltonian in Eq. (\ref{eq:H_low_2}), which stabilizes the magnetic order in Eq. (\ref{eq:spin_pattern_y_AFM_Gamma}) by lowering the energy. 


The spin ordering pattern for $\langle \mathcal{N}^{ y} \rangle\neq 0$ in the original frame can be obtained by performing $R(\hat{z},\pi)(U_6)^{-1}$ to Eq. (\ref{eq:spin_pattern_y_AFM_Gamma}), as
\bea
\vec{S}_1^{(m)}&=&(-)^m \langle \mathcal{N}^{ y}\rangle(-a,b,-c)^T\nn\\
\vec{S}_2^{(m)}&=&(-)^m \langle \mathcal{N}^{ y}\rangle(-d,d,0)^T\nn\\
\vec{S}_3^{(m)}&=&(-)^m \langle \mathcal{N}^{ y}\rangle(-b,a,c)^T\nn\\
\vec{S}_4^{(m)}&=&(-)^m \langle \mathcal{N}^{ y}\rangle(b,-a,-c)^T\nn\\
\vec{S}_5^{(m)}&=&(-)^m \langle \mathcal{N}^{ y}\rangle(d,-d,0)^T\nn\\
\vec{S}_6^{(m)}&=&(-)^m \langle \mathcal{N}^{ y}\rangle(a,-b,c)^T,
\label{eq:original_Ny_order_1_B}
\eea
which gives rise to a new type of magnetic order, named as the ``counter-rotating spiral order". 

Fig. (\ref{fig:spiral}) (a) and (b) show the directions of the spin orientations $\vec{S}^{(m)}_i$ in the rows $m=2n$ and $m=2n+1$, respectively, 
within the original frame for the counter-rotating spiral order, 
where $i$ ($1\leq i \leq 6$) is the sublattice index in the six-sublattice division defined in Fig. \ref{fig:honeycomb} (a).
There are five other degenerate spin configurations which can be obtained by applying 
$[U_6R(\hat{z}^\prime,\pi)]^{-1}T[U_6R(\hat{z}^\prime,\pi)]$ and $[U_6R(\hat{z}^{\prime\prime},\pi)]^{-1}R(\hat{z}^{\prime\prime},-\frac{2\pi}{3})T_a[U_6R(\hat{z}^{\prime\prime},\pi)]$ to the spin orientations in Eq. (\ref{eq:original_Ny_order_1_B}). 
In fact, Fig. (\ref{fig:spiral}) can be obtained by first performing a rotation $R(\hat{z},\pi)$ to Fig. \ref{fig:Ny_order_original},
then splitting the even and odd rows, and finally permuting the spins with odd row indices.
The other five degenerate configurations can be obtained in a similar way from Fig. \ref{fig:Ny_order_original_app} (b-f) in Appendix \ref{app:figures}.

The spin ordering pattern on the honeycomb lattice corresponding to Eq. (\ref{eq:original_Ny_order_1_B})
is shown in Fig. \ref{fig:orders} (b).
The spins in Fig. \ref{fig:orders} (b) are approximately coplanar  (exactly coplanar in the small $J$ limit, i.e., $a\approx c\approx d$, $b\approx 0$), 
and the normal direction of the plane is along the $(1,1,-1)$-direction in the small $J$ limit.  

It can be verified from Fig. \ref{fig:orders} (b) that for each horizontal zigzag chain formed by the $x$- and $y$-bonds,
the spin orientations rotate in a clockwise (counter-clockwise) way in the half chain formed by the odd (even) sites, where the numbering of the sites in the chain is given by the column indices in the equivalent brick wall lattice in Fig. \ref{fig:honeycomb} (b).
Hence, this is exactly a counter-rotating spiral order with a wavevector $(\frac{1}{3},0)$ in the 2D reciprocal lattice,
in which the units for the horizontal and vertical components of the wavevector are $2\pi/a$ and $2\pi/b$, respectively, 
where $a$ and $b$ are defined by the lengths of the green arrows in Fig. \ref{fig:honeycomb_original}.

The $g<0$ case can be similarly discussed. 
The spin orientations in the six sublattices of the six-sublattice division (defined according to Fig. \ref{fig:honeycomb} (a)) can be obtained from the configurations in Fig. \ref{fig:Nx_order_original_app} (a-f) in a similar way as the $g>0$ case, i.e., rotation around $R(\hat{z},\pi)$, split of even and odd rows, and permutation of spins in odd rows.

We note that 
by assuming again an absence of phase transition from $\alpha_z\ll 1$ to $\alpha_z=1$, 
there are in total three different degenerate types of counter-rotating spiral orders for the isotropic  spin-1/2 $KJ\Gamma$ model on the honeycomb lattice,
in which the common normal directions of the approximately coplanar spins are along $(1,1,-1)$-, $(-1,1,1)$- and $(1,-1,1)$-directions in the small $J$ limit. 

\subsection{$u_C\sim0$ and the zigzag order}
\label{subsec:uC_0}

When $u_C=0$, neither the 120$^\circ$ nor the counter rotating spiral order is able to lower the ground state energy. 
In this subsection, we show that when $u_C$ is small, a subdominant channel of instability drives the system into a zigzag order,
where subdominant means that the corresponding operator is less relevant than the operators in the dominant channel  in the sense of renormalization group. 
As a consequence, the corresponding AFM region is divided into three subregions as shown in Fig. \ref{fig:phase-Gamma} (b),
which are separated by weak first order phase transitions, represented by the two red lines in Fig. \ref{fig:phase-Gamma} (b). 
The zigzag phase lies in the narrow middle region among the three subregions

Returning back to an instability analysis of a decoupled 1D chain,
we note that $\mathcal{N}^{z}$ is also a relevant operator with scaling dimension $\kappa$.
However, since the scaling dimension of $\mathcal{N}^{ x}$ and $\mathcal{N}^{ y}$ is smaller than that of $\mathcal{N}^{z}$ when  $0.5\leq \kappa\leq 1$,
 the former two operators are more relevant than $\mathcal{N}^{z}$ in the sense of renormalization group. 
As a result, the  N\'eel order in $S^{\prime\prime x}S^{\prime\prime y}$-plane within the $OU_6R(\hat{z},\pi)$ frame  in general dominates over the N\'eel order along $S^{\prime\prime z}$ direction. 
However, this ceases to apply when $u_C=0$, where the instability in the $\mathcal{N}^{ x}$ and $\mathcal{N}^{ y}$ channels vanishes. 
In this case, the system is  prone to the $\mathcal{N}^{ z}$ order. 
In fact, it is expected that the N\'eel-$z^{\prime\prime}$ order should dominates over the N\'eel-$x^{\prime\prime}y^{\prime\prime}$ in a narrow range of $u_C$ around $u_C=0$, since the instability in the  N\'eel-$x^{\prime\prime}y^{\prime\prime}$ is very weak when $|u_C|$ is small. 

Next we perform a self-consistent mean field analysis to figure out the 2D order arising from the 1D N\'eel-$z^{\prime\prime}$ order. 
In this case, the mean field decoupling is still given by Eqs. (\ref{eq:Hcb},\ref{eq:Hcd}).
However, the spin expectation values should be obtained by assuming $\langle \mathcal{N}^{z}\rangle \neq 0$ in the bosonization formulas given by Eq. (\ref{eq:abelian_LL1_matrix_B}).
Letting 
\bea
r&=&\frac{1}{\sqrt{3}} \nu_C-\frac{1}{\sqrt{6}}\rho_C,\nn\\
s&=&\frac{1}{\sqrt{3}} \nu_C+\sqrt{\frac{2}{3}}\rho_C,
\label{eq:define_rs}
\eea
and plugging the expressions
\bea
\vec{S}^{\prime (m)}_1&=&(-)^{m}\langle \mathcal{N}^{z}\rangle(-r,-r,-s)^T\nn\\
\vec{S}^{\prime (m)}_2&=&(-)^{m}\langle \mathcal{N}^{z}\rangle(r,s,r)^T\nn\\
\vec{S}^{\prime (m)}_3&=&(-)^{m}\langle \mathcal{N}^{z}\rangle(-s,-r,-r)^T\nn\\
\vec{S}^{\prime (m)}_4&=&(-)^{m}\langle \mathcal{N}^{z}\rangle(r,r,s)^T\nn\\
\vec{S}^{\prime (m)}_5&=&(-)^{m}\langle \mathcal{N}^{z}\rangle(-r,-s,-r)^T\nn\\
\vec{S}^{\prime (m)}_6&=&(-)^{m}\langle \mathcal{N}^{z}\rangle(s,r,r)^T
\label{eq:Sdouble_prime_Nz}
\eea
into the mean field Hamiltonian in Eq. (\ref{eq:H_c}), we obtain the interchain Hamiltonian as
\bea
H_{cb}+H_{cd}=-\alpha_z \frac{1}{6a}\int dx \mathcal{H}_z,
\label{eq:interchain_MF_z}
\eea
in which 
\bea
\mathcal{H}_z=\langle \mathcal{N}^{z}\rangle[ (\mathcal{J}^{ x},\mathcal{J}^{ y},\mathcal{J}^{ z})A_z+(\mathcal{N}^{ x},\mathcal{N}^{ y},\mathcal{N}^{ z})B_z],
\label{eq:H_AB_z}
\eea
and the column vectors $A_z$  and $B_z$ are given by
\bea
A_z&=&(0,v_D,0)^T,\nn\\
B_z&=&(0,v_C,0)^T,
\label{eq:AB_column_z}
\eea
where
\bea
v_D&=& 0,\nn\\
v_C&=&
2(K+2\Gamma+3J)(\nu_C)^2+4\sqrt{2} (K-\Gamma) \nu_C\rho_C \nn\\
&&+2(2K+\Gamma+3J)(\rho_C)^2.
\label{eq:vC_vD_Neely}
\eea
Detailed derivations of $A_z$ and $B_z$ are included in Appendix \ref{app:derivation_low_MF}.

As is clear from Eq. (\ref{eq:vC_vD_Neely}),
when $K+2\Gamma=0$, 
$v_C$ is positive, approximately equal to $6J(\nu_C)^2$ in the small $\rho_C$ limit (i.e., small $J$).
Hence the energy is lowered by $\langle \mathcal{N}^{z}\rangle\neq 0$. 
As a result, when $u_C$ vanishes, although the system does not develop a N\'eel order in the $S^{\prime \prime x}S^{\prime\prime y}$-plane, 
a magnetic order with $\langle \mathcal{N}^{z}\rangle\neq 0$ lowers the  ground state energy. 
A self-consistent calculation gives (for details, see Appendix \ref{app:self_consistent})
\bea
\langle \mathcal{N}^z \rangle = \frac{1}{a}\big[\frac{\pi \alpha_z |v_C|}{6v\kappa \Lambda^2a^3}\big]^{\frac{\kappa}{2-2\kappa}}.
\label{eq:value_Nz}
\eea

By applying $[U_6R(\hat{z},\pi)]^{-1}$ to Eq. (\ref{eq:Sdouble_prime_Nz}), we obtain the spin orientations in the original frame as 
\bea
S_i^{(m)}=(-)^m \langle \mathcal{N}^z \rangle (r,r,-s)^T,~1\leq i\leq 6.
\label{eq:Sm_Nz}
\eea
In the small $J$ limit, Eq. (\ref{eq:Sm_Nz}) becomes
\bea
S_i^{(m)}\approx \frac{1}{\sqrt{3}}\nu_C (-)^m \langle \mathcal{N}^z \rangle (1,1,-1)^T,~1\leq i\leq 6,
\label{eq:Sm_Nz_3}
\eea
which is FM within the zigzag chains but AFM between the chains. 
This is exactly a zigzag order as shown in Fig. \ref{fig:orders} (c).

We note that in a narrow region away from $u_C=0$,
the zigzag order is energetically more favorable than both the 120$^\circ$ and counter-rotating spiral orders. 
To determine the range of the zigzag order, we compare the energies on a mean field level. 
The lowering of the free energy by the 120$^\circ$ or counter-rotating spiral order is given by 
\bea
\Delta E_{1}= -\frac{1}{6a}\alpha_z|u_C| \langle \mathcal{N}^y \rangle^2, 
\eea
whereas the energy lowered by the zigzag order is
\bea
\Delta E_{2}= -\frac{1}{6a}\alpha_z |v_C| \langle \mathcal{N}^z \rangle^2.
\eea
The condition $|\Delta E_2|>|\Delta E_1|$ reduces to
\bea
|\frac{u_C}{v_C}|<\frac{\langle \mathcal{N}^z \rangle^2}{\langle \mathcal{N}^y \rangle^2},
\label{eq:uC_vC_1}
\eea
in which $u_C$, $v_C$, $\langle \mathcal{N}^y \rangle$, and $\langle \mathcal{N}^z \rangle$ are given in Eq. (\ref{eq:uC_uD_Neely}) and Eq. (\ref{eq:vC_vD_Neely}), Eq. (\ref{eq:value_Ny}), and Eq. (\ref{eq:value_Nz}), respectively. 
Using Eq. (\ref{eq:value_Ny}) and Eq. (\ref{eq:value_Nz}), Eq. (\ref{eq:uC_vC_1}) becomes
\bea
|u_C|^{\frac{4\kappa}{4\kappa-1}}<(\Omega \alpha_z)^{\frac{4\kappa^2-1}{(4\kappa-1)(1-\kappa)}} |v_C|^{\frac{1}{1-\kappa}},
\label{eq:uC_vC_2}
\eea
in which 
\bea
\Omega=\frac{\pi}{6v\kappa\Lambda^2a^3}.
\eea
The phase transition between the zigzag phase and the 120$^\circ$ (or counter-rotating spiral) phase 
is determined by 
replacing ``$<$" with ``$=$" in Eq. (\ref{eq:uC_vC_2}),
which give rise to the red lines in Fig. \ref{fig:phase-Gamma} (b).
Since the order parameters do not vanish at the phase transition points when $J\neq 0$,
the two red lines in  Fig. \ref{fig:phase-Gamma} (b) are first order phase transitions.

In the small $J$ limit, $|u_C|\approx |K+2\Gamma| (\lambda_C)^2$ and $|v_C|\approx 6J(\nu_C)^2$.
Then Eq. (\ref{eq:uC_vC_2}) can be simplified as 
\bea
|K+2\Gamma|<\frac{1}{(\lambda_C)^2}(\Omega\alpha_z)^{\frac{4\kappa^2-1}{4\kappa(1-\kappa)}} 
[J(\nu_C)^2]^{\frac{4\kappa-1}{4\kappa(1-\kappa)}},
\eea
which is the condition for the zigzag order in the $J\ll 1$ limit.
Clearly, for $0.5\leq \kappa\leq 1$, 
the range of $K+2\Gamma$ is very small in the small $J$ limit,
since the exponent of $J$ is positive. 
Hence, the width of the  zigzag order is given by 
\bea
\Delta \phi\sim J^{\frac{4\kappa-1}{4\kappa(1-\kappa)}},
\label{eq:scaling_range_phi}
\eea
where $\phi$ is defined in Eq. (\ref{eq:parametrize_KJG}).
When $\kappa\approx 1/2$ (i.e., close to the equator in Fig. \ref{fig:phase-Gamma}),
the scaling in Eq. (\ref{eq:scaling_range_phi}) becomes $\Delta \phi\sim J$.

\subsection{Quantum critical point at $K=-2\Gamma$, $J=0$}

When $J=0$, it can be seen from 
Eq. (\ref{eq:uC_uD_Neely}) and Eq. (\ref{eq:vC_vD_Neely})  
that both $u_C$ and $v_C$ vanish at $K=-2\Gamma$.
This means that the two first order phase transition lines (represented by the two red lines in Fig. \ref{fig:phase-Gamma} (b)) merge into a single point 
at $J=0$, $K=-2\Gamma$.
According to Eq. (\ref{eq:value_Ny}) and Eq. (\ref{eq:value_Nz}),
the order parameters $\langle \mathcal{N}^y\rangle$ and $\langle \mathcal{N}^z\rangle$ vanish at $(K=-2\Gamma,J=0,\Gamma)$,
represented by the red solid circle in Fig. \ref{fig:phase-Gamma} (b).
Therefore, the point at $J=0$, $K=-2\Gamma$ is a quantum critical point, corresponding to a second order phase transition where the two first order phase transition lines terminate. 

We note that the zigzag, 120$^\circ$, and counter-rotating spiral orders in the AFM Gamma region all have distinct symmetry breaking patterns.
First, it is clear that the zigzag order is different from the 120$^\circ$ and counter-rotating spiral orders,
since the zigzag order does not break the translation symmetry of translating by a vector $\vec{a}$, 
whereas the latter two orders break this symmetry,
where $\vec{a}$ is defined in Fig. \ref{fig:honeycomb_original}. 
Next, we show that the 120$^\circ$ and counter-rotating spiral orders also have distinct symmetry breaking patterns.
Notice that both time reversal  $T$ and the translation operation $\mathcal{T}_{ab}=\frac{3}{2}\vec{a}-\vec{b}$ are symmetries of the 2D $KJ\Gamma$ model,
in which $\vec{a}$ and $\vec{b}$ are defined in Fig. \ref{fig:honeycomb_original}.
It can be verified  from Fig. \ref{fig:orders} (a,b) that the 120$^\circ$ II order  breaks the symmetry $T\mathcal{T}_{ab}$
whereas the counter-rotating spiral order breaks $\mathcal{T}_{ab}$,
indicating that the two orders break different symmetries.
Hence the quantum critical point at $J=0$, $K=-2\Gamma$  is a multi-critical point where several different ordered phases meet. 

On the other hand, it is known that certain continuous phase transitions between two ordered phases with distinct symmetry breaking patterns can be described by the deconfined quantum critical theory \cite{Senthil2004a,Senthil2004,Sandvik2007,Kuklov2008,Melko2008,Harada2013,Kaul2015,Nahum2015,Shao2016,WangC2017},
which is beyond the conventional Landau paradigm of second order phase transitions where the system transits from a disordered phase to an ordered phase when the critical point is traversed. 
Hence it may be worth to further investigate possible connections between  the quantum critical point in the AFM Gamma region and the deconfined  quantum critical theory beyond Landau paradigm from both analytical and numerical sides.
More numerical and analytical studies on the nature of this quantum critical point are valuable and desirable.

We make a comment on the relation between our results and the numerical studies in Ref. \cite{Rau2014}.
The phase diagram of the isotropic spin-1/2 $KJ\Gamma$ model on the honeycomb lattice has been studied in Ref. \cite{Rau2014} using a combination of  exact diagonalization and classical analysis. 
Both  methods revealed a 120$^\circ$ magnetic order in the region of FM Gamma, FM Kitaev, and AFM Heisenberg interactions,
which is consistent with the coupled-chain analysis. 
On the other hand, in the AFM Gamma region (still with FM Kitaev and AFM Heisenberg couplings),
no magnetic order commensurate with the lattice is found in Ref. \cite{Rau2014}. 
Exact diagonalization on a cluster of 24 sites in Ref. \cite{Rau2014} revealed a phase with an incommensurate spiral order in the AFM Gamma region, 
and the incommensurate wavevector of the spiral order varies continuously in the phase.
We note that  the correlation length in this parameter region can be very large, 
particularly  when the system is close to the quantum critical point located at $J=0$, $K=-2\Gamma$.
The large correlation length makes it difficult to determine the magnetic orders in numerical calculations,
which might be the reason why incommensurate behaviors are observed in the numerics in Ref. \cite{Rau2014}.

\subsection{Limitation of the theory and the $J\rightarrow 0$ regime}

Finally we briefly  discuss the limitation of our mean field theory, particularly in the $J\rightarrow 0$ regime. 
The coupled-chain analysis presented in this work crucially depends on the division of the instabilities in the underlying  Luttinger liquid theory into dominant and sub-dominant channels.  
However, these channels become degenerate in the $J=0$ case, which corresponds to a Kitaev-Gamma model. 
The degenerate and near-degenerate cases for small $J$  probably require an independent study, different from the current analysis. 

On the other hand, as discussed in Ref. \cite{Yang2019}, the 1D Kitaev-Gamma model has an intricate  symmetry group $G_0$, which is nonsymmorphic and satisfies $G_0/\mathbb{Z}\cong O_h$, where $O_h$ is the full octahedral group, 
or the largest 3D crystalline point group. 
Therefore, within a coupled-chain approach, there are many  more possibilities of magnetic orders and symmetry breaking patterns in 2D Kitaev-Gamma model because of the much larger symmetry group,
which is worth for future studies. 
Indeed, classical analysis and machine-learning-based method have revealed the great complexity of the phase diagram of the 2D Kitaev-Gamma model \cite{Liu2021,Rayyan2021},
where magnetic orders having a unit cell of $18$, $24$, or even $48$ sites are found.  
In addition, it cannot be ruled out the possibility  that the Kitaev-Gamma model hosts some disordered phases such as nematic paramagnets \cite{Rousochatzakis2017,Catuneanu2018,Gohlke2018,Gohlke2020,Liu2021}.

\section{Relations to $\text{Na$_2$IrO$_3$}$ and $\text{$\alpha$-Li$_2$IrO$_3$}$}
\label{sec:material}

$A_2$IrO$_3$ ($A$=Na, Li) is a family of intensively studied Kitaev candidate materials.
A collinear  zigzag  order has been established for Na$_2$IrO$_3$ experimentally \cite{Choi2012,Liu2011,Ye2012}. 
However, the magnetic order is drastically different when Na is replaced by Li \cite{Manni2014,Gao2013}. 
A counter-rotating spiral order has been experimentally observed in materials including $\alpha$-Li$_2$IrO$_3$ \cite{Williams2016}, $\beta$-Li$_2$IrO$_3$ \cite{Biffin2014}, and $\gamma$-Li$_2$IrO$_3$ \cite{Biffin2014_2}. 
Since $\beta$-Li$_2$IrO$_3$  and $\gamma$-Li$_2$IrO$_3$ have a 3D hyper-honeycomb and stripy-honeycomb lattice structure, respectively, they are not directly relevant to the geometry of our consideration.
On the other hand, $\alpha$-Li$_2$IrO$_3$ has a layered honeycomb structure,
which is relevant to our study. 
In this section, we compare our analytical predictions with the experimentally observed zigzag order in Na$_2$IrO$_3$ and the counter-rotating  spiral order in  $\alpha$-Li$_2$IrO$_3$.

In a recent work of Ref. \cite{Liu2022}, the exchange interactions in $d^5$ Kitaev materials have been analyzed in details,
which reveals that on the nearest neighboring level, the Kitaev, Heisenberg, and Gamma interactions in the Na$_2$IrO$_3$ material are 
FM, AFM, AFM, respectively. 
Assuming $\alpha$-Li$_2$IrO$_3$ to have the same signs of Kitaev, Heisenberg, and Gamma interactions,
our analysis of the spin-1/2 $KJ\Gamma$ model in the $K<0$, $J>0$, and $\Gamma>0$ region can be applied to both Na$_2$IrO$_3$ and $\alpha$-Li$_2$IrO$_3$.

We first compare the experimental results on Na$_2$IrO$_3$ with our analytical predictions.
The spin orientations in the zigzag order of Na$_2$IrO$_3$ are observed to be nearly along the $\hat{x}+\hat{y}$ direction \cite{Winter2016}. 
Comparing with Eq. (\ref{eq:Sm_Nz}), this corresponds to a small $s$,
where $s$ is defined in Eq. (\ref{eq:define_rs}).
Although $\rho_C\ll \nu_C$ in the small $J$ limit,
it is possible to satisfy the condition $\rho_C\simeq -\nu_C/\sqrt{2}$ (so that $s\sim 0$) when $J$ becomes large. 

Next, we discuss the counter-rotating spiral order in $\alpha$-Li$_2$IrO$_3$. 
Ref. \cite{Williams2016} observes a propagation wavevector for the  counter-rotating spiral  order as $(0.32,0,0)$, where the units for the three components in sequence are $2\pi/a$, $2\pi/b$ and $2\pi/c$, respectively. 
While the definitions of $a$ and $b$ are shown in Fig. \ref{fig:honeycomb_original}, $c$ is the lattice constant along the third direction which is perpendicular to the plane of the honeycomb lattice, hence not important for our consideration since the interactions between the layers are small. 
Notice that the coupled-chain analysis predicts a counter-rotating spiral order which is six-site periodic along the zigzag chains. 
Since the unit cell of a zigzag chain contains two sites, the predicted wavevector is  $2\pi/(3a)$, as can be seen by comparing the magnetic order in  Fig. \ref{fig:orders} (b) with the definition of $a$ given  in  Fig. \ref{fig:honeycomb_original}. 
Therefore, the predicted value $1/3$ is close to the experimentally observed value $0.32$,
though the experimental result is about $3\%$ away from commensuration.  

In addition, the experimentally observed  magnetic ordering structure within a honeycomb layer of $\alpha$-Li$_2$IrO$_3$ \cite{Williams2016}
exhibits the same pattern as our theoretical prediction,
except that  the experimental wavevector is slightly incommensurate. 
The expression for the spin ordering in $\alpha$-Li$_2$IrO$_3$ observed in experiments is summarized in Eq. (B4) in Ref. \cite{Williams2016}.
Denoting $\vec{S}_j$ ($j=a_1,a_2,a_3,a_4,a_5,a_6$) to be six consecutive sites within a zigzag chain and assuming a commensurate wavevector,
the pattern in Eq. (B4) in Ref. \cite{Williams2016} can be rewritten as (for details, see Appendix \ref{app:compare_experiment})
\begin{flalign}
&\vec{S}_{a_1}=\frac{1}{\sqrt{2}}(M_{z_o},-M_{z_o},0)^T,\nn\\
&\vec{S}_{a_2}=\frac{1}{2\sqrt{2}}(-\sqrt{3}M_{x_o}+M_{z_o},-\sqrt{3}M_{x_o}-M_{z_o},\sqrt{6}M_{y_o} )^T,\nn\\
&\vec{S}_{a_3}=\frac{1}{2\sqrt{2}}(\sqrt{3}M_{x_o}-M_{z_o},\sqrt{3}M_{x_o}+M_{z_o},-\sqrt{6}M_{y_o} )^T,\nn\\
&\vec{S}_{a_4}=\frac{1}{\sqrt{2}}(-M_{z_o},M_{z_o},0)^T,\nn\\
&\vec{S}_{a_5}=\frac{1}{2\sqrt{2}}(-\sqrt{3}M_{x_o}-M_{z_o},-\sqrt{3}M_{x_o}+M_{z_o},\sqrt{6}M_{y_o} )^T,\nn\\
&\vec{S}_{a_6}=\frac{1}{2\sqrt{2}}(\sqrt{3}M_{x_o}+M_{z_o},\sqrt{3}M_{x_o}-M_{z_o},-\sqrt{6}M_{y_o} )^T.
\label{eq:experiment_Li_1}
\end{flalign}
Comparing with Eq. (\ref{eq:original_Ny_order_1_B}), it is clear that the pattern in Eq. (\ref{eq:experiment_Li_1}) coincides exactly with the theoretical pattern via the following identifications,
\begin{flalign}
&a_1=5,~a_2=6,~a_3=1,~a_4=2,~a_5=3,~a_6=4,
\end{flalign}
and
\bea
a&=&\frac{1}{2\sqrt{2}} (-\sqrt{3}M_{x_o}+M_{z_o}),\nn\\
b&=&\frac{1}{2\sqrt{2}} (\sqrt{3}M_{x_o}+M_{z_o}),\nn\\ 
c&=&\frac{\sqrt{3}}{2} M_{y_o},\nn\\
d&=&\frac{1}{\sqrt{2}}M_{z_o}.
\eea

Notice that according to Eq. (\ref{eq:abcd}), the four parameters $a$, $b$, $c$, $d$ are dependent on  three bosonization coefficients $\lambda_C$, $\sigma_C$, and $\delta_C$, which contain the same number of degrees of freedom as $M_{x_o}$, $M_{y_o}$, and $M_{z_o}$.
Hence the three bosonization coefficients can be determined as
\bea
\lambda_C&=&\frac{1}{\sqrt{3}}(-M_{x_o}+\sqrt{2}M_{y_o}),\nn\\
\sigma_C&=&\frac{1}{\sqrt{3}}(\sqrt{2}M_{x_o}+M_{y_o}),\nn\\
\delta_C&=&\frac{1}{\sqrt{3}}(M_{x_o}-\sqrt{2}M_{y_o})+M_{z_o}.
\eea
Plugging in the experimentally determined ratios  
\bea
M_{x_o}:M_{y_o}:M_{z_o}=0.67:0.33:1,
\eea
we obtain
\bea
\lambda_C:\sigma_C:\delta_C=-0.117:0.738:1.117.
\label{eq:3bosonize_experiment}
\eea
Recall from Eq. (\ref{eq:abcd}) that in the small $J$ limit, we expect $\lambda_C:\sigma_C:\delta_C\approx 1:0:0$.
Hence the values in Eq. (\ref{eq:3bosonize_experiment})  indicate sizable Heisenberg interaction in the $\alpha$-Li$_2$IrO$_3$  material. 
On the other hand, it is worth to emphasize that the extracted values of $\lambda_C$, $\sigma_C$, and $\delta_C$ in Eq. (\ref{eq:3bosonize_experiment})
do not exactly coincide with the corresponding bosonization coefficients in a single decoupled chain,
since there can be renormalization effects due to inter-chain interactions. 
We note that the above discussion illustrates the crucial role played by the nonsymmorphic bosonization formulas in order to compare with experiments,
since the conventional U(1) symmetric bosonization formulas in Eq. (\ref{eq:abelian_bosonize}) are not able  to give a magnetic structure in Eq. (\ref{eq:experiment_Li_1}).

The above discussions confirm that the observed zigzag order in Na$_2$IrO$_3$ and the counter-rotating spiral order in $\alpha$-Li$_2$IrO$_3$ are consistent with the theoretically predicted ones. 
In particular, the monoclinic lattice structures of these materials in general  lead to anisotropies in the bond strengths \cite{Singh2010,Gretarsson2013,Chun2015,Kimchi2015},
which justifies the use of an anisotropic  $KJ\Gamma$ model as proposed in Eq. (\ref{eq:Ham_2D_orig}).
We note that the zigzag and counter-rotating spiral orders have been derived in previous theoretical works based on different generalized Kitaev spin models \cite{Kimchi2011,Singh2012,Foyevtsova2013,Sizyuk2014,Reuther2014,Kimchi2015},
which all involve beyond nearest neighboring interactions. 
Our work  indicates the alternative possibility  that a nearest neighboring spin-1/2 $KJ\Gamma$ model may  be used as a minimal model to capture the zigzag and counter-rotating spiral orders in the $A_2$IrO$_3$ materials ($A=$Na, Li).

In addition, our coupled-chain analysis takes into full account the intra-chain quantum fluctuations, which can be lost in the classical analysis since the latter is applicable in the large-$S$ limit.
By providing a clear physical picture for the origins of the magnetic  orders, 
the analysis based on coupled Luttinger liquid chains
gives a quasi-1D explanation to all the three different types of orders (i.e., the 120$^\circ$, counter-rotating spiral, and zigzag orders) on a theoretically sound ground.

\section{Summary}
\label{sec:summary}

In summary, starting from the Luttinger liquid phase of a decoupled spin-1/2 $KJ\Gamma$ chain with FM Kitaev and AFM Heisenberg  interactions,
we have studied the same model in two dimensions by coupling the Luttinger liquid chains together on the honeycomb lattice.
In addition to reproducing the 120$^\circ$ order in the FM Gamma region,
the coupled-chain analysis reveals three magnetic orders in the AFM Gamma region,
including 120$^\circ$, commensurate counter-rotating spiral, and  zigzag orders. 
While the two lines of phase transitions separating the three orders in the AFM Gamma region are first order phase transitions, 
they merge into a single point at $J=0$, $K=-2\Gamma$, which is predicted to be a quantum critical point. 
More analytical and numerical studies on the nature of this  quantum critical point are worth for future investigations. 
Furthermore, the uncovered zigzag and counter-rotating spiral orders may be applied to explain the observed magnetic orders in the Na$_2$IrO$_3$ and $\alpha$-Li$_2$IrO$_3$ materials. 
Our work reveals rich strongly correlated magnetic properties of the generalized  Kitaev spin models, 
and showcases the importance of the signs and magnitudes of the couplings in stabilizing various types of magnetic orders.

{\it Acknowledgments}
W.Y. and I.A. acknowledge support from NSERC Discovery Grant 04033-2016.
A.N. acknowledges computational resources and services provided by Compute Canada and Advanced Research Computing at the University of
British Columbia. A.N. acknowledges support from the Max Planck-UBC-UTokyo Center for Quantum Materials and the Canada First Research Excellence Fund
(CFREF) Quantum Materials and Future Technologies Program of the Stewart Blusson Quantum Matter Institute (SBQMI).
C.X. is partially supported by Strategic Priority Research Program of CAS (No. XDB28000000).
H.Y.K. is supported by the NSERC Discovery Grant No. 06089-2016, the Centre for Quantum Materials at the University of Toronto, the Canadian Institute for Advanced Research, and the Canada Research Chairs Program.

\appendix

\begin{widetext}
\section{Explicit forms of the Hamiltonians}
\label{sec:ham_explicit}

In this appendix, we give the explicit forms of the 1D Hamiltonian in both the original and the six-sublattice rotated frames.
We also give the forms of the interchain interactions. 

In the original frame, the Hamiltonian of the 1D spin-1/2 $KJ\Gamma$ model has a two-site periodicity.
Within a unit cell, the Hamiltonian is
\bea
H_{12}&=& KS_1^x S_2^x+J\vec{S}_1\cdot \vec{S}_2 +\Gamma (S_1^yS_2^z+S_1^zS_2^y),\nn\\
H_{23}&=& KS_2^y S_3^y+J\vec{S}_2\cdot \vec{S}_3 +\Gamma (S_2^zS_3^x+S_2^xS_3^z).
\eea
The interchain coupling in the original frame is given by
\bea
\alpha_z\big[KS_i^z S_j^z+J\vec{S}_i\cdot \vec{S}_j +\Gamma (S_i^xS_j^y+S_i^yS_j^x)\big]. 
\eea

In the $U_6$ frame, the Hamiltonian of the 1D spin-1/2 $KJ\Gamma$ model has a three-site periodicity.
Within a unit cell, the Hamiltonian is
\bea
H^\prime_{12}&=& (K+J)S_1^{\prime x} S_2^{\prime x} +\Gamma (S_1^{\prime y}S_2^{\prime y}+S_1^{\prime z}S_2^{\prime z})+J (S_1^{\prime y} S_2^{\prime z}+S_1^{\prime z} S_2^{\prime y}),\nn\\
 H^\prime_{23}&=& (K+J)S_2^{\prime z} S_3^{\prime z} +\Gamma (S_2^{\prime x}S_3^{\prime x}+S_2^{\prime y}S_3^{\prime y})+J (S_2^{\prime x} S_3^{\prime y}+S_2^{\prime y} S_3^{\prime x}),\nn\\
 H^\prime_{34}&=& (K+J)S_2^{\prime y} S_3^{\prime y} +\Gamma (S_2^{\prime z}S_3^{\prime z}+S_2^{\prime x}S_3^{\prime x})+J (S_2^{\prime z} S_3^{\prime x}+S_2^{\prime x} S_3^{\prime z}).
\eea
The interchain coupling in the $U_6$ frame for bond $\gamma$ is
\begin{flalign}
H^\prime_{ij}=\alpha_z\big[ (K+J)S_i^{\prime \gamma} S_j^{\prime \gamma} +\Gamma (S_i^{\prime \alpha} S_j^{\prime \alpha}+S_i^{\prime\beta} S_j^{\prime\beta})+J (S_i^{\prime\alpha} S_j^{\prime\beta}+S_i^{\prime\beta} S_j^{\prime\alpha})\big].
\end{flalign}

\section{The $1+1$-dimensional  SU(2)$_1$ WZW model}
\label{app:transformation_WZW}

The low energy physics of the spin-1/2 SU(2) AFM chain  is known to be described by the SU(2)$_1$ Wess-Zumino-Witten (WZW) model \cite{Affleck1988},
defined by the following Sugawara Hamiltonian
\begin{eqnarray}
H=\frac{2\pi}{3}v\int dx (\vec{J}_L\cdot \vec{J}_L+\vec{J}_R\cdot \vec{J}_R)-g_c\int dx \vec{J}_L\cdot \vec{J}_R,
\label{eq:Low_ham_KG}
\end{eqnarray}
in which $v$ is the spin velocity, $g_c>0$ is the marginally irrelevant coupling,
\bea
\vec{J_L}&=&-\frac{1}{4\pi}\text{tr} [(\partial_z g) g^{-1} \vec{\sigma}],\nn\\
\vec{J_R}&=&\frac{1}{4\pi}\text{tr} [g^{-1} (\partial_{\bar{z}} g) \vec{\sigma}],
\label{eq:def_WZW_current}
\eea
where the SU(2) matrix $g$ is the SU(2)$_1$ primary field.

Denoting $\epsilon(x)=\text{tr}g(x)$  and $\vec{N}(x)=i\text{tr}(g(x)\vec{\sigma})$,
the transformation properties of the SU(2)$_1$ WZW fields under time reversal, spatial translation, spatial inversion, and the global spin rotation are given by \cite{Yang2020}
\begin{eqnarray}
T: &\epsilon(x)\rightarrow \epsilon(x), &\vec{N}(x)\rightarrow -\vec{N}(x),\nn\\
&\vec{J}_L(x)\rightarrow -\vec{J}_R(x), &\vec{J}_R(x)\rightarrow -\vec{J}_L(x), 
\label{eq:transformT}
\end{eqnarray}
\begin{eqnarray}
T_a: &\epsilon(x)\rightarrow -\epsilon(x), &\vec{N}(x)\rightarrow -\vec{N}(x),\nn\\
&\vec{J}_L(x)\rightarrow \vec{J}_L(x), &\vec{J}_R(x)\rightarrow \vec{J}_R(x), 
\label{eq:transformTa}
\end{eqnarray}
\begin{eqnarray}
I: & \epsilon(x)\rightarrow -\epsilon(-x), &\vec{N}(x)\rightarrow \vec{N}(-x),\nn\\
&\vec{J}_L(x)\rightarrow \vec{J}_R(-x), &\vec{J}_R(x)\rightarrow \vec{J}_L(-x), 
\label{eq:transformI}
\end{eqnarray}
\begin{eqnarray}
R: &\epsilon(x)\rightarrow \epsilon(x), &N^\alpha(x)\rightarrow R^{\alpha}_{\,\,\beta}N^\beta(x),\nn\\
&J^\alpha_L(x)\rightarrow R^{\alpha}_{\,\,\beta} J^\beta_L(x), &J^\alpha_R(x)\rightarrow R^{\alpha}_{\,\,\beta}J^\beta_R(x), 
\label{eq:transformR}
\end{eqnarray}
in which $x$ is the spatial coordinate; $R^{\alpha}_{\,\,\beta}$ ($\alpha,\beta=x,y,z$) is the matrix element of the $3\times 3$ rotation matrix $R$.

\section{Symmetry analysis of the low energy field theory in the Luttinger liquid phase}
\label{app:sym_analysis_low}

Based on a symmetry analysis, the low energy Luttinger liquid theory in Eq. (\ref{eq:LL_liquid})
can be derived by performing a field theory perturbation on the couplings $K-\Gamma$ and $J$ as discussed in Ref. \cite{Yang2020},
where the unperturbed system is taken as the the hidden SU(2) symmetric AFM point $(K=\Gamma<0, J=0)$ \cite{Yang2019},
and $K-\Gamma$, $J$ are assumed to be small.
Particularly, the Luttinger liquid theory in Eq. (\ref{eq:LL_liquid}) has an emergent U(1) symmetry corresponding to the translation invariance of the $\varphi$ field (i.e., $\varphi\rightarrow \varphi+\beta$ where $\beta\in\mathbb{R}$).
In this appendix, we briefly review the perturbative analysis in Ref. \cite{Yang2020}. 

For $K<0$, there is an extended Luttinger liquid phase in the phase diagram discussed in Ref. \cite{Yang2020}.
In this subsection, we give a quick review of the derivation of the Luttinger liquid phase based on a symmetry analysis.
Because of Eq. (\ref{eq:equiv_1D}), we consider the $\Gamma<0$ region. 

When $K=\Gamma<0$, $J=0$, the system acquires the form of the SU(2) AFM model in the $U_6$ frame.
The SU(2)$_1$ WZW model can be taken as the unperturbed system, and the low energy physics for small $K-\Gamma$ and $J$  can be analyzed by performing a field theory perturbation. 
Here we use a symmetry analysis to figure out all the symmetry allowed operators whose scaling dimensions are less than or equal to $2$ (which correspond to the relevant and marginal operators in $1+1$-dimension).
The transformation properties of the SU(2)$_1$ WZW primary field and current operators which will be used for symmetry analysis have been summarized in Appendix \ref{app:transformation_WZW}. 

First, the dimension $1/2$ operator $g$ and the dimension $3/2$ operators $\vec{J}_Lg$, $\vec{J}_Rg$ are forbidden since they change sign under $T_{3a}$.

Second, for the dimension $1$ operators $\vec{J}_L$, $\vec{J}_R$, time reversal symmetry requires the combination $J^\alpha_L-J^\alpha_R$ ($\alpha=x,y,z$),
among which only $J_L^{\prime z}-J_R^{\prime z}$ is invariant under the  $D_3$ group ($=G/\mathopen{<}T_{3a}\mathclose{>}$),
where $\hat{z}^\prime = \frac{1}{\sqrt{3}}(1,1,1)^T$, and $G=\mathopen{<} R(\hat{z}^\prime,-\frac{2\pi}{3})T_a,R(\hat{y}^\prime,\pi)I \mathclose{>}$.

Third, for the dimension $2$ operators $J^\alpha_L J^\beta_L$, $J^\alpha_R J^\beta_R$, $J^\alpha_L J^\beta_R$, 
time reversal symmetry requires the combinations $J^\alpha_L J^\beta_L+J^\alpha_R J^\beta_R$ and $J^\alpha_L J^\beta_R+ J^\beta_L J^\alpha_R$.
Notice that either $\{J_L^\alpha\}_{\alpha=x,y,z}$ or $\{J_R^\alpha\}_{\alpha=x,y,z}$ can be decomposed as $A_2\oplus E$ according to the irreducible representations of the $D_3$ group \cite{Coxeter1965}. 
Since $A_2\otimes A_2=A_1$, and $E\otimes E= A_1\oplus A_2 \oplus E$, 
the $A_1$ representations among all the dimension $2$ operators are given by 
$\vec{J}_L\cdot \vec{J}_L+\vec{J}_L\cdot \vec{J}_R,~\vec{J}_L\cdot \vec{J}_R$, 
and
$J_L^{\prime z} J_L^{\prime z}+J_R^{\prime z} J_R^{\prime z},~J_L^{\prime z} J_R^{\prime z}$. 

In summary, the low energy field theory compatible with the nonsymmorphic symmetry group is 
\bea
H=\frac{2\pi}{3}v^\prime \int dx (\vec{J}_L\cdot \vec{J}_L+\vec{J}_R\cdot \vec{J}_R)-g^\prime_c\int dx \vec{J}_L\cdot \vec{J}_R
+\int dx [g_1(J_L^0-J_R^0) 
+g_2J^0_LJ^0_R
+g_3(J_L^0J_L^0+J_R^0J_R^0)].
\label{eq:Low_ham_KG_2}
\eea

On the other hand, 
$J_\lambda^{\prime z} J_\lambda^{\prime z}=\frac{1}{3}\vec{J}_\lambda\cdot \vec{J}_\lambda$
in the SU(2)$_1$ WZW model, where $\lambda=L,R$ \cite{Yang2020}.
In addition, by performing a chiral rotation \cite{Garate2010,Gangadharaiah2008,Schnyder2008} the chiral term  $J_L^0-J_R^0$ 
can be eliminated.
Hence Eq. (\ref{eq:Low_ham_KG_2}) can be further simplified into
\bea
H=\frac{2\pi}{3}v^{\prime\prime} \int dx (\vec{J}_L\cdot \vec{J}_L+\vec{J}_R\cdot \vec{J}_R)-g^\prime_c\int dx \big[ (J_L^{\prime x} J_R^{\prime x}+J_L^{\prime y} J_R^{\prime y})+(1+\delta_c)J_L^{\prime z} J_R^{\prime z}\big],
\label{eq:Low_ham_KG_3}
\eea
which is the same as the low energy field theory of the XXZ model.
Whether the anisotropy in Eq. (\ref{eq:Low_ham_KG_3}) is easy-plane or easy-axis depends on the sign of $\delta_c$.
In fact, for the spin-1/2 Kitaev-Heisenberg-Gamma model in the FM Kitaev region, the system has easy-plane anisotropy (i.e., gapless at low energies) when $J>0$ \cite{Yang2020}.

\section{DMRG numerics on the symmetry axis of emergent U(1) symmetry} 
\label{app:DMRG_sym_axis}

\begin{figure*}[htbp]
\includegraphics[width=5.8cm]{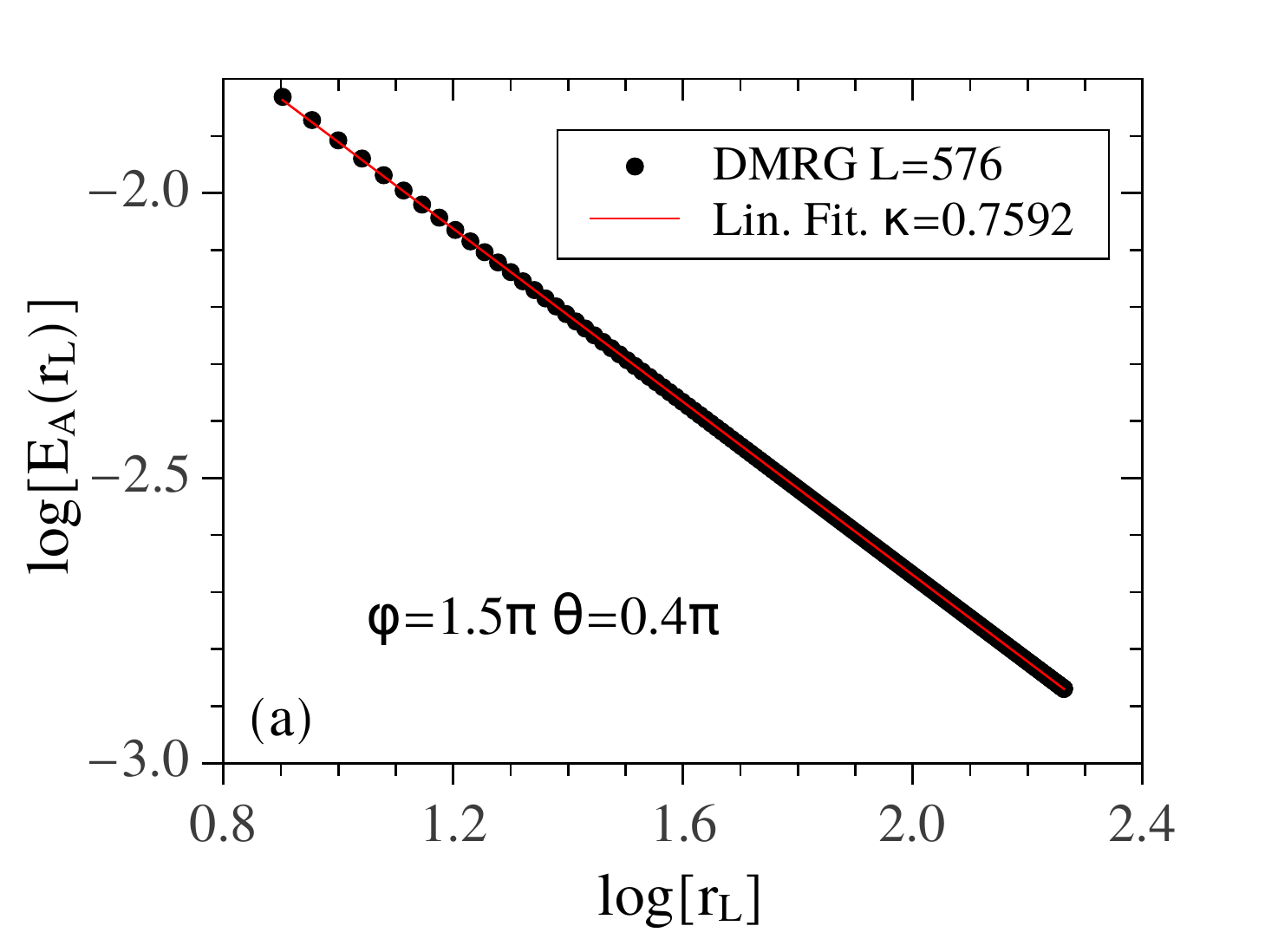}
\includegraphics[width=5.8cm]{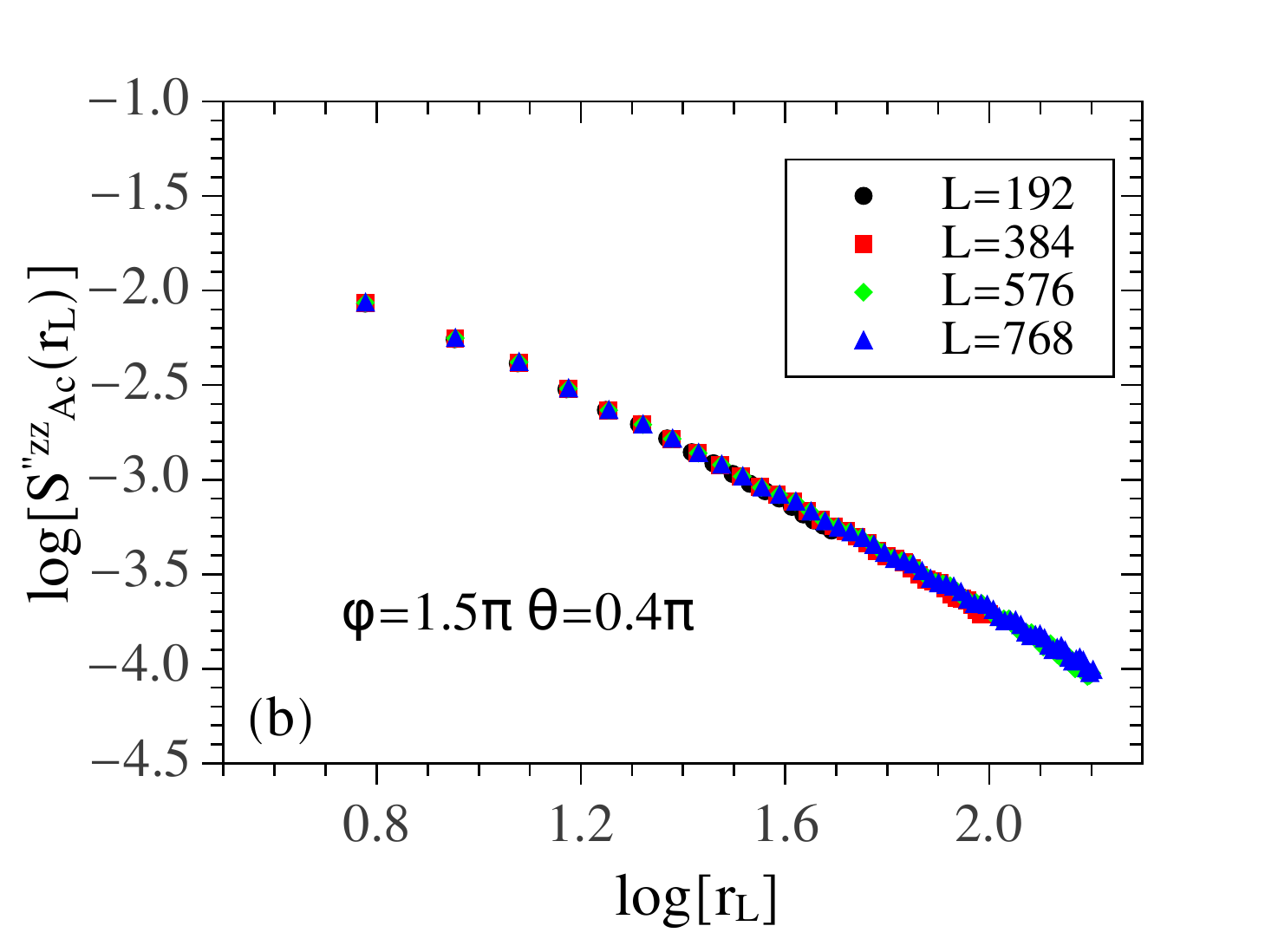}
\includegraphics[width=5.8cm]{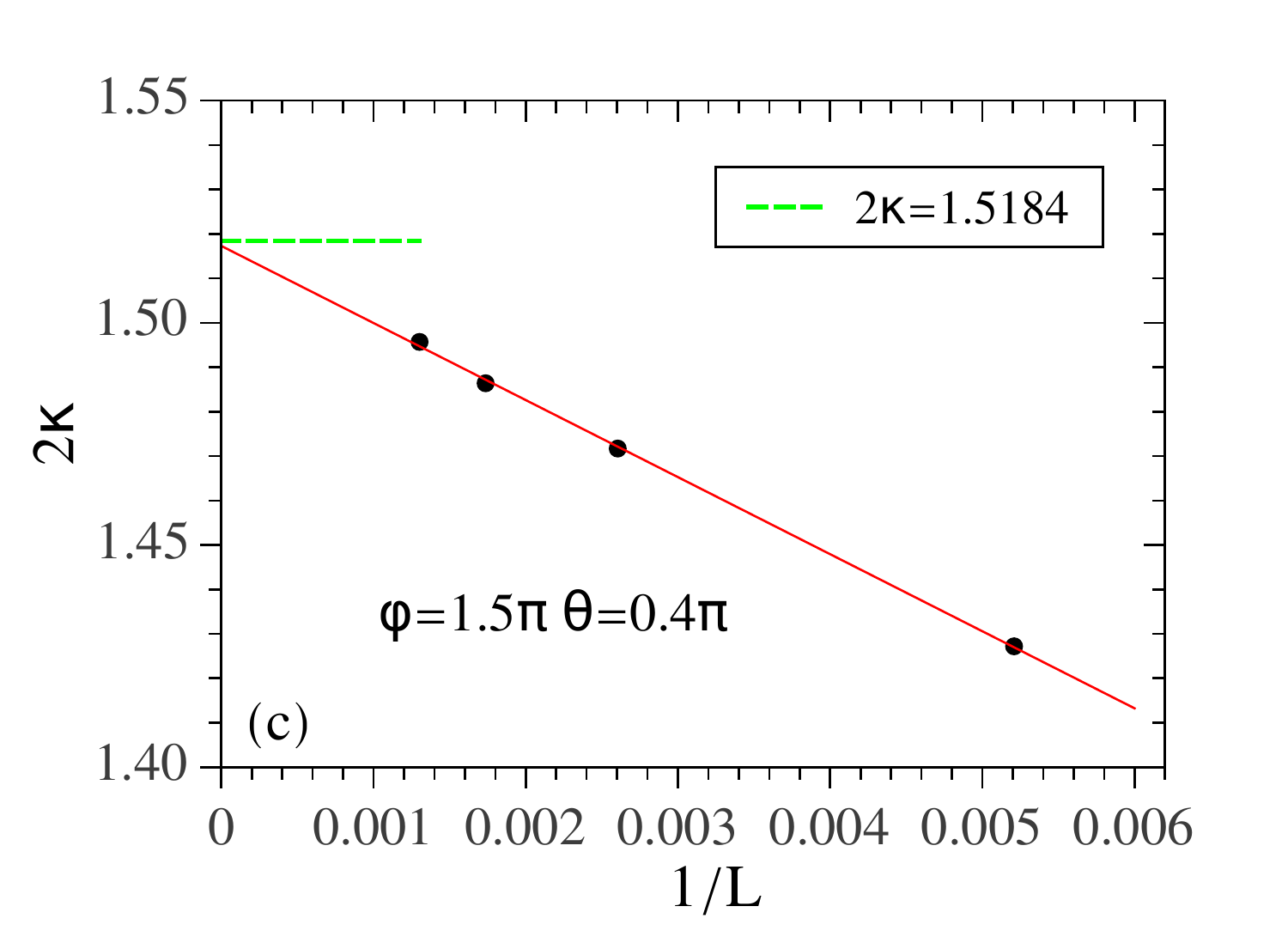}
\includegraphics[width=5.8cm]{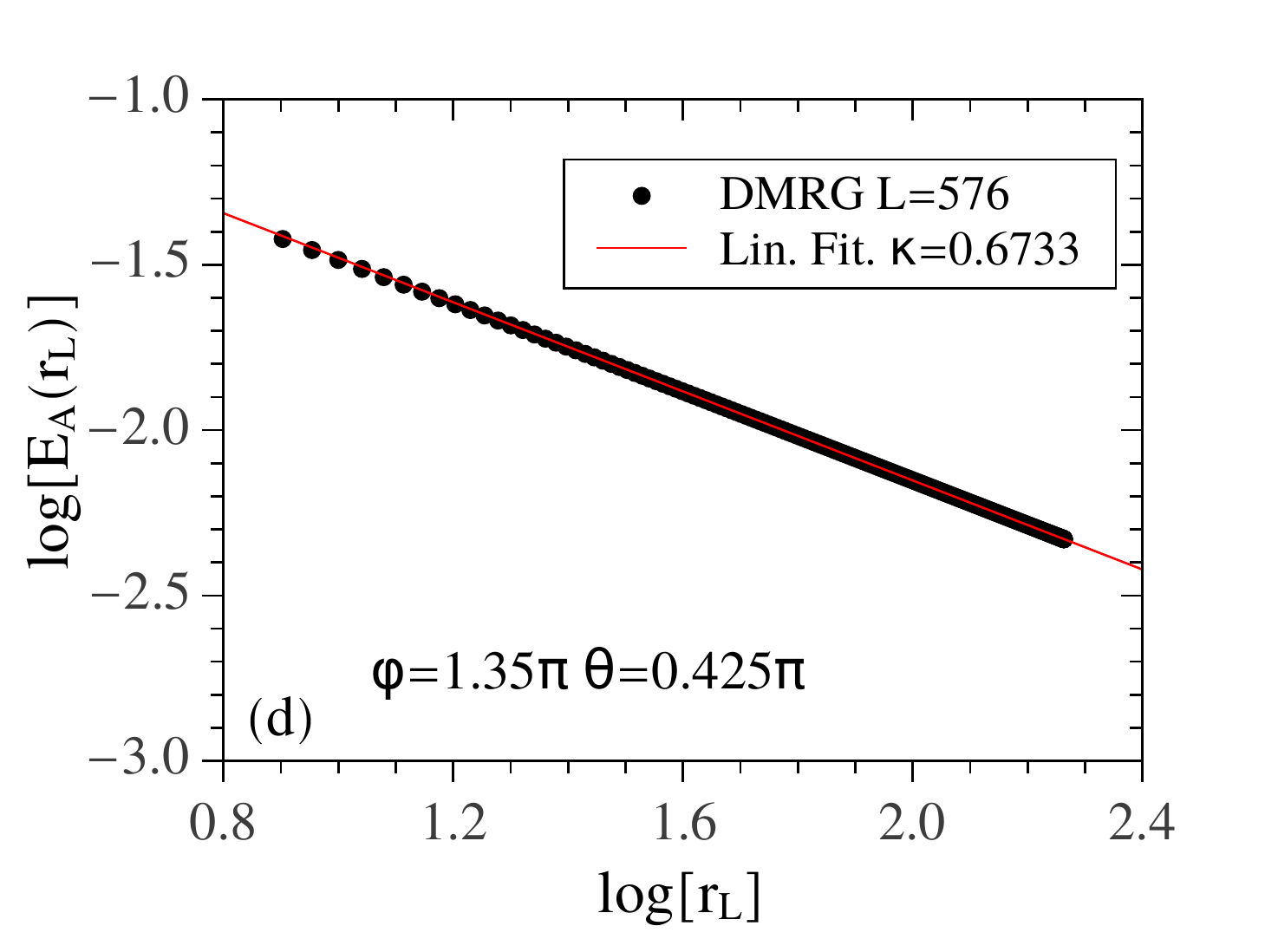}
\includegraphics[width=5.8cm]{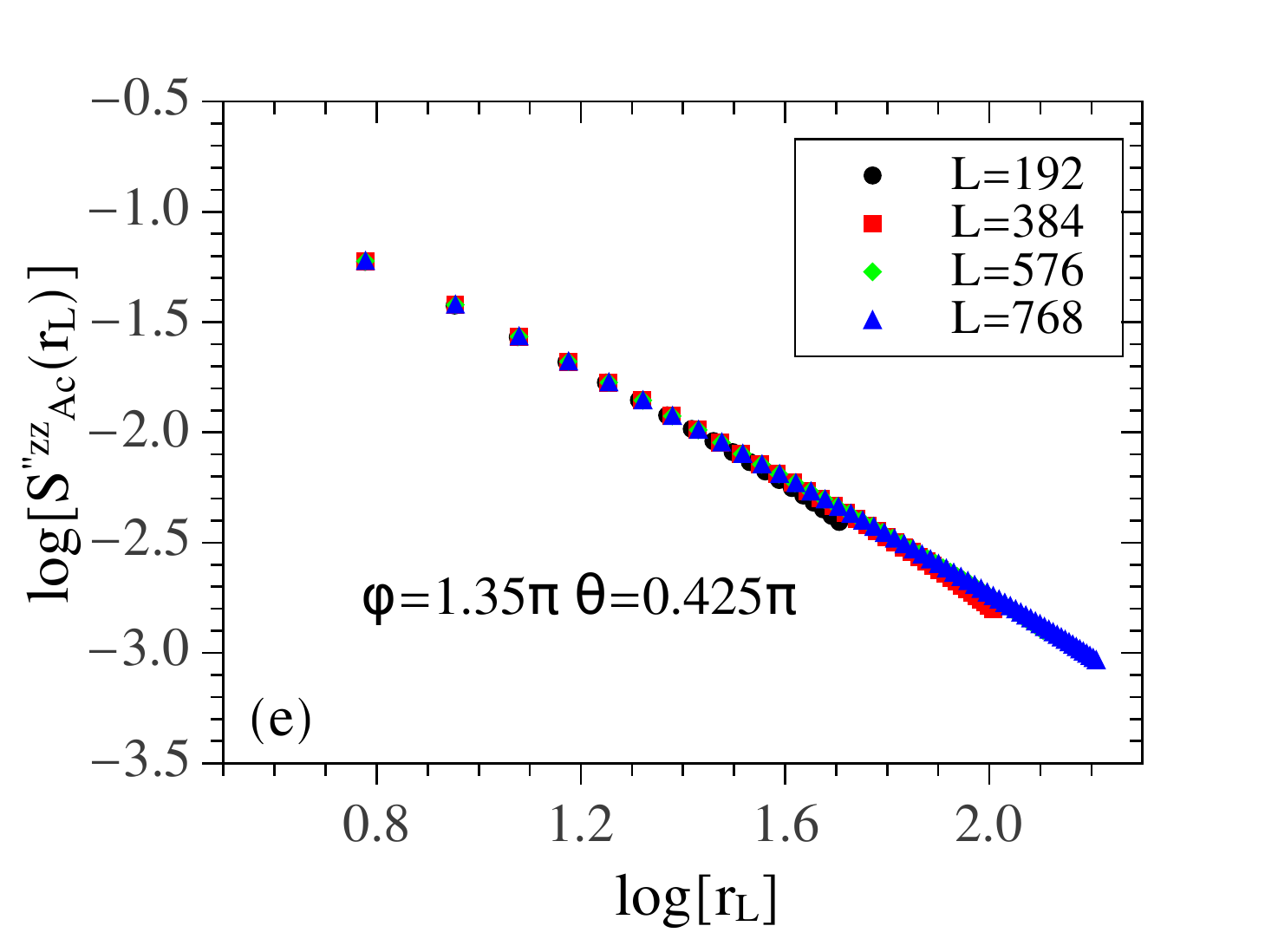}
\includegraphics[width=5.8cm]{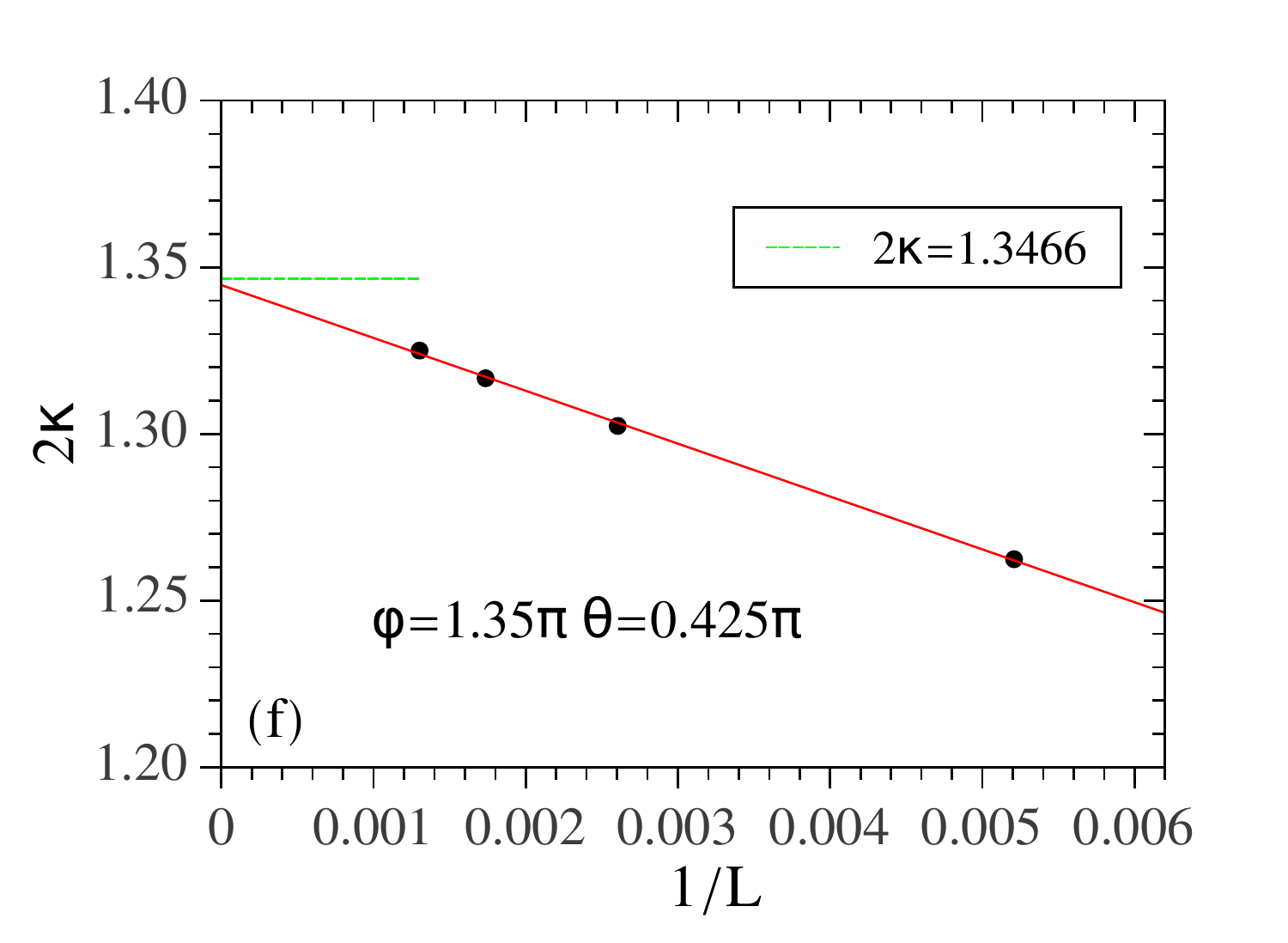}
\caption{(a,d) Staggered energy density $E_A(r)$ vs. $r_L=\frac{L}{\pi}\sin(\frac{\pi r}{L})$ on a log-log scale,
(b,e) $S_{Ac}^{\prime\prime zz}(r)$ vs.   $r_L$ on a log-log scale for a variety of system sizes $L$, 
(c,f) values of twice of the Luttinger parameter $2\kappa$ as functions of the inverse system size $1/L$ extracted  from the correlation functions  and the extrapolations of $2\kappa$ to $L\rightarrow \infty$,
where the parameters in the spin-1/2 $KJ\Gamma$ Hamiltonian are taken as ($\phi=1.5\pi$, $\theta=0.4\pi$) in (a,b,c) and ($\phi=1.35\pi$, $\theta=0.425\pi$) in (d,e,f), where $\phi$ and $\theta$ are defined in Eq. (\ref{eq:parametrize_KJG}).
In (a,d), DMRG calculations are performed on a system of $L=576$ sites with open boundary conditions within the original frame.
In (b,c,e,f), DMRG calculations are performed on systems of $L=192,384,576,768$ sites with open boundary conditions within the $OU_6$ frame.
In (a-f), the truncation error $\epsilon$ and bond dimension $m$ in DMRG calculations are taken as $\epsilon=10^{-10}$ and $m=1000$.
} \label{fig:DMRG_LL1} 
\end{figure*}

In principle, the ten bosonization coefficients  can be determined numerically by comparing numerical results with analytical predictions.
However,  a complete determination of all the ten parameters is difficult. 
Here we only study the symmetry axis of the emergent U(1) symmetry, and numerically verify that it is indeed along the $(1,1,1)$-direction.
DMRG calculations are performed on a variety of system sizes (including $L=192,384,576,768$ sites) using open boundary conditions.
In all  DMRG simulations, several sweeps were performed keeping the truncation error $\epsilon$ below $10^{-10}$, while the maximum bond dimension allowed is $m=1000$.

Using DMRG simulations, we first determine the Luttinger parameter $\kappa$, which is expected to vary continuously in the Luttinger liquid phase in Fig. \ref{fig:phase_KHG_1D}.  
We introduce the local energy density $E(r)$ as the expectation value of the Hamiltonian density $h(r)$,
where $h(r)$ is defined in the original frame as
\begin{flalign}
h(r)=KS_r^\gamma S_{r+1}^\gamma+ J\vec{S}_r\cdot \vec{S}_{r+1}+\Gamma (S_r^\alpha S_{r+1}^\beta+S_r^\beta S_{r+1}^\alpha).
\end{flalign}
According to Ref. \cite{Laflorencie2006}, 
the energy density $E(r)$ can be separated into a uniform component $E_U(r)$ and a staggered component $E_A(r)$ in the long distance limit $r\gg 1$ if open boundary conditions are used,
i.e.,
\bea
E(r)=E_U(r)+(-)^r E_A(r),
\label{eq:E_UA}
\eea
in which $r$ is the distance measured from one of the two boundaries of the system,
and both $E_U(r)$ and $E_A(r)$ are smooth functions of $r$ on a length scale much larger than the lattice constant.  
In the Luttinger liquid phase, $E_A(r)$ is predicted to behave in the long distance limit as 
\bea
E_A(r)\propto (r_L)^{-\kappa},
\label{eq:LL_EA}
\eea
in which $\kappa$ is the Luttinger parameter, and 
\bea
r_L=\frac{L}{\pi}\sin(\frac{\pi r}{L})
\eea
in accordance with conformal field theory on finite size systems with periodic boundary conditions. 
Eq. (\ref{eq:LL_EA}) can be used to numerically determine the Luttinger parameter $\kappa$,
provided the data points are far away from the boundaries so that there is essentially no difference between open and periodic boundary conditions.
We note that $E_A(r)$ can be extracted from the data of $E(r)$ using a three-point formula as discussed in Ref. \cite{Yang2020}.

We have computed $E_A(r)$ using DMRG simulations for an open system of $L=576$ sites, at two representative points.
Fig. \ref{fig:DMRG_LL1} (a) and (d) show the numerical results of $E_A(r)$ vs. $r_L=\frac{L}{\pi}\sin (\frac{\pi r}{L})$ on a log-log scale,
for $(\phi=1.5\pi,\theta=0.4\pi)$ and $(\phi=1.35\pi,\theta=0.425\pi)$, respectively,
where $\phi$ and $\theta$ are defined in Eq. (\ref{eq:parametrize_KJG}).
By fitting the numerical data with the formula in Eq. (\ref{eq:LL_EA}),
excellent linear relations are obtained which give $\kappa=0.7592$  and $\kappa=0.6733$ for Fig. \ref{fig:DMRG_LL1} (a) and (d), respectively.

To determine the symmetry axis of the emergent U(1) symmetry, we consider the static correlation function $S_c^{\prime\prime zz}(r)$ of the ``staggered center of mass" spin $S_{s.c.}^{\prime\prime z}(n)$ along $z^{\prime\prime}$-direction in the $OU_6$ frame, where $S_{s.c.}^{\prime\prime z}(n)$ and $S_c^{\prime\prime zz}(r)$ are defined as
\bea
S^{\prime\prime z}_{s.c.}(n)=-S^{\prime\prime z}_{1+3n}+S^{\prime\prime z}_{2+3n}-S^{\prime\prime z}_{3+3n}
\eea
and 
\bea
S_c^{\prime\prime zz}(r=3n)=\langle S^{\prime\prime z}_{s.c.}(1) S^{\prime\prime z}_{s.c.}(n)\rangle.  
\eea
Similar to Eq. (\ref{eq:E_UA}), $S_c^{\prime\prime zz}(r)$ can be decomposed into a uniform and staggered part as
\bea
S_c^{\prime\prime zz}(r)=S_{Uc}^{\prime\prime zz}(r) +(-)^r S_{Ac}^{\prime\prime zz}(r),
\eea
where both $S_{Uc}^{\prime\prime zz}(r)$ and $S_{Ac}^{\prime\prime zz}(r)$ are smooth functions of $r$ on a length scale much larger than the lattice constant. 
As can be verified, the nonsymmorphic bosonization formula in Eq. (\ref{eq:abelian_LL1_matrix_B}) predicts
\bea
S_{Ac}^{\prime\prime zz}(r)=\frac{9(\nu_C)^2}{(r_L)^{2\kappa}},\text{ $r\gg 1$}.
\label{eq:S_Ac}
\eea
Eq. (\ref{eq:S_Ac}) can be used to numerically check whether the symmetry axis of the emergent U(1) symmetry is along the $\hat{z}^{\prime\prime}=\frac{1}{\sqrt{3}}(1,1,1)^T$ direction in the $U_6$ frame.

Fig. \ref{fig:DMRG_LL1} (b) and (e) show the numerical results of $S_{Ac}^{\prime\prime zz}(r)$ vs. $r_L$ on a log-log scale,
for $(\phi=1.5\pi,\theta=0.4\pi)$ and $(\phi=1.35\pi,\theta=0.425\pi)$, respectively,
computed in open systems in the $OU_6$ frame for a variety of system sizes $L$ including  $L=192,384,576,768$.
It can be seen from Fig. \ref{fig:DMRG_LL1} (b,e) that the relations are very linear.

The Luttinger parameters extracted from correlation functions are shown in Fig. \ref{fig:DMRG_LL1} (c) for $(\phi=1.5\pi,\theta=0.4\pi)$, 
and Fig. \ref{fig:DMRG_LL1} (f) for $(\phi=1.35\pi,\theta=0.425\pi)$. 
In Fig. \ref{fig:DMRG_LL1} (c,f), the values of the Luttinger parameters for different system sizes are plotted against $1/L$.
Extrapolating to the $L\rightarrow \infty$ limit shown by the red dashed lines in Fig. \ref{fig:DMRG_LL1} (c,f), we obtain
$2\kappa=1.5184$ for $(\phi=1.5\pi,\theta=0.4\pi)$,
and $2\kappa=1.3466$ for $(\phi=1.35\pi,\theta=0.425\pi)$,
which are fully consistent with the  Luttinger parameters obtained in Fig. \ref{fig:DMRG_LL1} (a) and (d) to a high degree of accuracy. 
The excellent agreements between the values of the Luttinger parameter extracted from two independent methods confirm that the symmetry axis for the emergent U(1) symmetry in the $U_6$ frame is indeed along the $(1,1,1)$-direction in the Luttinger liquid phase in Fig. \ref{fig:phase_KHG_1D} (a).

\section{The nonsymmorphic bosonization formulas}
\label{app:bosonization_formulas}

\subsection{Derivation of the nonsymmorphic bosonization formulas}

In the Luttinger liquid phase shown in Fig. (\ref{fig:phase_KHG_1D}), the U(1) symmetry is  emergent  at low energies, while the microscopic Hamiltonian only has discrete symmetries.
Therefore, although the low energy field theory of the Luttinger liquid phase is given by Eq. (\ref{eq:LL_liquid}), the abelian bosonization formulas 
can break the emergent U(1) symmetry and are only required to respect the exact nonsymmorphic symmetry group of the system.
In fact, from a renormalization group (RG) point of view, such U(1) breaking effects originate from the multiplicative wavefunction  renormalizations of the spin operators  in the high energy region along the RG flow where the discrete lattice structure is still visible,
which has been discussed in detail in  \cite{Yang2022} for a parameter region of the spin-1/2 $KJ\Gamma$ chain different from this work.

Next we derive the most general form of the nonsymmorphic bosonization formulas compatible with the discrete nonsymmorphic symmetry group.
We note that the transformation properties of $\mathcal{J}^{ \alpha}$ and $\mathcal{N}^{ \alpha }$ ($\alpha=x,y,z$) under the symmetry operations of the system can be derived from Eq. (\ref{eq:transformation_theta_phi}). 
We will work out the most general forms of $\mathcal{C}_i$ ($i=1,2,3$) allowed by symmetries,
and the discussion for $\mathcal{D}_i$ is exactly similar. 

First consider the symmetry operation $R(\hat{z}^{\prime\prime},-\frac{2\pi}{3})T_a$.
Using the transformation properties 
\begin{flalign}
&[R(\hat{z}^{\prime\prime},-\frac{2\pi}{3})T_a](S^{\prime\prime x}_{i+3n}~S^{\prime\prime y}_{i+3n}~S^{\prime\prime z}_{i+3n})[R(\hat{z}^{\prime\prime},-\frac{2\pi}{3})T_a]^{-1}=(S^{\prime\prime x}_{i+1+3n}~S^{\prime\prime y}_{i+1+3n}~S^{\prime\prime z}_{i+1+3n})M_z,\nn\\
&[R(\hat{z}^{\prime\prime},-\frac{2\pi}{3})T_a](\mathcal{N}^{ x}~\mathcal{N}^{ y}~\mathcal{N}^{ z})[R(\hat{z}^{\prime\prime},-\frac{2\pi}{3})T_a]^{-1}=(-)(\mathcal{N}^{ x}~\mathcal{N}^{ y}~\mathcal{N}^{ z})M_z,
\label{eq:abe_Rz}
\end{flalign}
we obtain
\bea
\mathcal{C}_1&=&M_z^{-1}\mathcal{C}_2M_z,\nn\\
\mathcal{C}_3&=&M_z\mathcal{C}_2M_z^{-1},
\label{eq:Cs_z_2}
\eea
in which 
\bea
M_z=\left(\begin{array}{ccc}
-\frac{1}{2} & \frac{\sqrt{3}}{2} & 0\\
-\frac{\sqrt{3}}{2} & -\frac{1}{2} & 0\\
0&0&1
\end{array}
\right).
\eea
We note that there is an overall minus sign in the right hand side of the second equation in Eq. (\ref{eq:abe_Rz}) since $N^{\prime\prime \alpha}$ changes sign under $T_a$ according to Eq. (\ref{eq:transformation_theta_phi}).

Then consider the symmetry operation $R(\hat{y}^{\prime\prime},\pi)I$.
Using the transformation properties
\begin{flalign}
&[R(\hat{y}^{\prime\prime},\pi)I](S^{\prime\prime x}_{i+3n}~S^{\prime\prime y}_{i+3n}~S^{\prime\prime z}_{i+3n})[R(\hat{y}^{\prime\prime},\pi)I]^{-1}=(S^{\prime\prime x}_{10-i-3n}~S^{\prime\prime y}_{10-i-3n}~S^{\prime\prime z}_{10-i-3n})M_y,\nn\\
&[R(\hat{y}^{\prime\prime},\pi)I](\mathcal{N}^{ x}(r)~\mathcal{N}^{ y}(r)~\mathcal{N}^{ z}(r))[R(\hat{y}^{\prime\prime},\pi)I]^{-1}=(\mathcal{N}^{ x}(-r)~\mathcal{N}^{ y}(-r)~\mathcal{N}^{ z}(-r))M_y,
\end{flalign}
we obtain
\bea
\mathcal{C}_1&=& M_y \mathcal{C}_3 M_y^{-1},\nn\\
\mathcal{C}_2&=& M_y \mathcal{C}_2 M_y^{-1},
\label{eq:Cs_y_2}
\eea
in which the coordinate in $\mathcal{N}^\alpha$ is $r=(i+3n)a$ and the matrix $M_y$ is
\bea
M_y=\left(\begin{array}{ccc}
-1 & 0 & 0\\
0 & 1 & 0\\
0&0&-1
\end{array}
\right).
\eea
Because of the relation $(M_yM_z)^2=1$, Eqs. (\ref{eq:Cs_z_2},\ref{eq:Cs_y_2}) leads to a single independent constraint:
$\mathcal{C}_2=M_y \mathcal{C}_2M_y^{-1}$,
which can be easily solved and gives the expression of $\mathcal{C}_2$ in Eq. (\ref{eq:CD_mat_mathcal}).
Then the matrices $\mathcal{C}_1$ and $\mathcal{C}_3$ can be obtained from Eq. (\ref{eq:Cs_z_2}).
The discussions on the matrices $\mathcal{D}_i$ ($i=1,2,3$) are exactly similar.

\subsection{Explicit form of the bosonization formulas}

We give the explicit form of the nonsymmorphic bosonization formulas for the spin operators in both the $OU_6$ frame as well as the $U_6$ frame in the $\Gamma<0$ region. 
Notice that by performing a global spin rotation $R(\hat{z},\pi)$, these formulas also apply to the AFM $\Gamma$ region. 

The nonsymmorphic bosonization formulas in the $OU_6$ frame are given by
\bea
S_{1+3n}^{\prime\prime x}&=& (\lambda_D+\frac{3}{4}\delta_D) \mathcal{J}^x+\frac{\sqrt{3}}{4}\delta_D\mathcal{J}^y-\frac{1}{2}\rho_D\mathcal{J}^z+(-)^{1+n} \big[
(\lambda_C+\frac{3}{4}\delta_C) \mathcal{N}^x+\frac{\sqrt{3}}{4}\delta_C\mathcal{N}^y-\frac{1}{2}\rho_C\mathcal{N}^z
\big]
,\nn\\
S_{1+3n}^{\prime\prime y}&=& \frac{\sqrt{3}}{4}\delta_D \mathcal{J}^x+(\lambda_D+\frac{1}{4}\delta_D)\mathcal{J}^y+\frac{\sqrt{3}}{2}\rho_D\mathcal{J}^z+(-)^{1+n} \big[
\frac{\sqrt{3}}{4}\delta_C \mathcal{N}^x+(\lambda_C+\frac{1}{4}\delta_C)\mathcal{N}^y+\frac{\sqrt{3}}{2}\rho_C\mathcal{N}^z
\big]
,\nn\\
S_{1+3n}^{\prime\prime z}&=&-\frac{1}{2}\sigma_D \mathcal{J}^x+\frac{\sqrt{3}}{2}\sigma_D\mathcal{J}^y +\nu_D \mathcal{J}^z+(-)^{1+n}\big[
-\frac{1}{2}\sigma_C \mathcal{N}^x+\frac{\sqrt{3}}{2}\sigma_C\mathcal{N}^y +\nu_C \mathcal{N}^z
\big],
\label{eq:bosonize_LL1_S1_a}
\eea
\bea
S^{\prime\prime x}_{2+3n}&=& \lambda_D \mathcal{J}^x +\rho_D \mathcal{J}^z+(-)^{n} \big[ \lambda_C \mathcal{N}^x +\rho_C \mathcal{N}^z\big],\nn\\
S^{\prime\prime y}_{2+3n}&=& (\lambda_D+\delta_D) \mathcal{J}^y+(-)^{n} (\lambda_C+\delta_C) \mathcal{N}^y,\nn\\
S^{\prime\prime z}_{2+3n}&=& \sigma_D \mathcal{J}^x +\nu_D \mathcal{J}^z+(-)^{n} \big[ \sigma_C \mathcal{N}^x +\nu_C \mathcal{N}^z \big],
\label{eq:bosonize_LL1_S2_a}
\eea
\bea
S_{3+3n}^{\prime\prime x}&=& (\lambda_D+\frac{3}{4}\delta_D) \mathcal{J}^x-\frac{\sqrt{3}}{4}\delta_D\mathcal{J}^y-\frac{1}{2}\rho_D\mathcal{J}^z+(-)^{1+n} \big[
(\lambda_C+\frac{3}{4}\delta_C) \mathcal{N}^x-\frac{\sqrt{3}}{4}\delta_C\mathcal{N}^y-\frac{1}{2}\rho_C\mathcal{N}^z
\big]
,\nn\\
S_{3+3n}^{\prime\prime y}&=& -\frac{\sqrt{3}}{4}\delta_D \mathcal{J}^x+(\lambda_D+\frac{1}{4}\delta_D)\mathcal{J}^y-\frac{\sqrt{3}}{2}\rho_D\mathcal{J}^z+(-)^{1+n} \big[
-\frac{\sqrt{3}}{4}\delta_C \mathcal{N}^x+(\lambda_C+\frac{1}{4}\delta_C)\mathcal{N}^y-\frac{\sqrt{3}}{2}\rho_C\mathcal{N}^z
\big]
,\nn\\
S_{3+3n}^{\prime\prime z}&=&-\frac{1}{2}\sigma_D \mathcal{J}^x+\frac{\sqrt{3}}{2}\sigma_D\mathcal{J}^y +\nu_D \mathcal{J}^z+(-)^{1+n}\big[
-\frac{1}{2}\sigma_C \mathcal{N}^x+\frac{\sqrt{3}}{2}\sigma_C\mathcal{N}^y +\nu_C \mathcal{N}^z
\big].
\label{eq:bosonize_LL1_S3_a}
\eea

The nonsymmorphic bosonization formulas in the $U_6$ frame are given by
\begin{flalign}
&S_{1+3n}^{\prime x}=(-\frac{1}{\sqrt{6}}\lambda_D-\frac{\sqrt{3}}{2\sqrt{2}}\delta_D-\frac{1}{2\sqrt{3}}\sigma_D)\mathcal{J}^{ x}+(-\frac{1}{\sqrt{2}}\lambda_D-\frac{1}{2\sqrt{2}}\delta_D+\frac{1}{2}\sigma_D)\mathcal{J}^{ y}+(\frac{1}{\sqrt{3}}\nu_D-\frac{1}{\sqrt{6}}\rho_D)\mathcal{J}^{ z}\nn\\
&+(-)^{1+n}\big[
(-\frac{1}{\sqrt{6}}\lambda_C-\frac{\sqrt{3}}{2\sqrt{2}}\delta_C-\frac{1}{2\sqrt{3}}\sigma_C)\mathcal{N}^{ x}+(-\frac{1}{\sqrt{2}}\lambda_C-\frac{1}{2\sqrt{2}}\delta_C+\frac{1}{2}\sigma_C)\mathcal{N}^{ y}+(\frac{1}{\sqrt{3}}\nu_C-\frac{1}{\sqrt{6}}\rho_C)\mathcal{N}^{ z}
\big]\nn\\
&S_{1+3n}^{\prime y}=(\sqrt{\frac{2}{3}} \lambda_D+\frac{\sqrt{3}}{2\sqrt{2}}\delta_D-\frac{1}{2\sqrt{3}}\sigma_D)\mathcal{J}^{ x}+(\frac{1}{2\sqrt{2}} \delta_D+\frac{1}{2}\sigma_D)\mathcal{J}^{ y}+(\frac{1}{\sqrt{3}}\nu_D-\frac{1}{\sqrt{6}}\rho_D)\mathcal{J}^{ z}\nn\\
&+(-)^n \big[(\sqrt{\frac{2}{3}} \lambda_C+\frac{\sqrt{3}}{2\sqrt{2}}\delta_C-\frac{1}{2\sqrt{3}}\sigma_C)\mathcal{N}^{ x}+(\frac{1}{2\sqrt{2}} \delta_C+\frac{1}{2}\sigma_C)\mathcal{N}^{ y}+(\frac{1}{\sqrt{3}}\nu_C-\frac{1}{\sqrt{6}}\rho_C)\mathcal{N}^{ z}\big]\nn\\
&S_{1+3n}^{\prime z}= (-\frac{1}{\sqrt{6}}\lambda_D-\frac{1}{2\sqrt{3}} \sigma_D )\mathcal{J}^{ x}+(\frac{1}{\sqrt{2}}\lambda_D+\frac{1}{2}\sigma_D)\mathcal{J}^{ y}+(\frac{1}{\sqrt{3}} \nu_D+\sqrt{\frac{2}{3}}\rho_D)\mathcal{J}^{ z}\nn\\
&+(-)^{1+n}\big[
(-\frac{1}{\sqrt{6}}\lambda_C-\frac{1}{2\sqrt{3}} \sigma_C )\mathcal{N}^{ x}+(\frac{1}{\sqrt{2}}\lambda_C+\frac{1}{2}\sigma_C)\mathcal{N}^{ y}+(\frac{1}{\sqrt{3}} \nu_C+\sqrt{\frac{2}{3}}\rho_C)\mathcal{N}^{ z}
\big]
\label{eq:bosonize_LL1_S1_b}
\end{flalign}
\begin{flalign}
&S_{2+3n}^{\prime x}= (-\frac{1}{\sqrt{6}}\lambda_D+\frac{1}{\sqrt{3}}\sigma_D) \mathcal{J}^{ x}-\frac{1}{\sqrt{2}}(\lambda_D+\delta_D)\mathcal{J}^{ y}+(\frac{1}{\sqrt{3}}\nu_D-\frac{1}{\sqrt{6}}\rho_D)\mathcal{J}^{ z}\nn\\
&+(-)^n\big[
(-\frac{1}{\sqrt{6}}\lambda_C+\frac{1}{\sqrt{3}}\sigma_C) \mathcal{N}^{ x}-\frac{1}{\sqrt{2}}(\lambda_C+\delta_C)\mathcal{N}^{ y}+(\frac{1}{\sqrt{3}}\nu_C-\frac{1}{\sqrt{6}}\rho_C)\mathcal{N}^{ z}
\big]\nn\\
&S_{2+3n}^{\prime y}=(\sqrt{\frac{2}{3}}\lambda_D+\frac{1}{\sqrt{3}}\sigma_D)\mathcal{J}^{ x}+(\frac{1}{\sqrt{3}}\nu_D+\sqrt{\frac{2}{3}} \rho_D)\mathcal{J}^{ z}\nn\\
&+(-)^n\big[
(\sqrt{\frac{2}{3}}\lambda_C+\frac{1}{\sqrt{3}}\sigma_C)\mathcal{N}^{ x}+(\frac{1}{\sqrt{3}}\nu_C+\sqrt{\frac{2}{3}} \rho_C)\mathcal{N}^{ z}
\big]\nn\\
&S_{2+3n}^{\prime z}= (-\frac{1}{\sqrt{6}}\lambda_D+\frac{1}{\sqrt{3}}\sigma_D) \mathcal{J}^{ x}+\frac{1}{\sqrt{2}}(\lambda_D+\delta_D)\mathcal{J}^{ y}+(\frac{1}{\sqrt{3}}\nu_D-\frac{1}{\sqrt{6}}\rho_D)\mathcal{J}^{ z}\nn\\
&+(-)^n\big[
(-\frac{1}{\sqrt{6}}\lambda_C+\frac{1}{\sqrt{3}}\sigma_C) \mathcal{N}^{ x}+\frac{1}{\sqrt{2}}(\lambda_C+\delta_C)\mathcal{N}^{ y}+(\frac{1}{\sqrt{3}}\nu_C-\frac{1}{\sqrt{6}}\rho_C)\mathcal{N}^{ z}
\big]
\label{eq:bosonize_LL1_S2_b}
\end{flalign}
\begin{flalign}
&S_{3+3n}^{\prime x}= (-\frac{1}{\sqrt{6}}\lambda_D-\frac{1}{2\sqrt{3}} \sigma_D )\mathcal{J}^{ x}-(\frac{1}{\sqrt{2}}\lambda_D+\frac{1}{2}\sigma_D)\mathcal{J}^{ y}+(\frac{1}{\sqrt{3}} \nu_D+\sqrt{\frac{2}{3}}\rho_D)\mathcal{J}^{ z}\nn\\
&+(-)^{1+n}\big[
(-\frac{1}{\sqrt{6}}\lambda_C-\frac{1}{2\sqrt{3}} \sigma_C )\mathcal{N}^{ x}-(\frac{1}{\sqrt{2}}\lambda_C+\frac{1}{2}\sigma_C)\mathcal{N}^{ y}+(\frac{1}{\sqrt{3}} \nu_C+\sqrt{\frac{2}{3}}\rho_C)\mathcal{N}^{ z}
\big]\nn\\
&S_{3+3n}^{\prime y}=(\sqrt{\frac{2}{3}} \lambda_D+\frac{\sqrt{3}}{2\sqrt{2}}\delta_D-\frac{1}{2\sqrt{3}}\sigma_D)\mathcal{J}^{ x}-(\frac{1}{2\sqrt{2}} \delta_D+\frac{1}{2}\sigma_D)\mathcal{J}^{ y}+(\frac{1}{\sqrt{3}}\nu_D-\frac{1}{\sqrt{6}}\rho_D)\mathcal{J}^{ z}\nn\\
&+(-)^n \big[
(\sqrt{\frac{2}{3}} \lambda_C+\frac{\sqrt{3}}{2\sqrt{2}}\delta_C-\frac{1}{2\sqrt{3}}\sigma_C)\mathcal{N}^{ x}-(\frac{1}{2\sqrt{2}} \delta_C+\frac{1}{2}\sigma_C)\mathcal{N}^{ y}+(\frac{1}{\sqrt{3}}\nu_C-\frac{1}{\sqrt{6}}\rho_C)\mathcal{N}^{ z}
\big]\nn\\
&S_{3+3n}^{\prime z}=(-\frac{1}{\sqrt{6}}\lambda_D-\frac{\sqrt{3}}{2\sqrt{2}}\delta_D-\frac{1}{2\sqrt{3}}\sigma_D)\mathcal{J}^{ x}+(\frac{1}{\sqrt{2}}\lambda_D+\frac{1}{2\sqrt{2}}\delta_D-\frac{1}{2}\sigma_D)\mathcal{J}^{ y}+(\frac{1}{\sqrt{3}}\nu_D-\frac{1}{\sqrt{6}}\rho_D)\mathcal{J}^{ z}\nn\\
&+(-)^{1+n}\big[
(-\frac{1}{\sqrt{6}}\lambda_C-\frac{\sqrt{3}}{2\sqrt{2}}\delta_C-\frac{1}{2\sqrt{3}}\sigma_C)\mathcal{N}^{ x}+(\frac{1}{\sqrt{2}}\lambda_C+\frac{1}{2\sqrt{2}}\delta_C-\frac{1}{2}\sigma_C)\mathcal{N}^{ y}+(\frac{1}{\sqrt{3}}\nu_C-\frac{1}{\sqrt{6}}\rho_C)\mathcal{N}^{ z}
\big].
\label{eq:bosonize_LL1_S3_b}
\end{flalign}

\section{Derivation of low energy mean field Hamiltonian}
\label{app:derivation_low_MF}

In this appendix, we derive  the inter-chain low energy mean field Hamiltonian $\mathcal{H}$ defined in Eq. (\ref{eq:interchain_MF}).
Expanding Eqs. (\ref{eq:Hcb},\ref{eq:Hcd}), we obtain
\begin{flalign}
&H_{cb}=-\alpha_z\sum_n\big[\nn\\
&(K+J)S^{\prime z}_{c,2+6n}\langle S^{\prime z}_{b,2+6n}\rangle+\Gamma (S^{\prime x}_{c,2+6n}\langle S^{\prime x}_{b,2+6n} \rangle+S^{\prime y}_{c,2+6n} \langle S^{\prime y}_{b,2+6n}) \rangle+J (S^{\prime x}_{c,2+6n}\langle S^{\prime y}_{b,2+6n}\rangle+S^{\prime y}_{c,2+6n}\langle S^{\prime x}_{b,2+6n})\rangle+\nn\\
&(K+J)S^{\prime x}_{4,2+6n}\langle S^{\prime x}_{b,4+6n}\rangle+\Gamma (S^{\prime y}_{c,4+6n}\langle S^{\prime y}_{b,4+6n} \rangle+S^{\prime z}_{c,4+6n}\langle S^{\prime z}_{b,4+6n})\rangle+J (S^{\prime y}_{c,4+6n}\langle S^{\prime z}_{b,4+6n}\rangle+S^{\prime z}_{c,4+6n}\langle S^{\prime y}_{b,4+6n})\rangle+\nn\\
&(K+J)S^{\prime y}_{6,2+6n} \langle S^{\prime y}_{b,6+6n}\rangle+\Gamma (S^{\prime z}_{c,6+6n}\langle S^{\prime z}_{b,6+6n}\rangle+S^{\prime x}_{c,6+6n}\langle S^{\prime x}_{b,6+6n})\rangle+J (S^{\prime z}_{c,6+6n}\langle S^{\prime x}_{b,6+6n}\rangle+S^{\prime x}_{c,6+6n}\langle S^{\prime z}_{b,6+6n})\rangle
\big],
\label{eq:Hcb_app}
\end{flalign}
\begin{flalign}
&H_{cd}=-\alpha_z\sum_n\big[\nn\\
&(K+J)S^{\prime x}_{c,1+6n}\langle S^{\prime x}_{d,1+6n}\rangle +\Gamma (S^{\prime y}_{c,1+6n}\langle S^{\prime y}_{d,1+6n} \rangle+S^{\prime z}_{c,1+6n}\langle S^{\prime z}_{d,1+6n}) \rangle+J (S^{\prime y}_{c,1+6n}\langle S^{\prime z}_{d,1+6n}\rangle+S^{\prime z}_{c,1+6n}\langle S^{\prime y}_{d,1+6n})\rangle+\nn\\
&(K+J)S^{y\prime }_{3,2+6n}\langle S^{\prime y}_{d,3+6n}\rangle+\Gamma (S^{\prime z}_{c,3+6n}\langle S^{\prime z}_{d,3+6n}\rangle+S^{\prime x}_{c,3+6n}\langle S^{\prime x}_{d,3+6n})\rangle+J (S^{\prime z}_{c,3+6n}\langle S^{\prime x}_{d,3+6n}\rangle+S^{\prime x}_{c,3+6n}\langle S^{\prime z}_{d,3+6n})\rangle+\nn\\
&(K+J)S^{\prime z}_{5,2+6n}\langle S^{\prime z}_{d,5+6n}\rangle+\Gamma (S^{\prime x}_{c,5+6n}\langle S^{\prime x}_{d,5+6n}\rangle+S^{\prime y}_{c,5+6n}\langle S^{\prime y}_{d,5+6n})\rangle+J (S^{\prime x}_{c,5+6n}\langle S^{\prime y}_{d,5+6n}\rangle+S^{\prime y}_{c,5+6n}\langle S^{\prime x}_{d,5+6n})\rangle
\big].
\label{eq:Hcd_app}
\end{flalign}
The mean field Hamiltonian $\mathcal{H}=H_{cb}+H_{cd}$ can be rewritten as
\begin{flalign}
&\mathcal{H}=
-(S_1^{\prime x}, S_1^{\prime y}, S_1^{\prime z})
\left(\begin{array}{ccc}
\Gamma & J&0 \\
 J& \Gamma & 0 \\ 
 0&0 & K+J
\end{array}\right)
\left(\begin{array}{c}
\langle S_4^{\prime x}\rangle\\
\langle S_4^{\prime y}\rangle\\
\langle S_4^{\prime z}\rangle
\end{array}\right)
-(S_2^{\prime x}, S_2^{\prime y}, S_2^{\prime z})
\left(\begin{array}{ccc}
\Gamma & 0&J \\
 0& K+J & 0 \\ 
 J&0 & \Gamma
\end{array}\right)
\left(\begin{array}{c}
\langle S_5^x\rangle\\
\langle S_5^y\rangle\\
\langle S_5^z\rangle
\end{array}\right)\nn\\
&-(S_3^{\prime x}, S_3^{\prime y}, S_3^{\prime z})
\left(\begin{array}{ccc}
K+J & 0&0 \\
 0& \Gamma & J \\ 
 0&J & \Gamma 
\end{array}\right)
\left(\begin{array}{c}
\langle S_6^{\prime x}\rangle\\
\langle S_6^{\prime y}\rangle\\
\langle S_6^{\prime z}\rangle
\end{array}\right)-
(S_4^{\prime x}, S_4^{\prime y}, S_4^{\prime z})
\left(\begin{array}{ccc}
\Gamma & J&0 \\
 J& \Gamma & 0 \\ 
 0&0 & K+J
\end{array}\right)
\left(\begin{array}{c}
\langle S_1^{\prime x}\rangle\\
\langle S_1^{\prime y}\rangle\\
\langle S_1^{\prime z}\rangle
\end{array}\right)\nn\\
&-(S_5^{\prime x}, S_5^{\prime y}, S_5^{\prime z})
\left(\begin{array}{ccc}
\Gamma & 0&J \\
 0& K+J & 0 \\ 
 J&0 & \Gamma
\end{array}\right)
\left(\begin{array}{c}
\langle S_2^{\prime x}\rangle\\
\langle S_2^{\prime y}\rangle\\
\langle S_2^{\prime z}\rangle
\end{array}\right)
-(S_6^{\prime x}, S_6^{\prime y}, S_6^{\prime z})
\left(\begin{array}{ccc}
K+J & 0&0 \\
 0& \Gamma & J \\ 
 0&J & \Gamma
\end{array}\right)
\left(\begin{array}{c}
\langle S_3^{\prime x}\rangle\\
\langle S_3^{\prime y}\rangle\\
\langle S_3^{\prime z}\rangle
\end{array}\right)
\end{flalign}
where $\vec{S}^\prime_i$ ($1\leq i\leq 6$) should be replaced with $\vec{\mathcal{J}}$ and $\vec{\mathcal{N}}$  using the bosonization formulas in Eqs. (\ref{eq:bosonize_LL1_S1_b},\ref{eq:bosonize_LL1_S2_b},\ref{eq:bosonize_LL1_S3_b}).
By doing so, we obtain Eq. (\ref{eq:H_AB})
in which the $3\times 1$ column vectors $A$ and $B$ are given by
\begin{flalign}
&A=-D_1 H_z [(C_4)^{row}_y]^T-D_2 H_y [(C_5)^{row}_y]^T-D_3 H_x [(C_6)^{row}_y]^T\nn\\
&-D_4 H_z [(C_1)^{row}_y]^T-D_5 H_y [(C_2)^{row}_y]^T-D_6 H_x [(C_3)^{row}_y]^T,
\end{flalign}
\begin{flalign}
&B=-C_1 H_z [(C_4)^{row}_y]^T-C_2 H_y [(C_5)^{row}_y]^T-C_3 H_x [(C_6)^{row}_y]^T\nn\\
&-C_4 H_z [(C_1)^{row}_y]^T-C_5 H_y [(C_2)^{row}_y]^T-C_6 H_x [(C_3)^{row}_y]^T,
\end{flalign}
where the matrices $C_i$, $D_i$ are defined in Eq. (\ref{eq:new_C_D}), 
$(C_i)^{row}_y$ represents the second row of the matrix $C_i$,
and
\bea
H_x=\left(\begin{array}{ccc}
K+J & 0&0 \\
 0& \Gamma & J \\ 
 0&J & \Gamma 
\end{array}\right),~
H_y=\left(\begin{array}{ccc}
\Gamma & 0&J \\
 0& K+J & 0 \\ 
 J&0 & \Gamma
\end{array}\right),~
H_z=\left(\begin{array}{ccc}
\Gamma & J&0 \\
 J& \Gamma & 0 \\ 
 0&0 & K+J
\end{array}\right).
\eea
Using the relations $D_{i+3}=D_i$, $C_{i+3}=-C_i$ ($1\leq i \leq 3$),
we obtain 
\bea
A&=&0,\nn\\
B&=&2\big(C_1 H_z [(C_1)^{row}_y]^T+C_2 H_y [(C_2)^{row}_y]^T+C_3 H_x [(C_3)^{row}_y]^T\big).
\eea
Straightforward calculations lead to Eq. (\ref{eq:AB_column}) and 
Eq. (\ref{eq:uC_uD_Neely}).

Similarly, the expressions of $A_z$, $B_z$ in Eq. (\ref{eq:AB_column_z}) can be derived as
\begin{flalign}
&A_z=-D_1 H_z [(C_4)^{row}_z]^T-D_2 H_y [(C_5)^{row}_z]^T-D_3 H_x [(C_6)^{row}_z]^T\nn\\
&-D_4 H_z [(C_1)^{row}_z]^T-D_5 H_y [(C_2)^{row}_z]^T-D_6 H_x [(C_3)^{row}_z]^T,
\label{eq:Az_app}
\end{flalign}
\begin{flalign}
&B_z=-C_1 H_z [(C_4)^{row}_z]^T-C_2 H_y [(C_5)^{row}_z]^T-C_3 H_x [(C_6)^{row}_z]^T\nn\\
&-C_4 H_z [(C_1)^{row}_z]^T-C_5 H_y [(C_2)^{row}_z]^T-C_6 H_x [(C_3)^{row}_z]^T,
\label{eq:Bz_app}
\end{flalign}
in which $(C_i)^{row}_z$ represents the third row of the matrix $C_i$.
Evaluations of Eq. (\ref{eq:Az_app}) and Eq. (\ref{eq:Bz_app})
give Eq. (\ref{eq:AB_column_z}) and Eq. (\ref{eq:vC_vD_Neely}).


\section{Self-consistent mean field solution}
\label{app:self_consistent}

The self-consistent mean field solution has been discussed in Ref. \cite{Yang2022}.
For completeness, we briefly review the solution in this appendix.

\subsection{Sine-Gordon model with $\cos(\sqrt{\pi}\theta)$}

The 1+1-dimensional massive sine-Gordon model can be approximately solved using the variational method discussed in Ref. \cite{Giamarchi2004_b}. 
After integrating out the $\varphi$-field, the action in the imaginary time becomes
\bea
S=\frac{\kappa}{2} \int dxd\tau [\frac{1}{v} (\partial_\tau \theta)^2+v(\partial_x \theta)^2]-\frac{\lambda}{a^3} \langle \cos(\sqrt{\pi}\theta) \rangle \int dx \cos(\sqrt{\pi}\theta).
\label{eq:action_S_LL_sine_gordan}
\eea
In the variational method, the action is rewritten as
\bea
S=S_0+(S-S_0),
\eea
in which $S_0$ is a free part given by 
\bea
S_0=\frac{\kappa}{2} \int dxd\tau [\frac{1}{v} (\partial_\tau \theta)^2+v(\partial_x \theta)^2+\frac{1}{v}\Delta^2\theta^2],
\eea
where $\Delta$ is the variational mass of $S_0$.
Performing a perturbative expansion over $S-S_0$, the partition function becomes
\bea
Z=\int D\theta e^{-S}=Z_0 \langle e^{-(S-S_0)} \rangle_0,
\eea
in which $Z_0=\int D\theta e^{-S_0}$ and the expectation value $\langle ... \rangle_0$ is defined as $\frac{1}{Z_0}\int D\theta e^{-S_0} (...)$.
Expanding up to lowest order in $S-S_0$, the free energy is 
\bea
F^\prime = F_0 +\frac{1}{\beta} \langle S-S_0\rangle_0,
\eea
which can be evaluated as
\bea
F^\prime=-\frac{1}{\beta} \sum_{\vec{q},k>0} \log [G(\vec{q})]+\frac{\kappa}{2}\frac{1}{\beta} \sum_{\vec{q}} (\frac{1}{v}\omega_n^2+vk^2) G(\vec{q})-\frac{1}{\beta}\frac{\lambda}{a^3} \langle \cos(\sqrt{\pi}\theta) \rangle \beta L e^{-\frac{\pi}{2\beta L}\sum_{\vec{q}} G(\vec{q})},
\eea
where 
\bea
G(\vec{q})=\frac{\kappa^{-1}}{\frac{1}{v}\omega_n^2+vk^2+\frac{1}{v}\Delta^2}.
\eea

The parameter $\Delta$ can be determined by minimizing $F^\prime$, i.e., solving 
\bea
\frac{\partial F^\prime}{\partial G(\vec{q})}=0,
\eea
which yields
\bea
G^{-1}(\vec{q})=\kappa (\frac{1}{v}\omega_n^2+vk^2+\frac{\Delta^2}{v}),
\eea
where
\bea
\frac{\kappa\Delta^2}{v}=\frac{\pi\lambda}{a^3} \langle \cos(\sqrt{\pi}\theta) \rangle e^{-\frac{\pi}{2\beta L}\sum_{\vec{q}} \frac{v\kappa^{-1}}{\omega_n^2+v^2k^2+\Delta^2} }.
\label{eq:self_consistent_eq}
\eea

We will focus on the zero temperature case. 
In the weak coupling limit, $\Delta\ll \Lambda$, where $\Lambda$ is the UV cutoff in the Luttinger liquid theory,
which is on the same order as the inverse lattice constant. 
Performing the integral, 
\bea
\frac{\pi}{2\beta L}\sum_{\vec{q}} \frac{v\kappa^{-1}}{\omega_n^2+v^2k^2+\Delta^2}
&\simeq& (4\kappa)^{-1} \ln [v\Lambda/\Delta],
\eea
we can solve $\Delta$ from Eq. (\ref{eq:self_consistent_eq}) as
\bea
\Delta = v\Lambda \big[\frac{\pi\lambda\langle \cos(\sqrt{\pi}\theta) \rangle}{v\kappa \Lambda^2a^3}\big]^{\frac{1}{2-(4\kappa)^{-1}}}.
\eea

On the other hand, the expectation value $\cos(\sqrt{\pi}\theta)$ can be obtained from the action $S_0$ as
\bea
\langle \cos(\sqrt{\pi}\theta) \rangle=e^{-\frac{\pi}{2\beta L}\sum_{\vec{q}} G(\vec{q})}=\big(\frac{\Delta}{v\Lambda}\big)^{(4\kappa)^{-1}}.
\eea
Therefore,  self-consistency requires
\bea
\langle \cos(\sqrt{\pi}\theta) \rangle
=\big[\frac{\pi\lambda\langle \cos(\sqrt{\pi}\theta) \rangle}{v\kappa \Lambda^2a^3}\big]^{\frac{(4\kappa)^{-1}}{2-(4\kappa)^{-1}}},
\label{eq:value_cos}
\eea
which leads to
\bea
\langle \cos(\sqrt{\pi}\theta) \rangle=\big[\frac{\pi\lambda}{v\kappa \Lambda^2a^3}\big]^{\frac{1}{8\kappa-2}}\sim (\alpha_0)^{\frac{1}{8\kappa-2}}.
\label{eq:sol_Nx_2}
\eea
Notice that the Luttinger liquid Hamiltonian has an U(1) symmetry, hence the result is the same by replacing $\cos(\sqrt{\pi}\theta)$ with $\sin(\sqrt{\pi}\theta)$
in Eq. (\ref{eq:action_S_LL_sine_gordan}).

\subsection{Sine-Gordon model with $\sin(\sqrt{4\pi}\phi)$}

Instead of Eq. (\ref{eq:action_S_LL_sine_gordan}), we add $\sin(\sqrt{4\pi}\phi)$ to the Luttinger liquid Hamiltonian and consider the following action,
\bea
S=\frac{\kappa}{2} \int dxd\tau [\frac{1}{v} (\partial_\tau \theta)^2+v(\partial_x \theta)^2]-\frac{\lambda}{a^3} \langle \sin(\sqrt{4\pi}\phi) \rangle \int dx \sin(\sqrt{4\pi}\phi).
\label{eq:action_S_LL_sine_gordan_2}
\eea
The calculation is exactly similar as the previous subsection, and the only difference is that the exponent $1/(4\kappa)$ should be replaced by $\kappa$ in Eq. (\ref{eq:value_cos}), i.e., 
\bea
\langle \sin(\sqrt{4\pi}\phi) \rangle
=\big[\frac{\pi\lambda\langle \sin(\sqrt{4\pi}\phi) \rangle}{v\kappa \Lambda^2a^3}\big]^{\frac{\kappa}{2-\kappa}}.
\label{eq:value_sin}
\eea
Hence, we obtain
\bea
\langle \sin(\sqrt{4\pi}\phi) \rangle=\big[\frac{\pi\lambda}{v\kappa \Lambda^2a^3}\big]^{\frac{\kappa}{2-2\kappa}}.
\eea

\section{Degenerate symmetry breaking spin configurations} 
\label{app:figures}

In this appendix, we present the spin configurations in the six-degenerate symmetry ground states for both the $g>0$ and $g<0$ cases,
where $g$ is the coupling constant of the $\cos(6\sqrt{\pi}\theta)$ term defined in Eq. (\ref{eq:u}).

\subsection{The $g>0$ case}

\begin{figure}[h]
\begin{center}
\includegraphics[width=5cm]{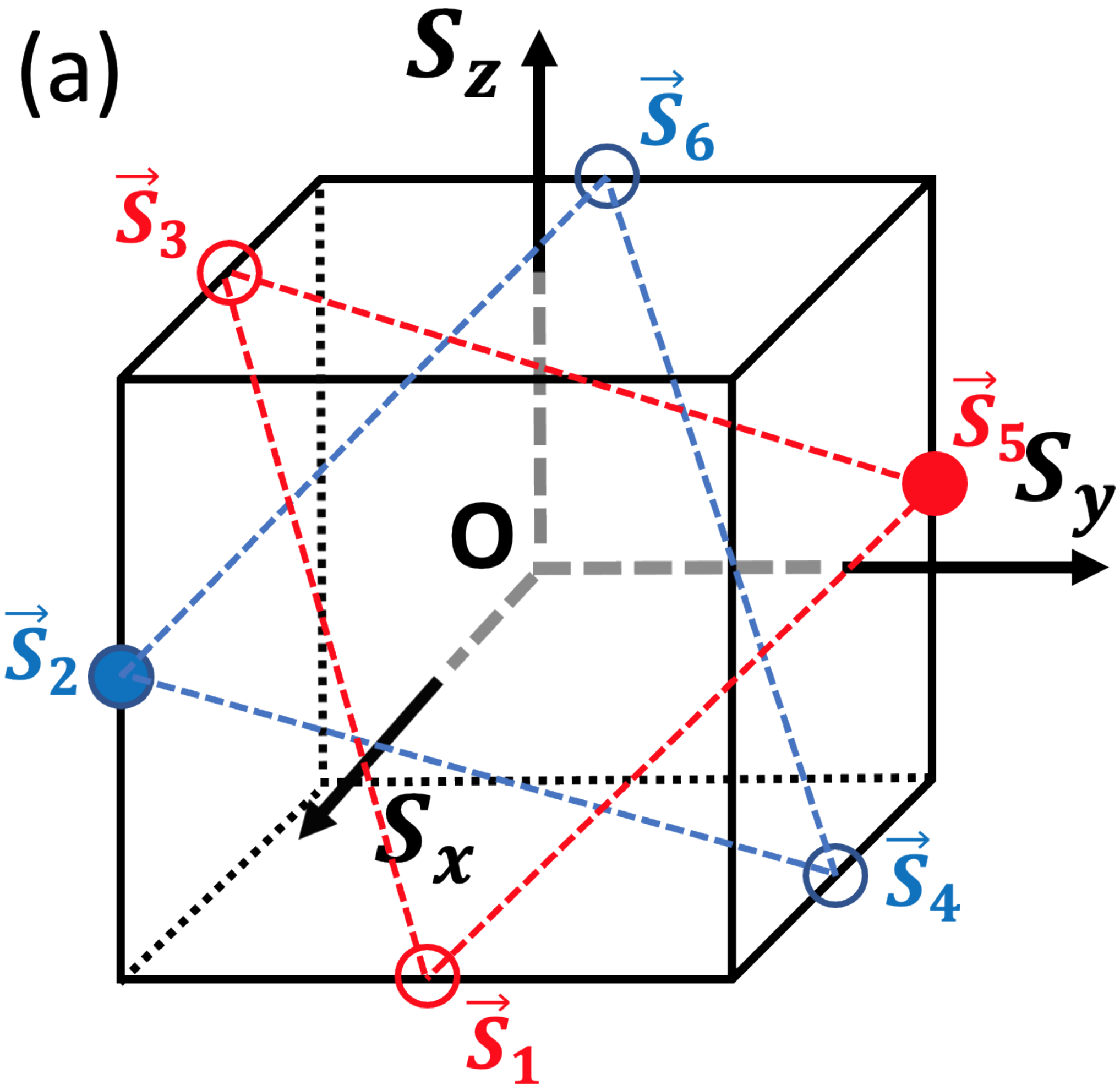}
\includegraphics[width=5cm]{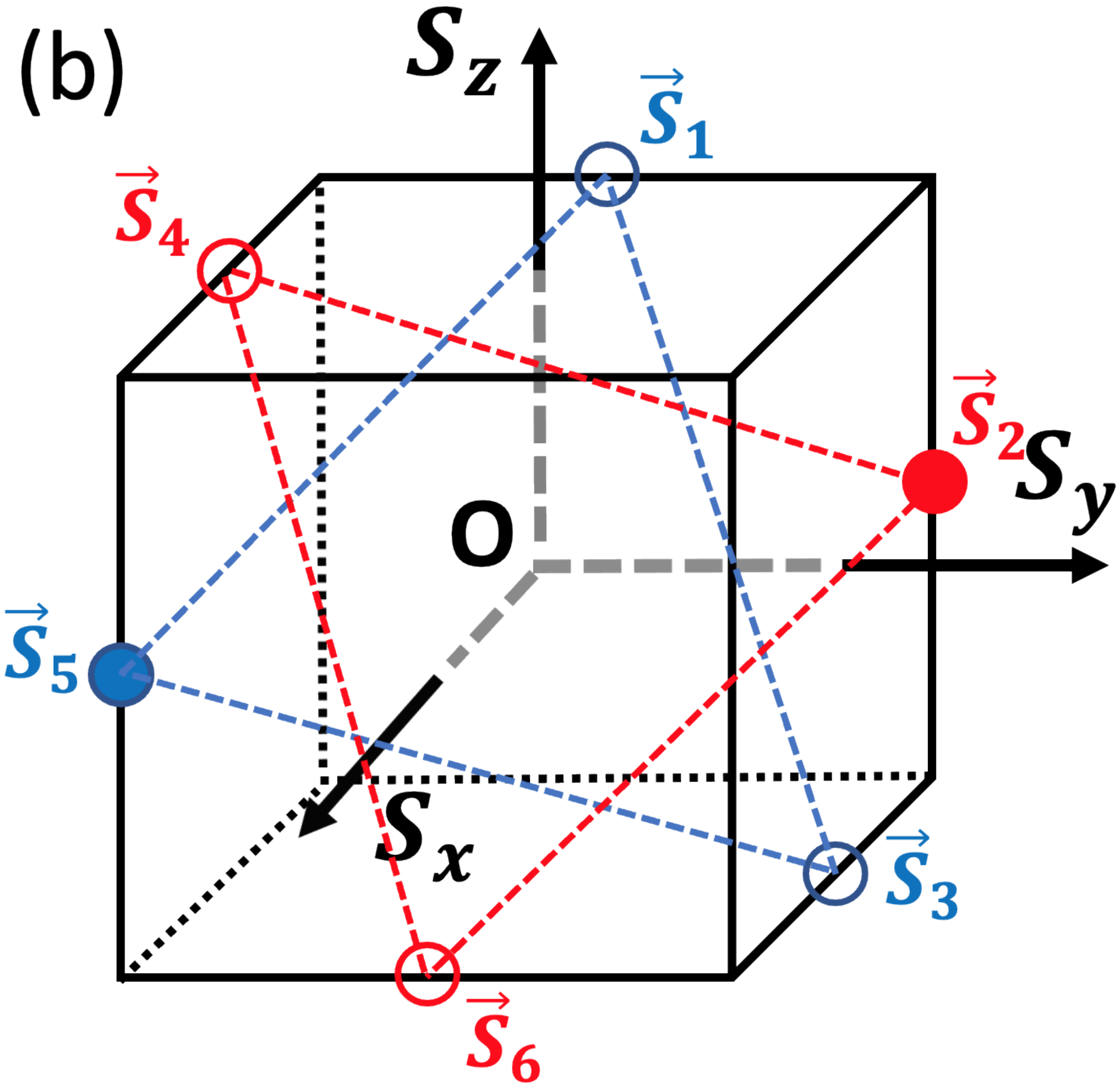}
\includegraphics[width=5cm]{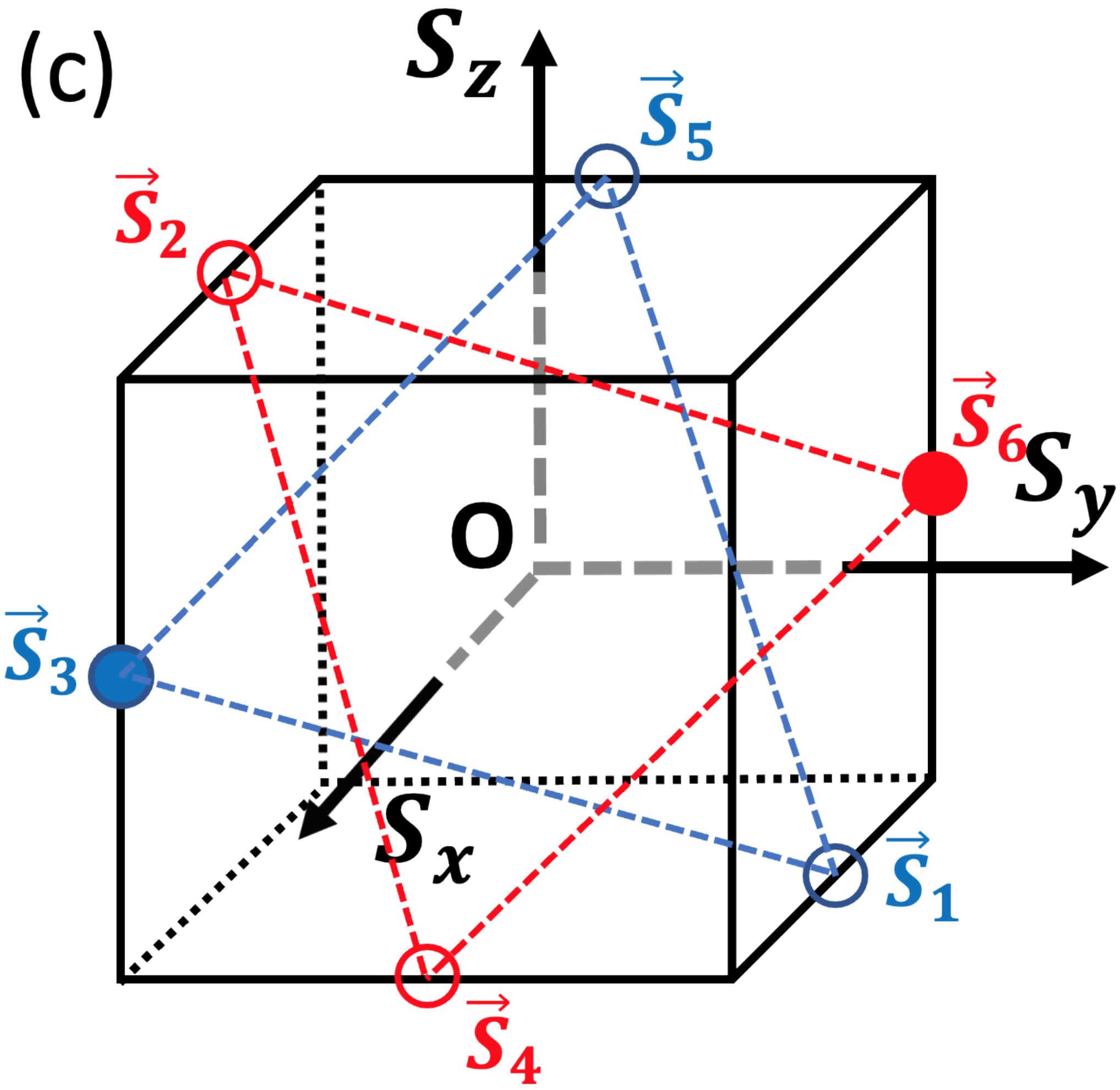}
\includegraphics[width=5cm]{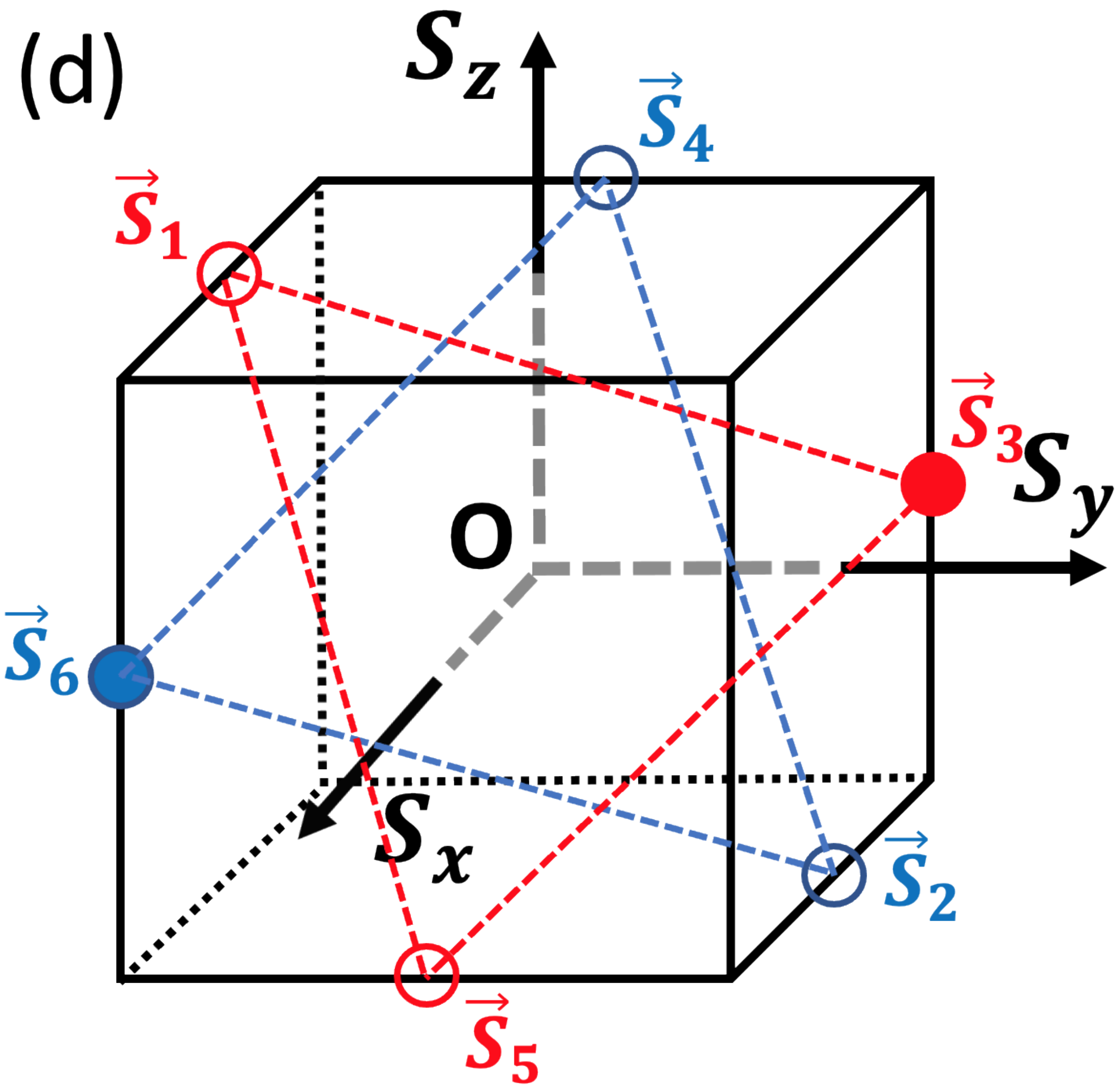}
\includegraphics[width=5cm]{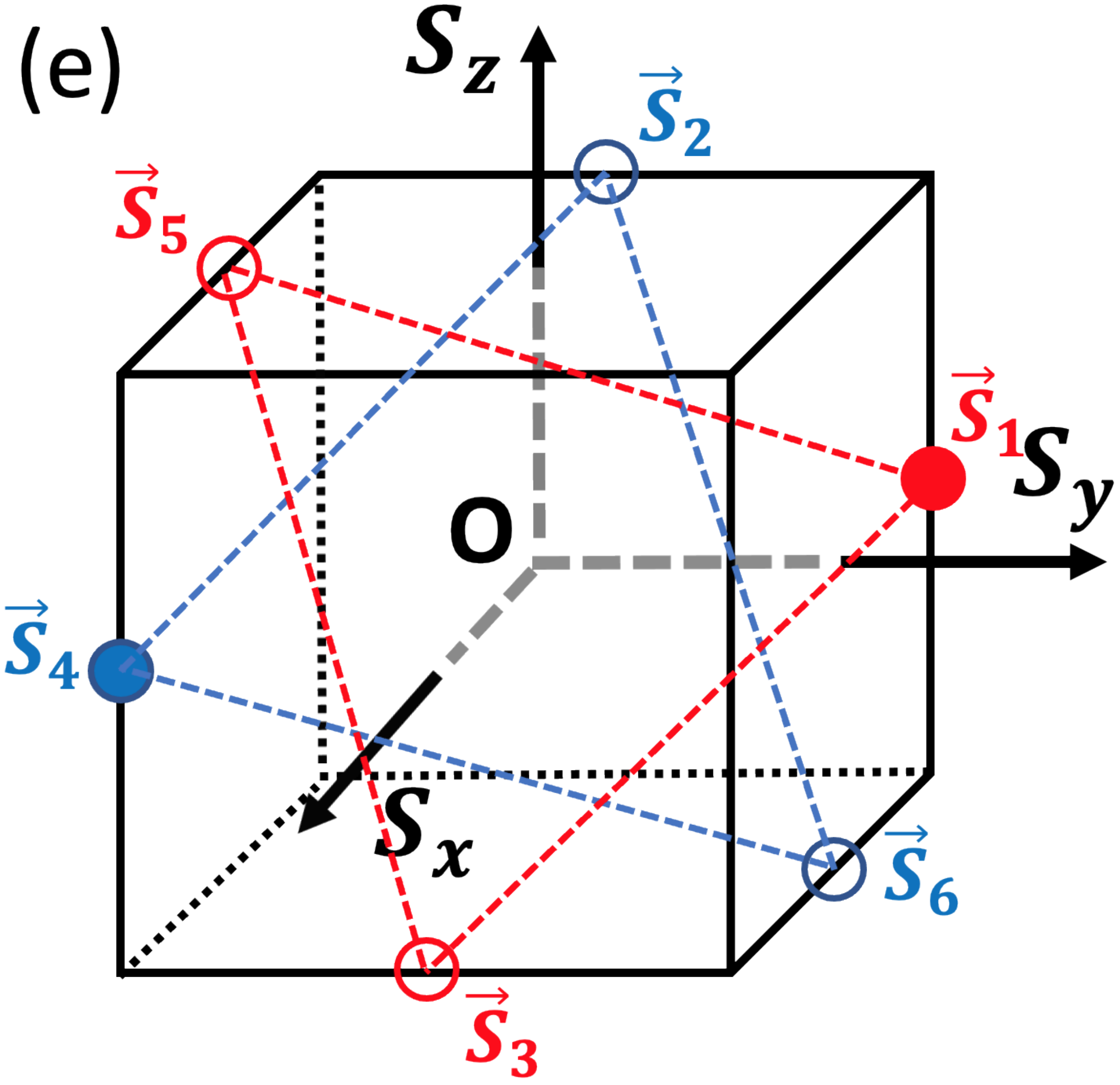}
\includegraphics[width=5cm]{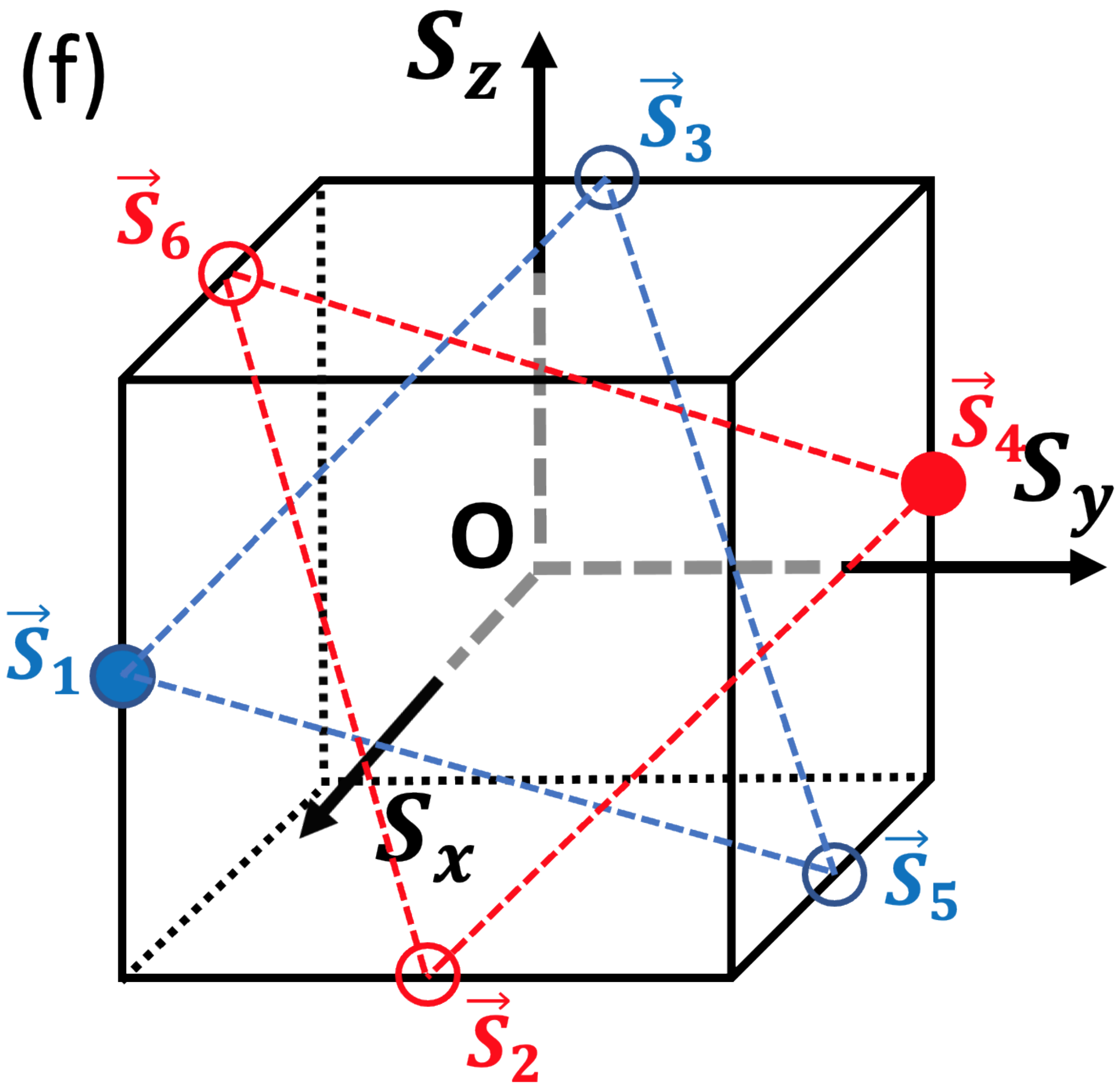}
\caption{(a-f) Directions of the spin orientations $\vec{S}_i$ within the original frame in the sublattice $i$ ($1\leq i \leq 6$) of the six-sublattice division (see Fig. \ref{fig:honeycomb} (a)) in the six degenerate symmetry breaking ground states for $g>0$ (where $g$ is defined in Eq. (\ref{eq:u})).
Subfigure (a) is the spin configuration in the original frame corresponding to the Ne\'el-$y^{\prime\prime}$ order,
and the spin configurations in (b-f) can be obtained from (a) by applying $T$, $(U_6)^{-1}R(\hat{z}^{\prime\prime},-\frac{2\pi}{3})T_aU_6$,
$T(U_6)^{-1}R(\hat{z}^{\prime\prime},-\frac{2\pi}{3})T_aU_6$,  $(U_6)^{-1}R(\hat{z}^{\prime\prime},\frac{2\pi}{3})T_{2a}U_6$, $T(U_6)^{-1}R(\hat{z}^{\prime\prime},\frac{2\pi}{3})T_{2a}U_6$, respectively.
The directions for the two solid circles are exact, whereas the remaining directions for the hollow circles are approximate.  
} \label{fig:Ny_order_original_app}
\end{center}
\end{figure}

Fig. \ref{fig:Ny_order_original_app} (a-f) show the directions of the spin orientations within the original frame in the six-degenerate symmetry breaking ground states for the $g>0$ case  (see Eq. (\ref{eq:u}) for the definition of $g$),
in which $\vec{S}_i$ represents the spin operator in sublattice $i$ of the six-sublattice division defined in Fig. \ref{fig:honeycomb} (a). 
Fig. \ref{fig:Ny_order_original_app} (a) is the spin configuration for the N\'eel-$\hat{y}^{\prime\prime}$ order,
and Fig. \ref{fig:Ny_order_original_app} (b-f) can be obtained from Fig. \ref{fig:Ny_order_original_app} (a) by applying the broken symmetries 
$T$, $(U_6)^{-1}R(\hat{z}^{\prime\prime},-\frac{2\pi}{3})T_aU_6$,
$T(U_6)^{-1}R(\hat{z}^{\prime\prime},-\frac{2\pi}{3})T_aU_6$,  $(U_6)^{-1}R(\hat{z}^{\prime\prime},\frac{2\pi}{3})T_{2a}U_6$, $T(U_6)^{-1}R(\hat{z}^{\prime\prime},\frac{2\pi}{3})T_{2a}U_6$, respectively.

\subsection{The $g<0$ case}

\begin{figure}[h]
\begin{center}
\includegraphics[width=5cm]{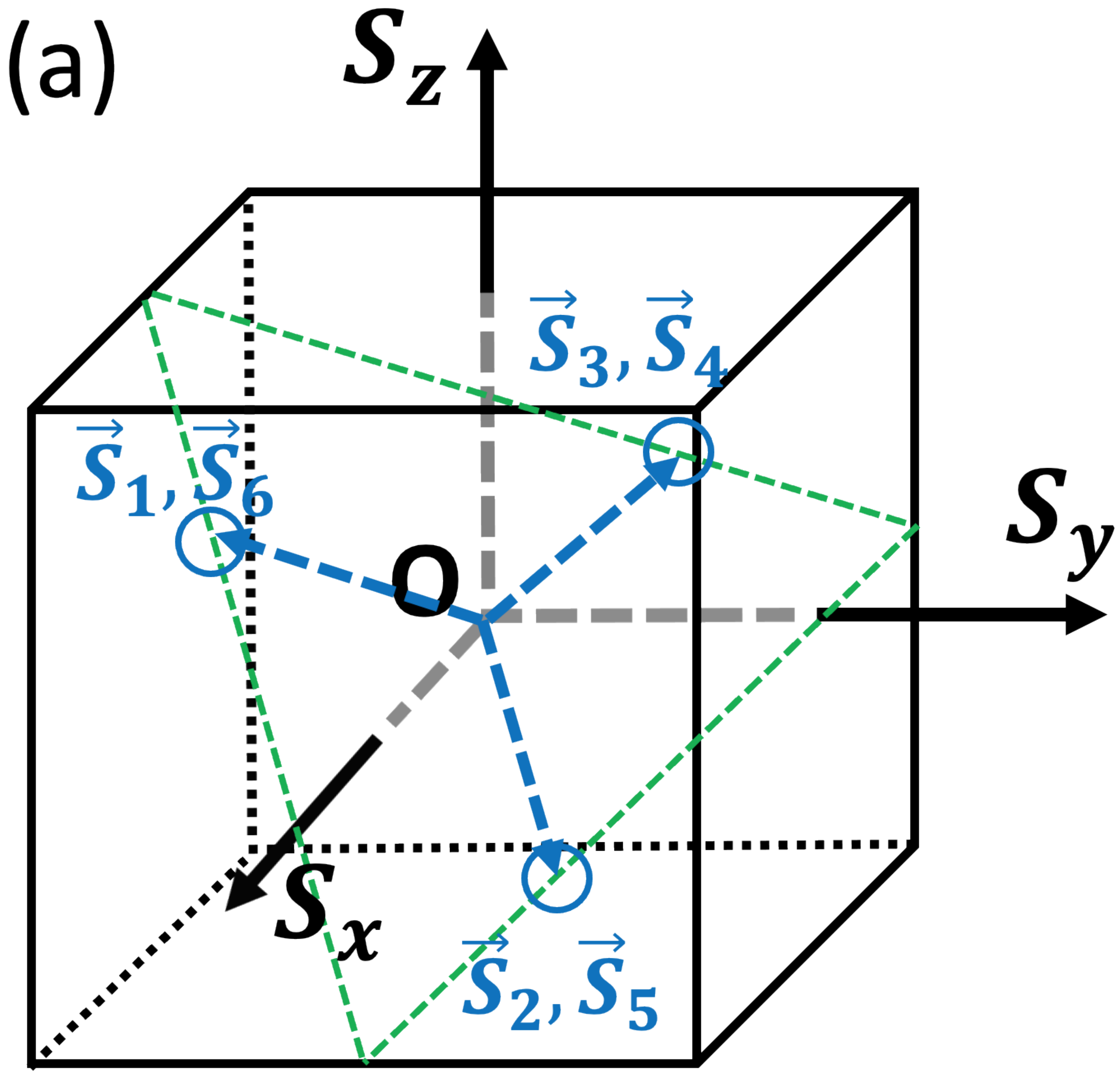}
\includegraphics[width=5cm]{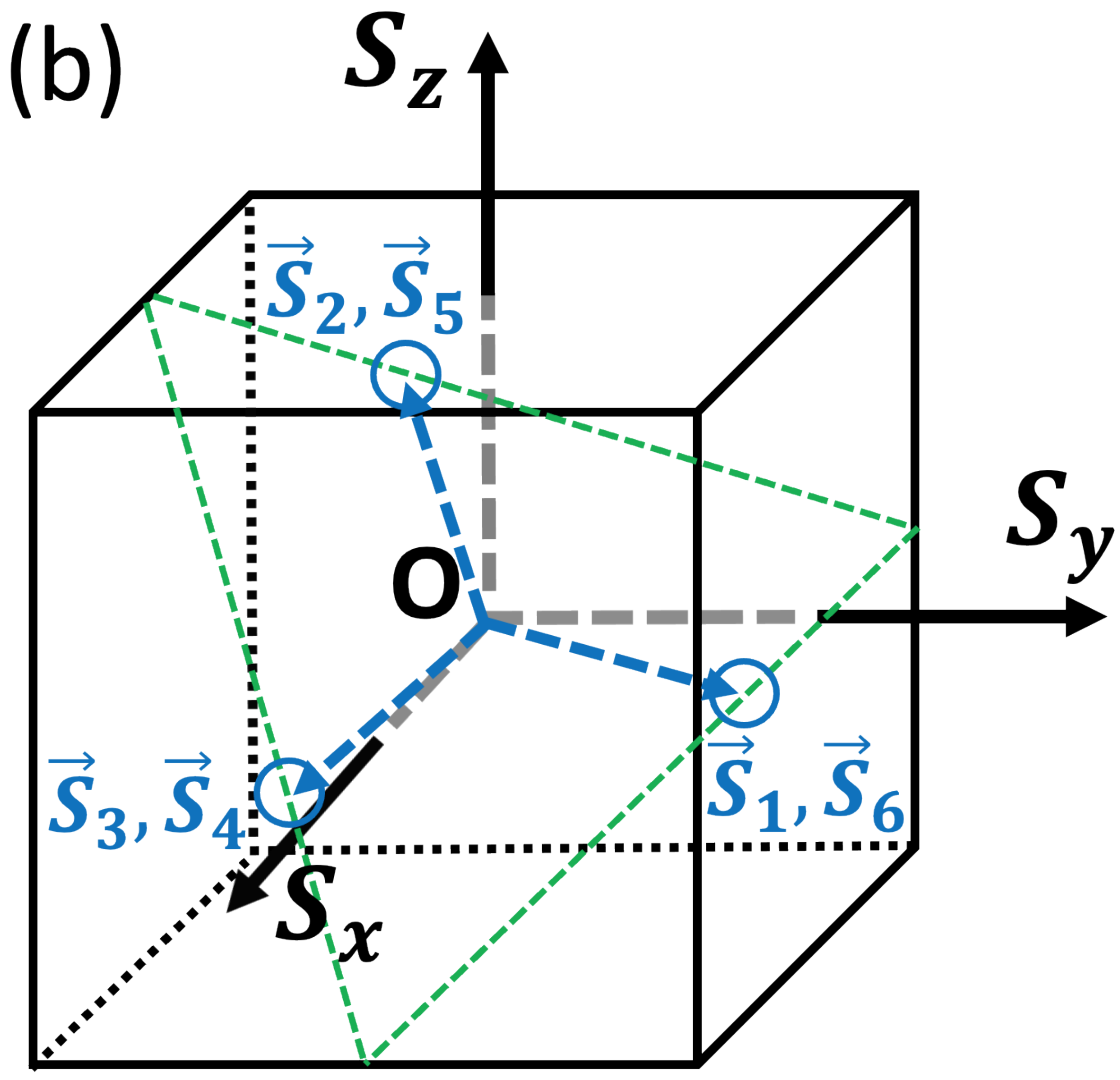}
\includegraphics[width=5cm]{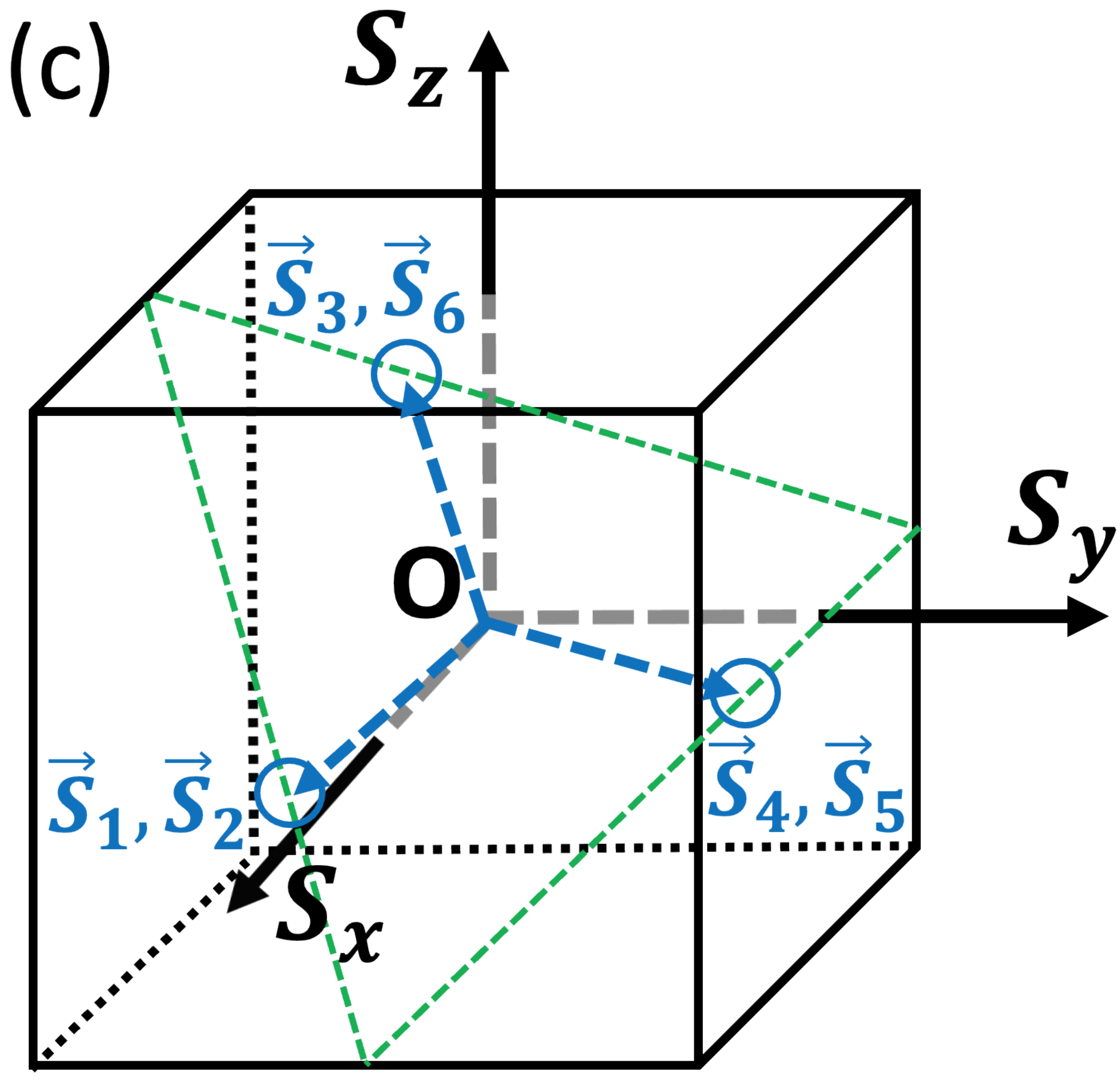}
\includegraphics[width=5cm]{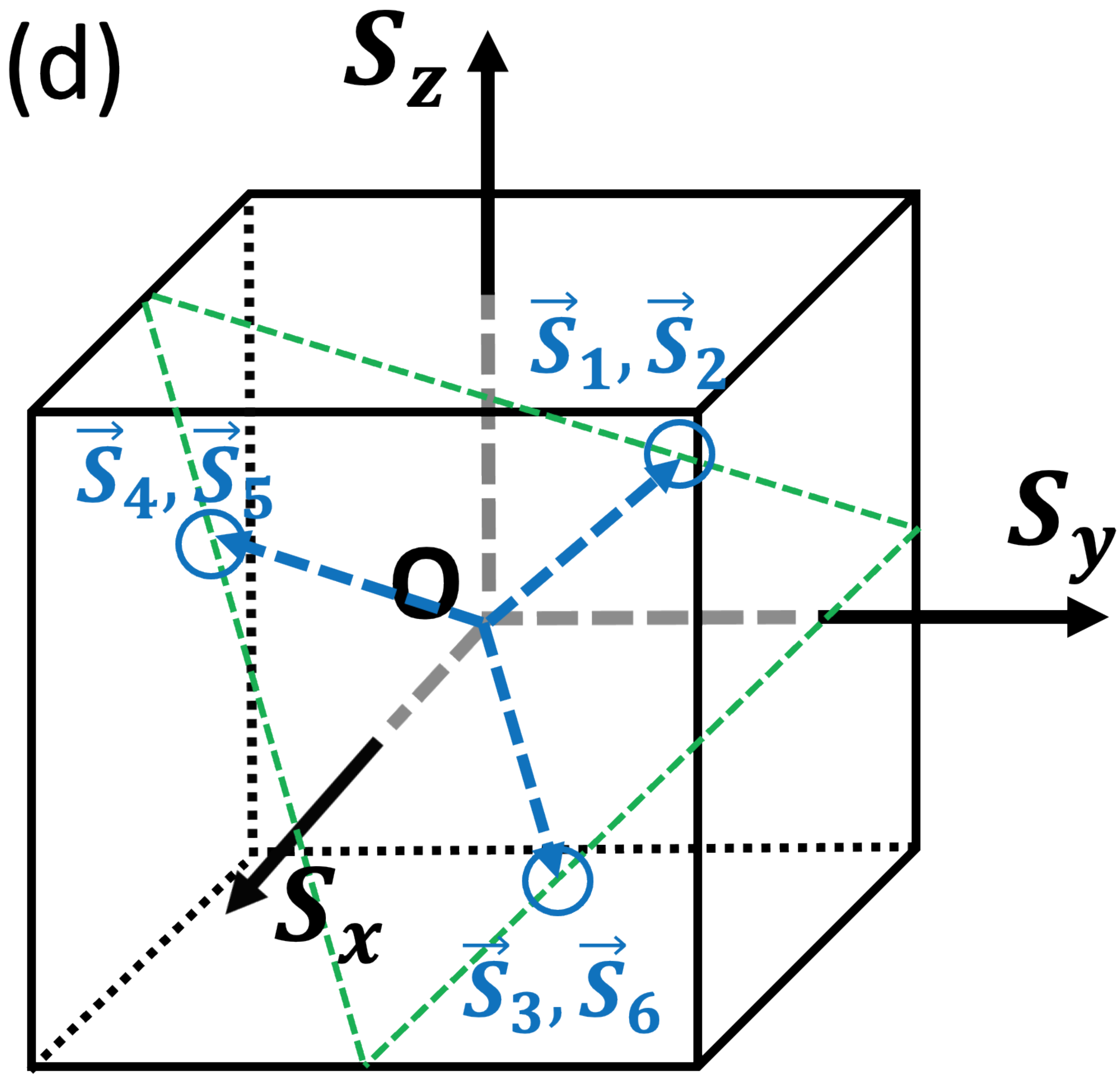}
\includegraphics[width=5cm]{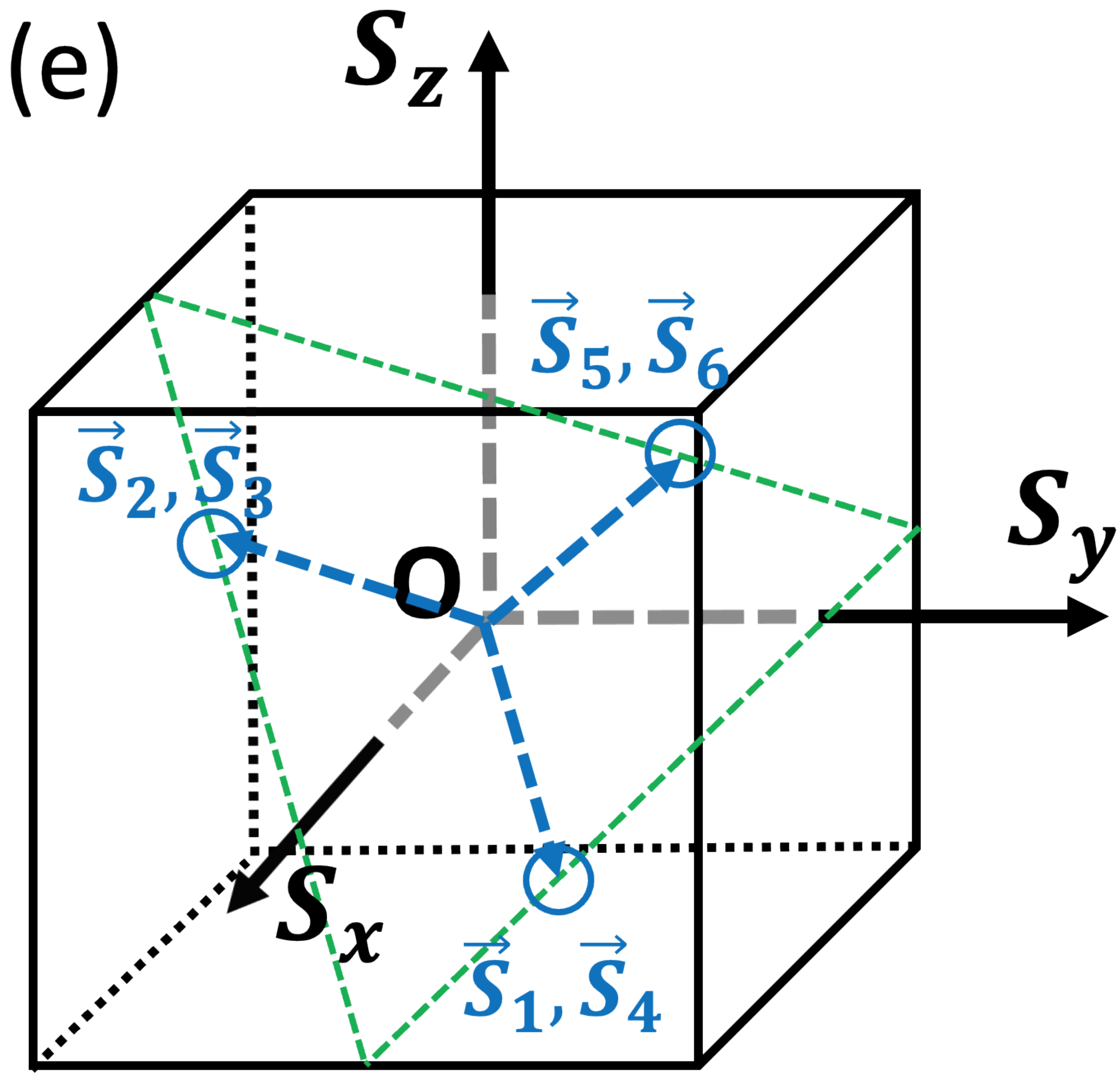}
\includegraphics[width=5cm]{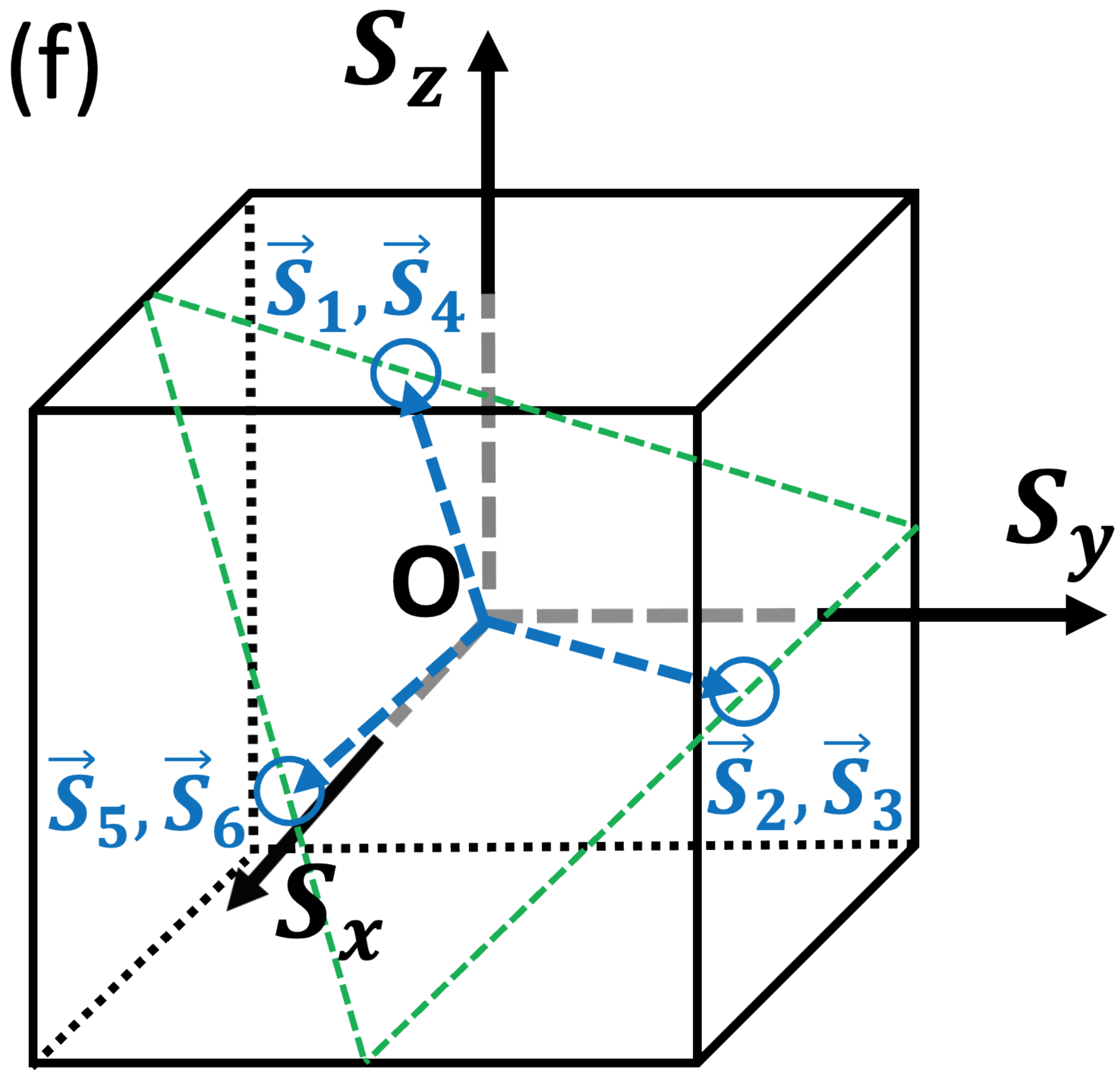}
\caption{
(a-f) Directions of the spin orientations $\vec{S}_i^{(m)}$ within the original frame in row $m$ and sublattice $i$ ($1\leq i \leq 6$) of the six-sublattice division (see Fig. \ref{fig:honeycomb} (a)) in the six degenerate symmetry breaking ground states for $g<0$ (where $g$ is defined in Eq. (\ref{eq:u})).
Subfigure (a) is the spin configuration in the original frame which corresponds to the Ne\'el-$x^{\prime\prime}$ order in the $U_6$ frame,
and the spin configurations in (b-f) can be obtained from (a) by applying $T$, $(U_6)^{-1}R(\hat{z}^{\prime\prime},-\frac{2\pi}{3})T_aU_6$,
$T(U_6)^{-1}R(\hat{z}^{\prime\prime},-\frac{2\pi}{3})T_aU_6$,  $(U_6)^{-1}R(\hat{z}^{\prime\prime},\frac{2\pi}{3})T_{2a}U_6$, $T(U_6)^{-1}R(\hat{z}^{\prime\prime},\frac{2\pi}{3})T_{2a}U_6$, respectively.
All the directions of the spins  are approximate due to the bosonization coefficients $\delta_C$, $\sigma_C$.
} \label{fig:Nx_order_original_app}
\end{center}
\end{figure}

Similarly, Fig. \ref{fig:Nx_order_original_app} (a-f) show the directions of the spin orientations within the original frame in the six-degenerate symmetry breaking ground states for the $g<0$ case (see Eq. (\ref{eq:u}) for the definition of $g$),
in which $\vec{S}_i$ represents the spin operator in sublattice $i$ of the six-sublattice division defined in Fig. \ref{fig:honeycomb} (a). 
Fig. \ref{fig:Nx_order_original_app} (a) is the spin configuration for the N\'eel-$\hat{x}^{\prime\prime}$ order,
and Fig. \ref{fig:Nx_order_original_app} (b-f) can be obtained from Fig. \ref{fig:Nx_order_original_app} (a) by applying the broken symmetries 
$T$, $(U_6)^{-1}R(\hat{z}^{\prime\prime},-\frac{2\pi}{3})T_aU_6$,
$T(U_6)^{-1}R(\hat{z}^{\prime\prime},-\frac{2\pi}{3})T_aU_6$,  $(U_6)^{-1}R(\hat{z}^{\prime\prime},\frac{2\pi}{3})T_{2a}U_6$, $T(U_6)^{-1}R(\hat{z}^{\prime\prime},\frac{2\pi}{3})T_{2a}U_6$, respectively.

\section{Counter-rotating spiral order: comparison with experimental results on $\alpha$-Li$_2$IrO$_3$} 
\label{app:compare_experiment}

In this appendix, we give the detailed derivations of Eq. (\ref{eq:experiment_Li_1}) from the experimental pattern in Eq. (B4) in  Ref. \cite{Williams2016}.

In Eq. (B4) in  Ref. \cite{Williams2016}, the pattern of counter-rotating spiral spin ordering in the real space is given by
\bea
\vec{M}_{\vec{r},n}=\mp (\hat{x}_o M_{x_o}-\hat{y}_o M_{y_o})\sin(\vec{q}\cdot \vec{r})+\hat{z}_o \cos(\vec{q}\cdot \vec{r}),
\label{eq:M_real_space}
\eea
in which the upper (lower) sign in the $\sin(\vec{q}\cdot \vec{r})$ term is for $n=1(2)$ sublattice.
The relation between the unit vectors $\{\hat{x}_o,\hat{y}_o,\hat{z}_o\}$ and the unit vectors $\{\hat{x},\hat{y},\hat{z}\}$ in the spin space is 
\bea
\hat{x}&=&\frac{1}{\sqrt{2}} (\hat{x}_o+\hat{z}_o),\nn\\
\hat{y}&=&\frac{1}{\sqrt{2}} (\hat{x}_o-\hat{z}_o),\nn\\
\hat{z}&=&\hat{y}_o.
\eea
In terms of the basis vectors $\{\hat{x},\hat{y},\hat{z}\}$, 
Eq. (\ref{eq:M_real_space}) becomes
\bea
\vec{M}_{\vec{r},n}&=& \left(\begin{array}{c}
\frac{1}{\sqrt{2}} [\mp M_{x_o}\sin(\vec{q}\cdot \vec{r}) +M_{z_o} \cos(\vec{q}\cdot \vec{r})]\\
\frac{1}{\sqrt{2}} [\mp M_{x_o}\sin(\vec{q}\cdot \vec{r}) -M_{z_o} \cos(\vec{q}\cdot \vec{r})]\\
\pm M_{y_o} \sin(\vec{q}\cdot \vec{r})
\end{array}\right).
\label{eq:M_real_space_2}
\eea

Although experiments observe a slightly incommensurate wavevector $q\sim 0.32\times 2\pi/a$, we will take the commensurate value $q=2\pi/(3a)$.
Taking $n=1$ in Eq. (\ref{eq:M_real_space_2}), and letting $\vec{r}=0$ (for site $a_1$), $\vec{r}=\vec{a}$ (for site $a_3$), $\vec{r}=2\vec{a}$ (for site $a_5$),
we obtain 
\begin{flalign}
&\vec{M}_{a_1}=\frac{1}{\sqrt{2}}(M_{z_o},-M_{z_o},0)^T,\nn\\
&\vec{M}_{a_3}=\frac{1}{2\sqrt{2}}(\sqrt{3}M_{x_o}-M_{z_o},\sqrt{3}M_{x_o}+M_{z_o},-\sqrt{6}M_{y_o} )^T,\nn\\
&\vec{M}_{a_5}=\frac{1}{2\sqrt{2}}(-\sqrt{3}M_{x_o}-M_{z_o},-\sqrt{3}M_{x_o}+M_{z_o},\sqrt{6}M_{y_o} )^T.
\label{eq:M_real_space_3a}
\end{flalign}
A comparison with Eq. (\ref{eq:original_Ny_order_1_B}) indicates that $a_1$ represents a site in an even row (i.e., $m\in 2\mathbb{Z}$ in Eq. (\ref{eq:original_Ny_order_1_B})) with sublattice index equal to $5$. 
According to Fig. \ref{fig:honeycomb} (a), $\vec{M}_{a_1}$ corresponds to $\vec{S}_{c,3}$ in Fig. \ref{fig:honeycomb} (b).
Therefore, Eq. (\ref{eq:M_real_space_3a}) becomes
\begin{flalign}
&\vec{S}_{c,3}=\frac{1}{\sqrt{2}}(M_{z_o},-M_{z_o},0)^T,\nn\\
&\vec{S}_{c,5}=\frac{1}{2\sqrt{2}}(\sqrt{3}M_{x_o}-M_{z_o},\sqrt{3}M_{x_o}+M_{z_o},-\sqrt{6}M_{y_o} )^T,\nn\\
&\vec{S}_{c,7}=\frac{1}{2\sqrt{2}}(-\sqrt{3}M_{x_o}-M_{z_o},-\sqrt{3}M_{x_o}+M_{z_o},\sqrt{6}M_{y_o} )^T.
\label{eq:M_real_space_3}
\end{flalign}

Next we consider $n=2$ in Eq. (\ref{eq:M_real_space}), which corresponds to the site located at row $d$, column $3$ in  Fig. \ref{fig:honeycomb} (b).
Taking minus sign in Eq. (\ref{eq:M_real_space_2}), and identifying $\vec{r}=0$ with $\vec{S}_{d,3}$, $\vec{r}=\vec{a}$ with $\vec{S}_{d,5}$, $\vec{r}=2\vec{a}$ with $\vec{S}_{d,7}$, we obtain
\begin{flalign}
&\vec{S}_{d,3}=\frac{1}{\sqrt{2}}(M_{z_o},-M_{z_o},0)^T,\nn\\
&\vec{S}_{d,5}=\frac{1}{2\sqrt{2}}(-\sqrt{3}M_{x_o}-M_{z_o},-\sqrt{3}M_{x_o}+M_{z_o},\sqrt{6}M_{y_o} )^T,\nn\\
&\vec{S}_{d,7}=\frac{1}{2\sqrt{2}}(\sqrt{3}M_{x_o}-M_{z_o},\sqrt{3}M_{x_o}+M_{z_o},-\sqrt{6}M_{y_o} )^T.
\label{eq:M_real_space_4}
\end{flalign}
On the other hand, according to Fig. \ref{fig:honeycomb} (a), $\vec{S}_{d,3}$, $\vec{S}_{d,5}$, $\vec{S}_{d,7}$
are identified with $\vec{S}^{(\text{odd})}_2$, $\vec{S}^{(\text{odd})}_4$, $\vec{S}^{(\text{odd})}_6$ in Eq. (\ref{eq:original_Ny_order_1_B}),
in which the upper index ``$(\text{odd})$" means that this is an odd row, i.e., $m\in 2\mathbb{Z}+1$ in Eq. (\ref{eq:original_Ny_order_1_B}).

Now we can use Eq. (\ref{eq:M_real_space_4}) to obtain the spin orderings for $\vec{S}^{(c)}_4$, $\vec{S}^{(c)}_6$, $\vec{S}^{(c)}_8$ in row $c$,
which have sublattice indices in the six-sublattice division as $6$, $2$, $4$, respectively.
Due to the $(-)^m$ sign in Eq. (\ref{eq:original_Ny_order_1_B}), we need to flip the signs in Eq. (\ref{eq:M_real_space_4}), which give
\begin{flalign}
&\vec{S}_{c,6}=\frac{1}{\sqrt{2}}(-M_{z_o},M_{z_o},0)^T,\nn\\
&\vec{S}_{c,8}=\frac{1}{2\sqrt{2}}(\sqrt{3}M_{x_o}+M_{z_o},\sqrt{3}M_{x_o}-M_{z_o},-\sqrt{6}M_{y_o} )^T,\nn\\
&\vec{S}_{c,4}=\frac{1}{2\sqrt{2}}(-\sqrt{3}M_{x_o}+M_{z_o},-\sqrt{3}M_{x_o}-M_{z_o},\sqrt{6}M_{y_o} )^T.
\label{eq:M_real_space_5}
\end{flalign}

Then clearly, Eq. (\ref{eq:experiment_Li_1}) in the main text can be obtained from Eq. (\ref{eq:M_real_space_3}) and Eq. (\ref{eq:M_real_space_5}) via the following identifications,
\bea
\vec{S}_{c,3}\rightarrow \vec{S}_{a_1},~\vec{S}_{c,4}\rightarrow \vec{S}_{a_2},~\vec{S}_{c,5}\rightarrow \vec{S}_{a_3},~
\vec{S}_{c,6}\rightarrow \vec{S}_{a_4},~\vec{S}_{c,7}\rightarrow \vec{S}_{a_5},~\vec{S}_{c,8}\rightarrow \vec{S}_{a_6}.
\eea

\end{widetext}

\end{document}